\documentclass[apj,twocolumn,appendixfloats]{openjournal}
\usepackage{natbib}

\usepackage{graphicx}	
\usepackage{amsmath}	
\usepackage{amssymb}	
\usepackage{bm}		
\usepackage{soul}
\usepackage{enumitem}
\usepackage[dvipsnames]{xcolor}
\usepackage[normalem]{ulem}
\usepackage{aas_macros} 
\usepackage{hyperref}	
\hypersetup{colorlinks=true,linkcolor=blue,citecolor=blue,filecolor=blue,urlcolor=blue}

\newcommand{\Mpc}{\mathrm{Mpc}}
\newcommand{\Myr}{\mathrm{Myr}}
\newcommand{\Gyr}{\mathrm{Gyr}}

\newcommand{\pc}{\mathrm{pc}}
\newcommand{\kmpers}{\mathrm{km} \, \mathrm{s}^{-1}}

\newcommand{\Msol}{\textup{M}_\mathrm{\sun}}
\newcommand{\Zsol}{\textup{Z}_\mathrm{\sun}}

\newcommand{\K}{\mathrm{K}}
\newcommand{\xMsol}[2]{\ensuremath{{#1}\times 10^{#2} \,\Msol}}
\newcommand{\xScientific}[2]{\ensuremath{{#1} \times 10^{#2}}}

\newcommand{\mdm}{m_{\mathrm{DM}}}
\newcommand{\softening}{\Delta x}
\newcommand{\epsilonff}{\epsilon_{\text{ff}}}

\newcommand{\Mvir}{M_{200}}

\newcommand{\Mstar}{M_{\star}}
\newcommand{\rvir}{r_{200}}
\newcommand{\magv}{\mathcal{M}_V}

\newcommand{\feh}{[\rm Fe / H]}
\newcommand{\cfe}{[\rm C / Fe]}
\newcommand{\mgfe}{[\rm Mg / Fe]}
\newcommand{\averagefeh}{\langle [\rm Fe / H] \rangle}

\newcommand{\hi}{\text{H}\,\textsc{i}}
\newcommand{\hmol}{\mathrm{H}_\mathrm{2}}

\graphicspath{{./}{figs/}}

\begin{document}
\title[Plateau in MZR]{MEGATRON: how the first stars can create an iron metallicity plateau\\in the smallest dwarf galaxies \vspace{-17mm}}
\author{Martin P. Rey$^{1*}$, Harley Katz$^{2,3}$, Corentin Cadiou$^4$, \\
Mahsa Sanati$^5$,
Oscar Agertz$^6$, Jeremy Blaizot$^{7}$, Alex J. Cameron$^{5}$, Nicholas Choustikov$^{5}$, Julien Devriendt$^{5}$, Uliana Hauk$^{2}$, Alexander P. Ji$^{2,3}$, Gareth C. Jones$^{8,9}$, Taysun Kimm$^{10}$, Isaac Laseter$^{11}$, Sergio Martin-Alvarez$^{12}$, Kosei Matsumoto$^{13}$, Autumn Pearce$^{2}$, Yves Revaz$^{14}$, Francisco Rodríguez Montero$^{2,3}$, Joki Rosdahl$^{7}$, Aayush Saxena$^{5}$, Adrianne Slyz$^{5}$, Richard Stiskalek$^{5}$, Anatole Storck$^{5}$, Oscar Veenema$^{5}$, Wonjae Yee$^{2}$}
\thanks{$^*$E-mail: \href{mpr47@bath.ac.uk}{mpr47@bath.ac.uk}}

\affiliation{$^{1}$University of Bath, Department of Physics, Claverton Down, Bath, BA2 7AY, UK}
\affiliation{$^{2}$Department of Astronomy \& Astrophysics, University of Chicago, 5640 S Ellis Avenue, Chicago, IL 60637, USA}
\affiliation{$^{3}$ Kavli Institute for Cosmological Physics, University of Chicago, Chicago IL 60637, USA}
\affiliation{$^{4}$Institut d’Astrophysique de Paris, Sorbonne Universités, CNRS, UMR 7095, 98 bis bd Arago, 75014 Paris, France}
\affiliation{$^{5}$Sub-department of Astrophysics, University of Oxford, Keble Road, Oxford OX1 3RH, United Kingdom}
\affiliation{$^{6}$Lund Observatory, Division of Astrophysics, Department of Physics, Lund University, Box 43, SE-221 00 Lund, Sweden}
\affiliation{$^{7}$ Université Claude Bernard Lyon 1, CRAL UMR5574, ENS de Lyon, CNRS, Villeurbanne, F-69622, France}
\affiliation{$^{8}$Kavli Institute for Cosmology, University of Cambridge, Madingley Road, Cambridge CB3 0HA, UK}
\affiliation{$^{9}$Cavendish Laboratory, University of Cambridge, 19 JJ Thomson Avenue, Cambridge CB3 0HE, UK}
\affiliation{$^{10}$Department of Astronomy, Yonsei University, 50 Yonsei-ro, Seodaemun-gu, Seoul 03722, Republic of Korea}
\affiliation{$^{11}$Department of Astronomy, University of Wisconsin-Madison, Madison, WI 53706, USA}
\affiliation{$^{12}$Kavli Institute for Particle Astrophysics \& Cosmology (KIPAC), Stanford University, Stanford, CA 94305, USA}
\affiliation{$^{13}$Sterrenkundig Observatorium Department of Physics and Astronomy Universiteit Gent, Krijgslaan 281 S9, B-9000 Gent, Belgium}
\affiliation{$^{14}$Institute of Physics, Laboratoire d’Astrophysique, École Polytechnique Fédérale de Lausanne (EPFL), CH-1015 Lausanne, Switzerland}

\begin{abstract}  
We study the stellar mass-iron metallicity relation of dwarf galaxies in the new high-resolution \textsc{megatron} cosmological radiation-hydrodynamics simulations. These simulations model galaxy formation up to $z\approx8$ in a region that will collapse into a Milky-Way-like galaxy at $z=0$, while self-consistently tracking Population III and II (Pop.~III, Pop.~II) star formation, feedback and chemical enrichment. \textsc{megatron} dwarf galaxies are in excellent agreement with the observed stellar mass-metallicity relation at $z=0$, including an over-abundance of dwarfs along a flat plateau in metallicity ($\langle [\rm{Fe}/\rm{H}] \rangle \approx -2.5$) at low stellar masses ($M_{\star} \leq 10^5 \, \rm{M}_{\odot}$). We tie this feature to the chemical enrichment of dwarf galaxies by Pop.~III pair-instability supernova (PISN) explosions. The strong Lyman-Werner background (LW) from the protogalaxy ensures that PISNe occur in haloes massive enough ($\approx 10^7\, \rm{M}_{\odot}$) to retain their ejecta. We also predict a tail of $\approx 20\%$ of iron-deficient ($\langle [\rm{Fe}/\rm{H}] \rangle \leq - 3$) dwarf galaxies. We show that both plateau and tail (i) are robust to large variations in Pop.~II feedback assumptions, and (ii) survive in bound satellites surrounding the central galaxy at $z=0$. 
\end{abstract}




\section{Introduction} \label{sec:intro}

Characterising how the first generation of stars enriched the pristine Universe with chemical elements remains a key question of modern astrophysics. The advent of the \textit{James Webb Space Telescope} (JWST) now allows us to observe this process in-situ. With its new infrared spectroscopic capabilities, we can observe emission lines originating in the interstellar medium (ISM) of high-redshift galaxies. This allows us to, for example, constrain the relationship between stellar mass and gas-phase oxygen abundance at $z\geq6$ (e.g.\ \citealt{Curti2023Te, Heintz2023a, Curti2024LowMassEnd, Laseter2024Calibration, Sanders2023DirectTe, Li2025MZR, Scholte2025MZR}), as well as the relative abundances between individual chemical elements (e.g.\ \citealt{Cameron2023GNz11, Isobe2023LowC/N, Stiavelli2023, Hsiao2024Carbon, Schaerer2024, Stanton2025, Topping2024NEmitterSample, Topping2024NEmitterSample2, Nakane2025}).

The same star formation episodes that spawn the short-lived ($\approx 10 \, \Myr$) massive stars which power the observed emission lines at $z\geq6$ also create long-lived ($\geq 10 \, \Gyr$) low-mass stars that survive to the present day. The chemistry of these metal-poor stars keeps a record of the gas conditions in which they formed, providing us with a complementary view of the chemical enrichment history of the Universe at high redshift. We have long been able to identify and characterize these remnant metal-poor stars in the Milky Way and nearby satellite galaxies (e.g.\ \citealt{Tolstoy2009, Frebel2015, Bonifacio2025} for reviews). In turn, the chemistry of $z=0$ metal-poor stars has allowed us to, for example, constrain the mass function of primordial Population III (Pop.~III) stars (e.g.\ \citealt{Hartwig2015, Ishigaki2018, Jiang2024b}) or the yields and explosion properties of metal-poor massive stars (e.g.\ \citealt{Kobayashi2014, Koutsouridou2023a}). 

However, the connection between the observed spectroscopic properties of $z>6$ galaxies and the local metal-poor stars is challenging. Not only because of the $\geq 13$ billion years gap in cosmic evolution between the two regimes, but also because the elements accessible to galaxy and stellar spectroscopy tend to differ (e.g.\ oxygen in the former, iron in the latter). Furthermore, metal-poor stars can only be resolved in the immediate vicinity of the Milky Way, providing us with an important, but biased, view of low-metallicity chemical enrichment in one specific environment. How to generalize Milky-Way data to the high-redshift Universe and vice-versa remains an open question. 

In this paper, we introduce the \textsc{megatron} project, which aims to clarify this connection between high-redshift emission lines and low-redshift metal-poor stars. 

This paper initiates this effort by focusing on the properties of ultra-faint dwarf galaxies (UFDs; $\magv \geq -6$; $\Mstar \leq 10^5 \, \Msol$ at $z=0$). These galaxies populate the faintest end of the galaxy luminosity function and are the most iron-poor galaxies known at $z=0$ ($\averagefeh \leq -2$; see \citealt{Simon2019} for a review). Their star formation histories almost invariably favour a high-redshift formation, with limited or non-existent star formation after $z\leq 4$ (e.g.\ \citealt{Okamoto2012, Brown2014, Weisz2014Reionization, Savino2025M31SFHs, Durbin2025}). This truncation in star formation is usually attributed to the low dynamical masses of UFDs. Once the cosmic ultraviolet (UV) background following cosmic reionization has heated the intergalactic medium (IGM), their gravitational potential wells are too shallow to accrete fresh gas from the now-warm IGM (\citealt{Efstathiou1992, Shapiro1994, Gnedin2000, Noh2014}), terminating star formation activity (e.g.\ \citealt{Bullock2000, Benson2002,Somerville2002}). As a result, UFDs are `fossil relics' of the high-redshift Universe (e.g.\ discussions in \citealt{Ricotti2005, Bovill2009, Salvadori2009, Bland-Hawthorn2015}), with their lack of star formation after $z\approx 3$\footnote{Reionization quenching is not instantaneous, and residual star formation can persist after reionization until galactic outflows have expelled self-shielded gas within haloes (e.g.\ \citealt{Susa2004, Onorbe2015, Rey2020})} making them ideal to isolate and constrain metal-free and metal-poor stellar evolution with observed abundances (e.g.\ \citealt{Magg2018, Rossi2021, Rossi2023, Rossi2025, Chiti2025}). 

In particular, the iron metallicity of UFDs is a sensitive probe of high-redshift star formation and feedback conditions. Observed UFDs around the Milky Way show a break in slope in their mass-metallicity relation\footnote{Hereafter, the mass-metallicity relation refers to the relationship between $\Mstar$ and stellar-based metallicities $\averagefeh$.}. All UFDs with $\Mstar \leq 10^5 \, \Msol$ cluster around $\averagefeh \approx -2.5$, in a flat `plateau' as stellar masses decrease to $\Mstar \leq 10^2 \, \Msol$ (e.g.\ \citealt{Simon2019, Fu2023}). This plateau has been a challenge for recent models of UFDs, which struggle to reproduce the comparatively-high iron content of UFDs given their very low stellar masses (e.g.\ \citealt{Munshi2019, Wheeler2019, Applebaum2021, Sanati2023, Go2025}). The origin of this plateau is debated, with studies showing that it could stem from modifications of the Pop.~II IMF (\citealt{Prgomet2022}), Pop.~III physics (e.g.\ \citealt{Jeon2017}), a threshold metallicity in the intergalactic medium set by external enrichment (e.g.\ \citealt{Jeon2017, Ahvazi2024}, though see \citealt{Wheeler2025IGMPlateau}), or the details of how galactic outflows transport iron out of the ISM of dwarf galaxies (e.g.\ \citealt{Agertz2020EDGE, Rey2025}).

In this paper, we show that a plateau in iron metallicity is naturally explained from internal chemical enrichment following the birth of high-mass Pop.~III stars. To demonstrate this, we leverage the population of faint dwarf galaxies formed in the new suite of high-redshift \textsc{megatron} simulations (\citealt{Katz2025MegP1}). 

The novelty of the \textsc{megatron} simulations is multifold: 
\begin{itemize}
  \item We solve for both radiative transfer and non-equilibrium chemistry of $\geq 80$ primordial species, molecules, and  metal ions. This allows us to make \textit{ab-initio}, robust predictions of emission lines and absorption lines (\citealt{Katz2025MegP1, Cadiou2025MegP1}), in turn allowing a direct quantification of how these observables respond to changes in feedback modelling (e.g.\ \citealt{Choustikov2025MegP1}).
  \item We model small enough spatial and mass scales to resolve most Pop.~III star formation sites in the smallest first galaxies (dark matter particle mass $\mdm = \xMsol{2.5}{4}$, spatial resolution $\softening \approx 3 \, \pc$), allowing us to quantify their birth properties and observables (\citealt{Storck2025MegP1})
  \item The simulation volume is chosen to be a proto-Milky-Way-like environment, with the final progenitor reaching a halo mass $\approx 10^{12} \, \Msol$ at $z=0$. These initial conditions form a large gas and stellar disc ($\Mstar \approx 5 \times 10^{10} \, \Msol$) when evolved to $z=0$ with well-tested low-redshift models (\citealt{Rey2022VintergatanGM, Joshi2025WholeSuite}). Leveraging this last aspect to bridge high- and low-redshift observables of early chemical enrichment is the focus of this paper.
\end{itemize}

We present the \textsc{megatron} simulations and our galaxy formation model in Section~\ref{sec:methods}. We then showcase the mass-metallicity relation of the population of simulated \textsc{megatron} dwarf galaxies ($\geq 500$) in Section~\ref{sec:mstar-feh}, highlighting the emergence of a plateau around $\averagefeh \approx -2.5$ at low stellar masses ($\Mstar \leq 10^5 \, \Msol$) in excellent agreement with local Universe observations. We attribute the origin of this plateau to Pop.~III physics in Section~\ref{sec:popiii} and justify the connection between high-redshift simulations and low-redshift data in Section~\ref{sec:lowzconnection}. We conclude in Section~\ref{sec:conclusion}.

\section{The Megatron high-redshift suite} \label{sec:methods}

We use the high-redshift \textsc{megatron} simulations that are evolved to $z\approx8$ (\citealt{Katz2025MegP1}). This suite consists of four simulations of the same initial conditions (ICs), each with a different Pop.~II star formation and feedback model (but identical Pop.~III assumptions). We describe the ICs and their link to the Milky Way in Section~\ref{sec:methods:ic}, and summarize the \textsc{megatron} galaxy formation model in Section~\ref{sec:methods:megatron} (see \citealt{Katz2024MegPilot,Katz2025MegP1} for a more in-depth description)

\subsection{Proto-Milky Way initial conditions} \label{sec:methods:ic}

The \textsc{megatron} suite is based on the ICs of the \textsc{vintergatan-gm} suite of hydrodynamical simulations (\citealt{Rey2022VintergatanGM, Joshi2025WholeSuite}). These initial conditions are selected to sample a range of Milky-Way-mass haloes ($\Mvir (z=0) \approx 10^{12} \, \Msol$), where $\Mvir$ defines the mass enclosed within the radius $\rvir$ where the density equals 200 times the critical density of the Universe. 

Out of the library of \textsc{vintergatan-gm} ICs, \textsc{megatron} uses an early-forming Milky-Way-mass halo (introduced as `Halo 599' in \citealt{Rey2022}). The mass accretion history is then modified using the `genetic modification' technique (\citealt{Roth2016, Rey2018, Stopyra2021}). First, we genetically modify the ICs to zero the velocity of the Lagrangian patch with respect to the grid. This enhances the precision of numerical integration with a grid code by reducing advection errors (\citealt{Pontzen2021}). Then, we engineer two genetically-modified variants to have more or less high-redshift substructures using quadratic modifications (\citealt{Rey2018}), but always maintaining a very similar mass growth between $z=2$ and $z=0$ (see \citealt{Katz2025MegP1}, fig.\ 1).  

To benchmark these ICs, the reference, unmodified formation history is evolved with two galaxy formation models that are well tested at $z=0$: the \textsc{vintergatan} model (\citealt{Agertz2020Vintergatan}) and the \textsc{IllustrisTNG} model (\citealt{Pillepich2018TNGModel}). With both models, the \textsc{megatron} ICs lead to a large disc galaxy at $z=0$ ($\Mstar = 3-7 \times 10^{10} \Msol$) hosted in a Milky-Way-mass halo (\citealt{Joshi2025WholeSuite}, fig.2, left column for a visual). These low-redshift runs serve two purposes: (i) they provide us with well-defined benchmarks on which to compare our high-redshift galaxy formation efficiencies for which data is currently sparse (see discussion in \citealt{Katz2025MegP1}); and (ii) they strongly establish that \textsc{megatron} is studying high-redshift galaxy formation in an environment that represents a reasonable realization of the proto-Milky Way.  

The \textsc{megatron} initial conditions currently use the same transfer function for dark matter and gas, i.e.\ neglecting the anti-correlation between the dark matter and gas density fields and the relative velocities between them (e.g.\ \citealt{Tseliakhovich2010}). Both these effects can have significant impact on high-redshift star formation. The density anti-correlation is expected to impact high-redshift gas fractions at the $\approx 10\%$ level (\citealt{Jessop2026}), but is likely trumped by the much larger uncertainties in the modelling of star formation and feedback at these redshifts. Streaming velocities between dark matter and gas can significantly reduce gas fractions in high-redshift mini-haloes and raise the critical halo mass necessary to host star formation (e.g.\ \citealt{Greif2011, Stacy2011, OLeary2012, Naoz2012, Naoz2013, Schauer2019, Schauer2022}). Given their large-scale coherence, they could be applied as a constant velocity offset across the volume. But the magnitude of this offset is sourced by modes on scales much larger ($\gtrsim 100 \, \Mpc$) than the zoom region (or even parent volume) of \textsc{megatron}, making a self-consistent implementation of these effects in our simulations challenging. We thus leave to future work an exploration of how different transfer functions for dark matter and gas could impact our results.

\begin{figure*}
  \centering
    \includegraphics[width=\textwidth, trim={0.5cm 3.5cm 0.5cm 1cm},clip]{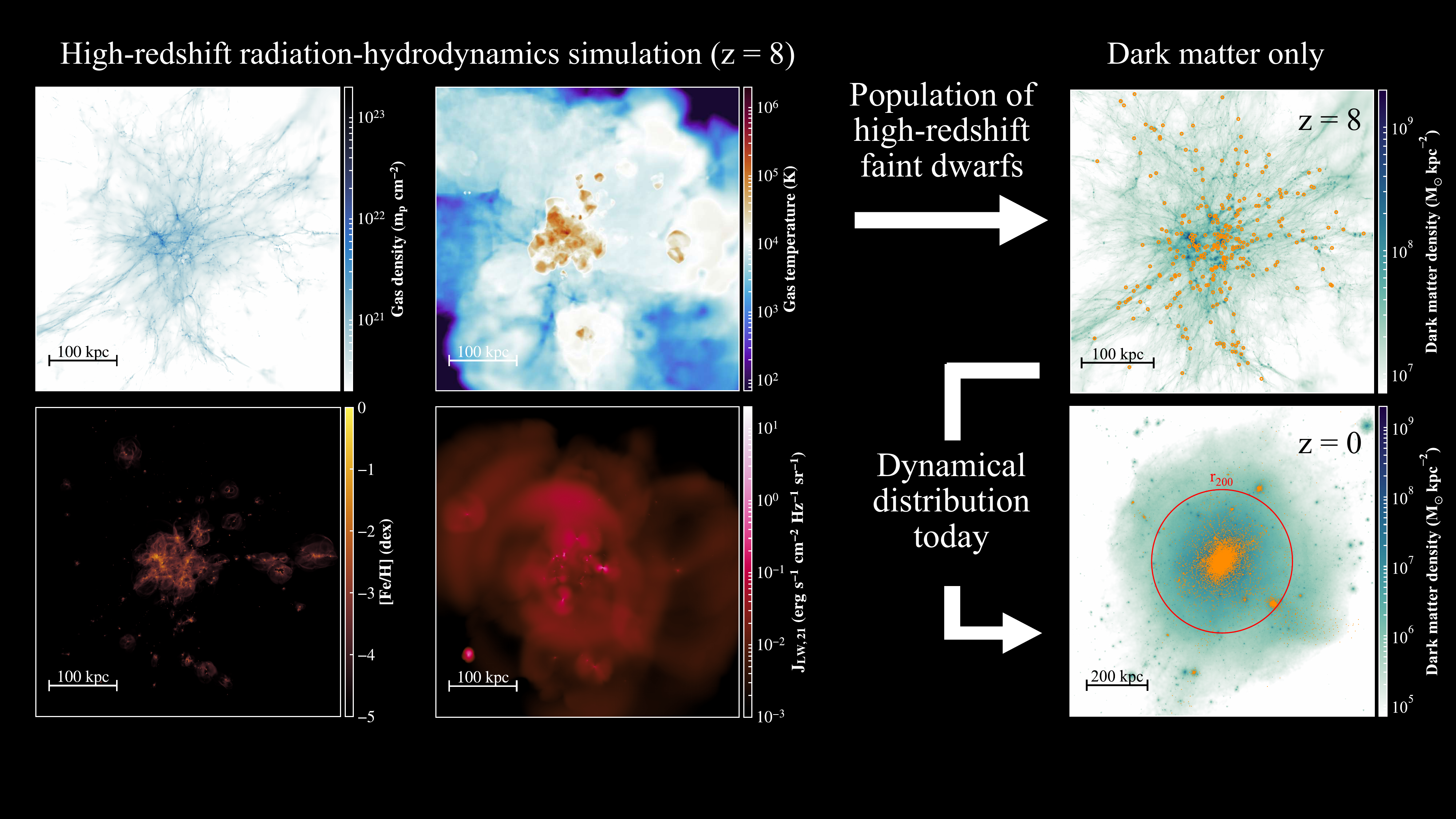}

    \caption{ \textsc{megatron} cosmological radiation-hydrodynamics simulations (left panels, gas density, gas temperature, gas iron metallicity, LW background at $z\approx8$) model galaxy formation of a proto-galaxy that will turn into a Milky-Way-like galaxy at $z=0$ (right panels). High-redshift dwarf galaxies in \textsc{megatron} are mapped onto a dark matter only simulation (right, top panel, orange circles highlighting dwarf galaxies with $\Mstar \leq 10^5 \, \Msol$) to predict their dynamical future around the Milky-Way-like galaxy at $z=0$ (e.g.\ first infall, disrupted in the stellar halo, bound in orbiting satellites, bottom right panel, see Section~\ref{sec:lowzconnection} for details of the particle tagging technique). 
    }
    \label{fig:hero}
\end{figure*}

\subsection{The high-redshift galaxy formation model}
\label{sec:methods:megatron}

Each \textsc{megatron} simulation is a zoomed, cosmological simulation with $\mdm = \xMsol{2.5}{4}$. The spatial resolution evolves with redshift, from $\softening = 2.5 \, \pc$ at $z=25$ to $\softening \approx 5 \, \pc$ at $z=8$ (see \citealt{Katz2025MegP1} for more details).

The simulations are evolved with the adaptive mesh refinement \textsc{ramses-rtz} code (\citealt{Teyssier2002, Rosdahl2013,Katz2022RTZ}). We solve for gravity, hydrodynamics, radiative transfer, non-equilibrium primordial chemistry (including $\hmol$), and non-equilibrium metal chemistry of $\geq 70$ heavy elements and molecules all coupled together on-the-fly. Non-equilibrium metal chemistry and radiative transfer are coupled through the \textsc{prism} model (\citealt{Katz2022PRISM}; see \citealt{Katz2024MegPilot} for updates). The radiation field observed in the left panels of Fig.~\ref{fig:hero} is produced by the star particles in the simulation. An external ultraviolet background is added to the simulations that continue beyond $z\approx6$ (e.g.\ \citealt{Cadiou2025MegP1}) but does not impact the results presented in this paper at $z\approx8$.

\textsc{megatron} employs a detailed galaxy formation model including the treatment of Pop.~III star formation, supernova feedback, radiative feedback, and metal enrichment from individual stars. Briefly, Pop.~III star formation is modelled as in \citet{Kimm2017, Katz2023PopIII}, spawning individual Pop.~III stars following a lognormal IMF with a characteristic mass $100 \, \Msol$ (\citealt{Wise2012}; see Appendix~\ref{app:popiii}). Pop.~III stars with masses $10 < m_{\star} \leq 20 \, \Msol$ explode as core-collapse supernovae (CCSNe), those with masses $20 < m_{\star} \leq 40 \, \Msol$ explode as hypernovae (HNe), and those with masses $140 \leq m_{\star} \leq 300 \, \Msol$ explode as PISNe. Each of these stellar types has a different explosion energy, namely $10^{51}$, $\approx 2 \times 10^{52}$, $ \approx 5 \times 10^{52}$ erg for CCSNe, HNe and PISNe, respectively (see \citealt{Kimm2017}, section 2.3.3 for the exact scaling with progenitor mass or core mass). Element-by-element Pop.~III yields are injected according to \citet{Nomoto2013}. Radiative feedback from Pop.~III stars follow the models of \citet{Schaerer2002}. Pop.~III stars whose progenitor masses are outside the mass range mentioned are assumed to collapse directly as a black hole (no explosion, no injection of chemical elements). More details of the implementation can be found in \citet{Katz2023PopIII, Storck2025MegP1}.

Once gas is above $Z \geq 10^{-6} \, \Zsol$ where $\Zsol = 0.02$ (the metallicity where gas cloud fragmentation becomes more efficient in the presence of dust; \citealt{Omukai2005}), we transition to a Pop.~II star formation mode. The star formation algorithm remains the same. We use a multi-freefall approach based on a local gravo-thermo-turbulent condition (\citealt{Kimm2017}), and stellar particles represent a simple stellar population rather than individual stars. We assume a fiducial Pop.~II IMF from \citet{Kroupa2001} and a minimum stellar particle mass of $500 \, \Msol$. Pop.~II populations eject stellar winds, undergo supernova explosions (CCSNe and Type Ia), and inject radiative feedback and metal enrichment following the approach in \citet{Agertz2020EDGE}. Briefly, supernova feedback is injected as energy ($10^{51} \, \mathrm{erg}$) if the cooling radius is resolved, and as momentum otherwise ($3 \times 10^5 \, \Msol \, \kmpers$). Radiative feedback is injected following the {\small BPASS v2.2.1} binary stellar evolution model \citep{Stanway2016}. CCSNe inject chemical elements following the yields of \citet{Limongi2018} assuming no rotation. Type Ia supernovae are injected according to the delay-time distribution of \citet{Maoz2017}, with yields from \citet{Seitenzahl2013}. We choose this Pop.~II feedback budget as it is well tested at $z=0$ in both Milky-Way galaxies (e.g.\ \citealt{Agertz2020Vintergatan}) and dwarf galaxies (e.g.\ \citealt{Rey2025}). 

Metals are tracked as passive scalars and advected on the grid as we solve for hydrodynamics. We do not account for any additional subresolution mixing (e.g.\ through unresolved turbulence \citealt{Escala2018}) or increased scatter (e.g.\ from sub-resolution inhomogeneities; \citealt{Tarumi2020}). The observed scatter in abundances thus results from the inhomogeneous distribution of metals in the ISM at $\approx 5 \, \pc$ scale and the stochasticity of star formation and chemical enrichment.

More details about the implementation of \textsc{megatron} can be found in \citealt{Katz2024MegPilot, Katz2025MegP1}. 

\subsection{The suite of four simulations}
\label{sec:methods:megatronsuite}

The \textsc{megatron} high-redshift suite consists of 4 simulations of the same initial conditions varying assumptions in Pop.~II star formation and feedback modelling (`Efficient SF', `Bursty SF', `Varying IMF' and `HN, $\epsilon_\mathrm{eff}$'). 

Briefly, the first two models differ in the strength of Pop.~II supernova feedback (the energy released per supernova is $10^{51} \, \mathrm{erg}$ for `Efficient SF' vs. $5 \times 10^{51} \, \mathrm{erg}$ for `Bursty SF'), leading to efficient gas conversion into stars in the first case and burstier star formation histories in the second case. The `Varying IMF' implements a (primarily) density-dependent Pop.~II IMF from \citet{Marks2012}, which becomes more top-heavy in higher-density star formation environments and converges to our fiducial \citet{Kroupa2001} otherwise. In addition to a varying-IMF, this run assumes that low-metallicity Pop.~II stars can undergo HN explosions, with a varying energy explosion depending on mass (\citealt{Nomoto2006}) and a fraction of HNe that strongly depends on metallicity \citep{Kobayashi2011} (50\% at $Z \leq 10^{-4}$; 1\% at $Z=\Zsol$). The last model changes parameters of the star formation algorithm to ensure that star formation occurs is more efficient (the star formation efficiency $\epsilonff$ is fixed at 100\% rather than varying with Mach number and turbulence) and more clustered (stellar particle mass of $2,000 \Msol $ instead of our fiducial $500 \Msol$). This last model also implements the same HN feedback for Pop.~II low-metallicity stars as the `Varying IMF'. 

We refer the reader to \citet{Katz2025MegP1} for the exact parameter choices of each model and to \citet{Choustikov2025MegP1} for how they affect the ISM structure at high redshift. In the context of this paper, comparing between these four simulations simply allows us to verify the robustness of our findings when making large variations to Pop.~II feedback implementation. 

\subsection{Halo finding and dynamical future of high-redshift dwarfs} \label{sec:methods:tagging}

\begin{figure}
  \centering
    \includegraphics[width=\columnwidth]{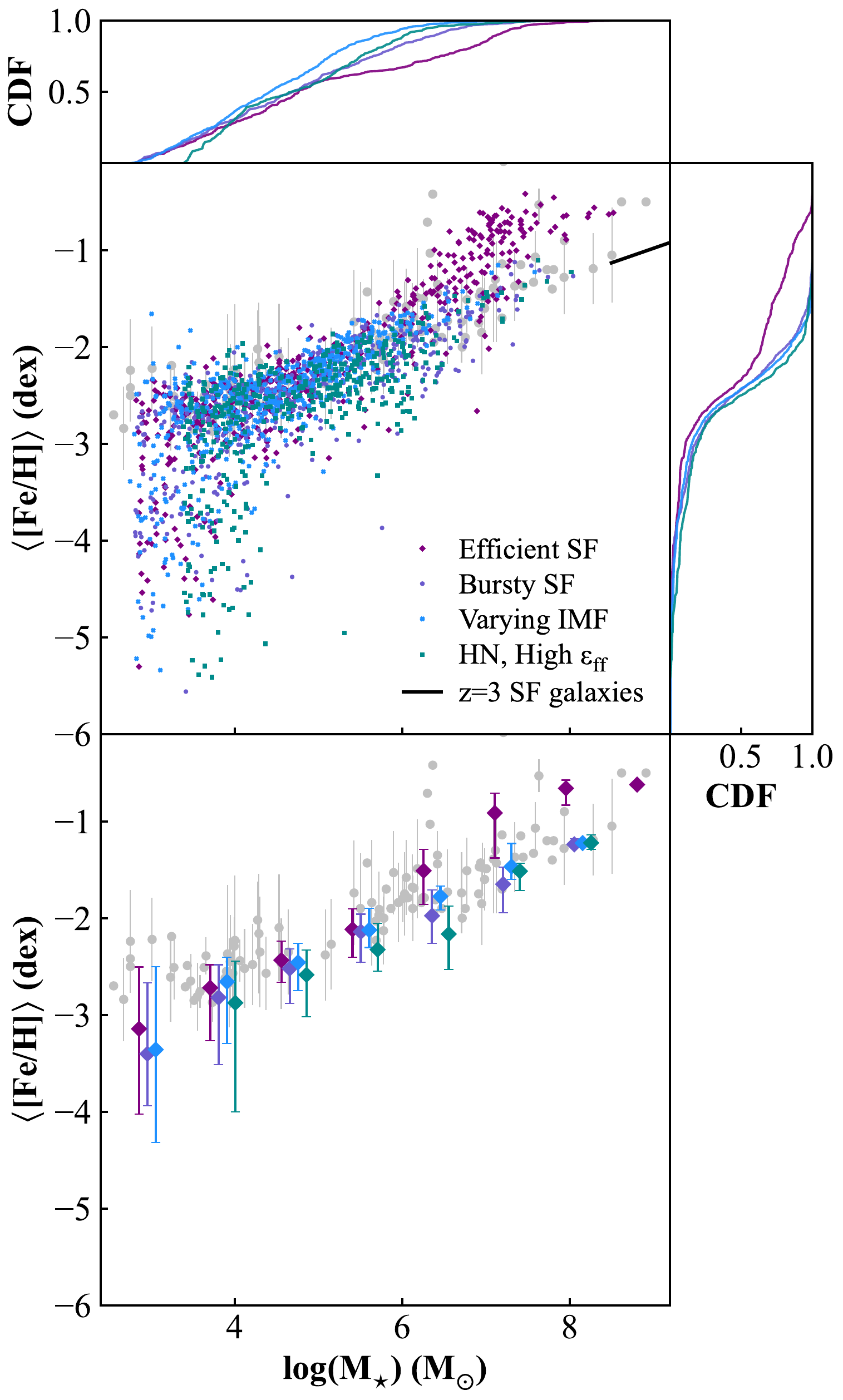}

    \caption{The stellar mass-metallicity relation of dwarf galaxies, observed at $z=0$ (grey points, \citealt{Pace2025LVDatabase}), observed in $z=3$ star-forming galaxies (\citealt{Stanton2024b}), and simulated \textsc{megatron} galaxies at $z\approx8$ (color points). The middle panel shows all individual simulated dwarfs, with cumulative distribution functions shown in top and right panels, and the mode in bins of $\Mstar$ shown in the bottom panel. Simulated and observed dwarfs are in agreement in slope, normalisation, and scatter over $10^5 \leq \Mstar \leq 10^8 \, \Msol$ for all simulations but one. The `Efficient SF' run (purple) overshoots the data, due to inefficient galactic outflows in higher-$\Mstar$ galaxies. All \textsc{megatron} simulations plateau in iron metallicities at low masses ($\Mstar \leq 10^5\, \Msol)$, with an extended tail of iron-poor galaxies (Figure~\ref{fig:ufdplateau}). 
    }
    \label{fig:mstar-feh-suite}
\end{figure}

The \textsc{megatron} simulations are only evolved to $z\approx8$, but we wish to connect the properties of their simulated dwarfs to local data around the Milky Way at $z=0$.

To perform this connection, we leverage the dark-matter-only counterpart of the \textsc{megatron} ICs, which we evolve to $z=0$ with the \textsc{ramses} code (\citealt{Teyssier2002}) under gravity and the same $\mdm$. Even though we have access to hydrodynamical runs of these initial conditions at $z=0$ (see Section~\ref{sec:methods:ic}), these runs are evolved with 8x coarser dark matter resolution than the fiducial \textsc{megatron} simulations ($\mdm \approx \xMsol{2}{5}$ rather than $\mdm = \xMsol{2.5}{4}$). As a result, the dark-matter-only run with matching resolution is better suited to track the fate of faint objects. 

In all simulations, we identify dark matter haloes and subhaloes with the \textsc{Rockstar} halo finder (\citealt{Behroozi2013haloes}). We then match galaxies in the hydrodynamical \textsc{megatron} simulations to their dark-matter-only counterparts at $z\approx8$ (Figure~\ref{fig:hero}, upper right panel). The IDs of dark matter particles are generated self-consistently across all \textsc{megatron} simulations, allowing us to cross-match the haloes with the most particles in common between simulations. We discard haloes if $\Mvir$ differ by more than a factor two between the hydrodynamical and dark-matter-only runs.

Once a halo is matched between the hydrodynamical and dark-matter-only runs, we use the particle-tagging approach from \citet{Rey2022} to estimate its dynamical fate at $z=0$. We identify the 5\% most bound dark matter particles of the host halo at $z \approx 8$ and `tag' them as dynamical proxies for the future evolution of the stars (\citealt{Rey2022}; see also \citealt{Bullock2005, DeLucia2008, Cooper2010, Cooper2017} for similar approaches). 

Tagged particles are then tracked to $z=0$, allowing us to predict whether the dwarf galaxy is disrupted in the stellar halo of the Milky Way, is on first infall, or is bound in orbiting satellites (Figure~\ref{fig:hero}, lower right panel). The dwarf is labelled as an identifiable $z=0$ structure if at least 50\% of the tagged particles belong to a bound structure identified by the \textsc{Rockstar} halo finder. We verified that using 2\% of the most bound particles does not qualitatively change our results (see also discussion in \citealt{LeBret2017, Rey2022}).

Another technicality in the comparison between high and low redshift is that stellar evolution will induce mass-loss over the Hubble time, reducing stellar masses between $z\approx8$ and $z=0$. To estimate the importance of this effect, we compute the projected stellar mass at $z=0$ for each dwarf galaxy by counting which stars would still be on the main sequence at $z=0$ given our stellar evolution model (mass-loss rate for low-mass stars and explodability range for massive stars; see \citealt{Katz2025MegP1}). Using the projected stellar mass at $z=0$ or the simulated one at $z\approx8$ does not impact our conclusions (see Section~\ref{sec:lowzconnection:sfhs}). 

\section{The stellar mass-metallicity relation of dwarf galaxies} \label{sec:mstar-feh}

Figure~\ref{fig:mstar-feh-suite} shows the average iron content in the stars of \textsc{megatron} galaxies at $z\approx8$ as a function of their stellar masses at $z\approx8$. We select only central galaxies (i.e.\ excluding galaxies that are already satellites at $z\approx8$), and galaxies that contain at least 1 Pop.~II star particle (excluding purely Pop.~III objects which are still forming at $z\approx8$; \citealt{Storck2025MegP1}, fig.2). We compute $\averagefeh$ as the mass-weighted average over stellar particles (see \citealt{Escala2018}, eq.\ 3 and 4) and verified that using the median does not impact our results. 

To obtain $\Mstar$, we sum the mass of all Pop.~II stellar particles within $\rvir$, removing the mass contributed by non-luminous stellar evolution remnants (e.g.\ black holes). $\averagefeh$ is computed by averaging the metallicity of all star particles within $\rvir$ weighted by the current mass of each star particle. We use solar abundance patterns from \citet{Asplund2009} to calculate $\averagefeh$. 

\begin{figure}
  \centering
    \includegraphics[width=\columnwidth]{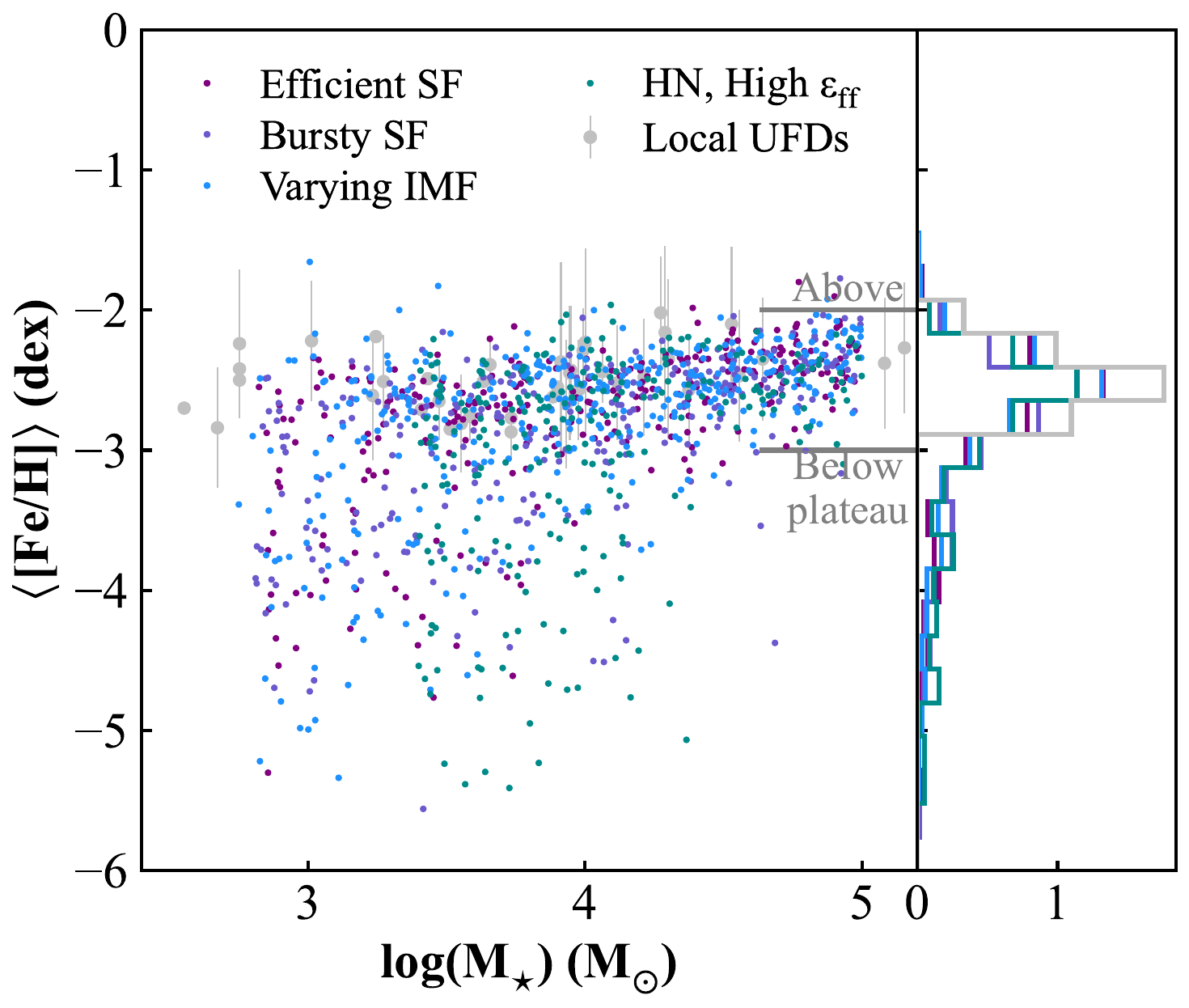}

    \caption{Same as Figure~\ref{fig:mstar-feh-suite} but zooming on the low-mass end of the stellar mass-metallicity relation. In all simulations, the distribution of faint dwarf galaxies is strongly peaked around $\averagefeh \approx -2.5$ over a broad range of $\Mstar$, reproducing the observed `plateau' of local dwarf galaxies (grey histogram and data points), and a long tail of iron-deficient galaxies. Both features emerge from our choices of Pop.~III star formation and stellar evolution (Figure~\ref{fig:popiiinumbers} and Figure~\ref{fig:popiiitype}).
    }
    \label{fig:ufdplateau}
\end{figure}

The middle panel shows all individual simulated dwarf galaxies per simulation, bottom panel shows the mode of the $\averagefeh$ distribution in bins of $\Mstar$ for each simulation (where individual simulations have been offset from the centre of the bin for clarity). We use the mode, rather than the median, to highlight where most dwarfs cluster. (We show the full shape of the distribution at the faint end in Figure~\ref{fig:ufdplateau}). The grey points show the observed stellar mass-metallicity in local dwarf galaxies taken from the database of \citet{Pace2025LVDatabase} (see Appendix~\ref{app:data} for individual citations). The black line shows the best-fit to the stellar mass-metallicity relation from $z=3$ star-forming galaxies from \citet{Cullen2021, Stanton2024b}.

The agreement between the simulated and observed stellar mass-metallicity relation is excellent. For all simulations, the population of simulated dwarf galaxies overlaps with the $z=0$ data, particularly in the range $10^4 \leq \Mstar \leq 10^6 \, \Msol$ where they match the normalisation and slope of the observed data and the CDFs (right and top panels) overlap. This is perhaps surprising, since the simulated dwarf galaxies are formed at $z\approx8$, while observed dwarf galaxies are at $z=0$ or $z=3$. In Section \ref{sec:lowzconnection}, we justify this connection between high-redshift simulations and low-redshift data. We show that the shape and normalisation of the stellar mass-metallicity relation in \textsc{megatron} remain when we forecast the dynamical and chemical evolution of the simulated dwarfs to $z=0$, and discuss the remaining open questions that need to be answered in this connection.

Towards the high-mass end ($\Mstar \geq 10^6 \, \Msol$), the `Efficient SF' simulation (purple) overshoots the relation. The `Efficient SF' simulation drives the weakest galactic outflows (\citealt{Katz2025MegP1}), allowing a higher iron retention in the ISM and iron over-abundance in the stars. The other three simulations, with much more efficient outflows, are in close agreement to the observed data. These results re-affirm that the slope and normalisation of the stellar-mass-metallicity relation of dwarf galaxies is particularly sensitive to the strength of galactic outflows (see also \citealt{Agertz2020EDGE, Rey2025}), for the first time demonstrated over a large population of simulated dwarf galaxies ($N\geq 200$ per simulation) rather than a handful of objects.

At the low-mass end of the relation, the observed data in the Local Volume show a break in slope, turning into a `plateau' of $\averagefeh \approx -2.5$ for $\Mstar \leq 10^5 \, \Msol$ and a dearth of iron-poor galaxies at low-$\Mstar$. All simulations show a similar break in slope in the mode, with the majority of simulated dwarf galaxies clustering around the same plateau as the data. However, the simulations also show a more continuous and scattered distribution of iron metallicities and a tail extending below the plateau. To visualize this distribution better, Figure~\ref{fig:ufdplateau} shows the stellar mass-metallicity zoomed in on the low-mass end of the relation with aggregated histograms (right panel). Note that the lack of very-low-mass dwarf galaxies in the `HN, $\epsilon_{\rm eff}$' simulation is due to the increased stellar particle mass in this run only allowing us to start resolving Pop.~II galaxies with $\Mstar \geq 10^{3.5} \, \Msol$. 

From Figure~\ref{fig:ufdplateau}, we quantitatively confirm that 78\% of simulated faint dwarf galaxies cluster between $-3 \leq \averagefeh \approx -2.0$ (hereafter referred to as the `iron plateau'), similar to the observed ultra-faint dwarfs in the Local Group (grey points and histogram). Unlike the data, however, $22\%$ of the dwarf population populates a long tail of iron-poor galaxies below $\averagefeh \leq -3.0$ with a handful of objects having higher-than-plateau iron metallicities. We discuss the observability of this tail around the Milky Way and its implications in Section~\ref{sec:conclusion}. 

All simulations showcase similar distributions of dwarf galaxies, despite large variations in Pop.~II assumptions within them. This hints that the iron distribution at the faintest end is rather set by Pop.~III modelling which stays the same across all four simulations. We explore this further in the next section.

\section{The origin of the iron plateau} \label{sec:popiii}

\subsection{The importance of Pop.~III microphysics} \label{sec:popiii:types}

\begin{figure}
  \centering
    \includegraphics[width=\columnwidth]{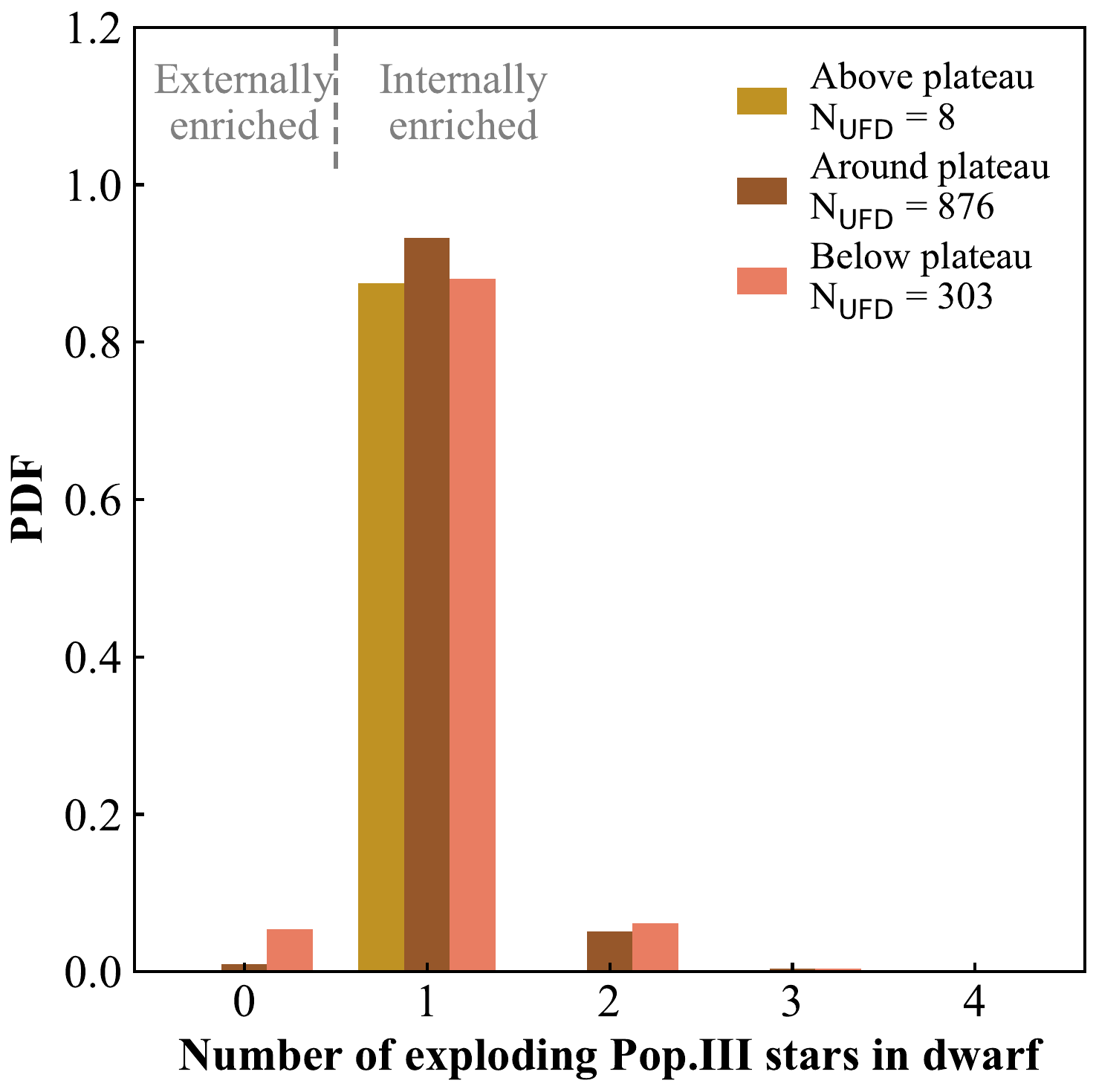}

    \caption{The number of Pop.~III explosions within \textsc{megatron} dwarf galaxies broken down by their position on, above, or below the plateau (red, brown, and gold, respectively). Almost all faint dwarf galaxies undergo exactly one Pop.~III explosion. A small fraction (2\% of the total population) of dwarf galaxies are externally enriched by diluted ejecta from nearby objects and have had no Pop.~III explosions. Those galaxies are exclusively found below the plateau but cannot explain the whole tail of iron-deficient dwarf galaxies.  
    }
    \label{fig:popiiinumbers}
\end{figure}

\begin{figure*}
  \centering
    \includegraphics[width=\textwidth]{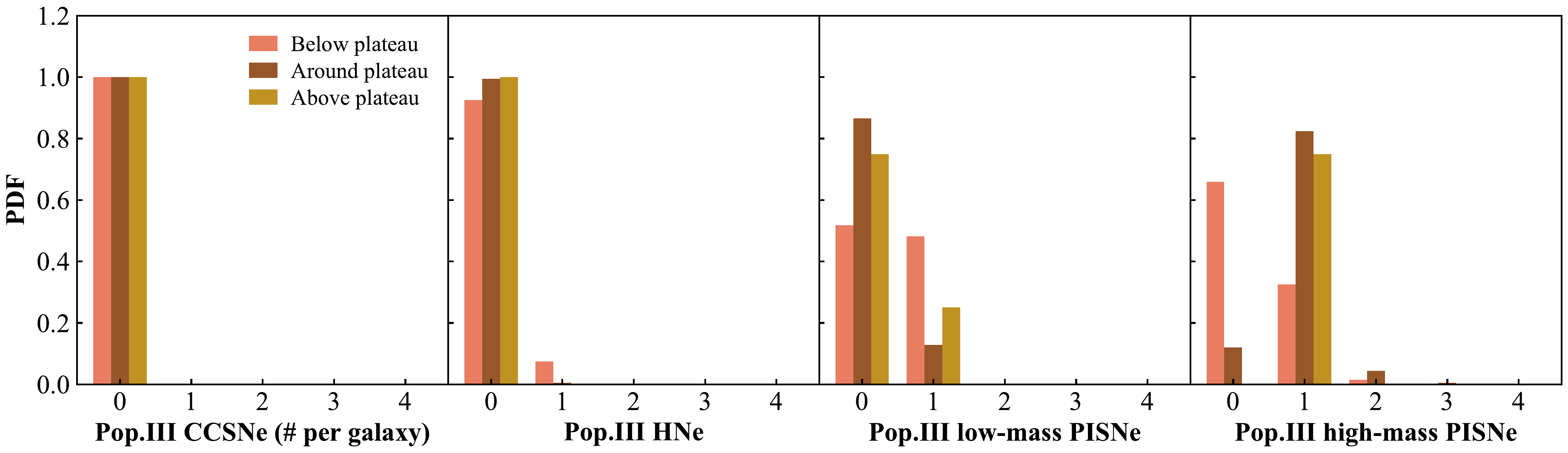}

    \caption{Same as Figure~\ref{fig:popiiinumbers}, but breaking down each sub-population of dwarfs by the exact progenitor of Pop.~III explosions. \textsc{megatron} galaxies experience almost no Pop.~III CCSNe and HNe explosions due to the choice of a top-heavy Pop.~III IMF (first and second panels). But dwarf galaxies on the metallicity plateau are mostly enriched by high-mass PISNe (fourth panel), while galaxies below the plateau have a significant contribution from low-mass PISNe (third panel). This split is naturally explained by the vastly different iron yields in explosions of low-mass and high-mass PISNe (see text for discussions).  
    }
    \label{fig:popiiitype}
\end{figure*}

To understand the origin of the distribution of $\averagefeh$, we break the sample of faint dwarf galaxies into three populations: above, on, and below the plateau. We define the boundaries of the plateau as $-3.0 \leq \averagefeh \leq -2.0$ (horizontal lines; Figure~\ref{fig:ufdplateau}). The lower boundary is somewhat arbitrary, since the simulations show a continuous tail of iron-poor dwarfs, but we check that all results are qualitatively unchanged if we instead define the lower limit of the plateau as $\averagefeh \geq -2.7$ or $\averagefeh \geq -3.2$.

Figure~\ref{fig:popiiinumbers} then shows the number of exploding Pop.~III stars in each subpopulation of dwarf galaxies stacked across all four simulations. Irrespective of whether they are on, above, or below the plateau, the majority of dwarf galaxies undergo precisely one Pop.~III explosion. As we will see in Section~\ref{sec:popiii:yieldminihalo}, chemical elements are being retained in the ISM of our Pop.~III formation sites, despite the large energies associated with Pop.~III explosions. This single explosion is thus enough to enrich the minihalo of the nascent galaxy above our critical metallicity threshold ($Z \geq 10^{-6} \Zsol$) and transition to Pop.~II star formation.  

A small fraction (6\%) of dwarf galaxies have transitioned to Pop.~II star formation with multiple Pop.~III explosions. On close inspection, we find that this arises roughly half of the time when Pop.~III stars are rapidly spawned in a small cluster with an age spread $\leq 1 \, \Myr$, leading to nearly simultaneous explosions and enrichment. The other half of the time, the age spreads are much larger ($\geq 5 \, \Myr$) which we hypothesize can occur either when all metals are vacated from the minihalo after the first Pop.~III explosion (requiring a second explosion before transitioning to Pop.~II star formation), or when Pop.~III form in two independent minihaloes that merge together into the faint dwarf we are studying at $z=8$. 

An even smaller fraction (2\%) of the dwarf galaxy population have transitioned to Pop.~II star formation without any Pop.~III explosions. These exclusively sit below the plateau of the mass-metallicity relation (red line) and have likely been enriched externally by diluted ejecta from nearby star formation. However, the small numbers of such externally-enriched dwarfs cannot explain the whole tail of iron-deficient dwarfs noted in Figure~\ref{fig:ufdplateau}. Furthermore, we find no difference in the spatial distribution or chemical abundances of externally-enriched dwarfs compared to the rest of the population (Appendix~\ref{app:externalpollution}). Given their low occurrence rate which will be further reduced by tidal disruption and stripping by $z=0$, this supports the picture that the chemical content of the Milky-Way's ultra-faint dwarfs is dominated by internal Pop.~III or Pop.~II events, rather than external contamination (see also discussion in \citealt{Griffen2018, Wheeler2025IGMPlateau}).

Next, we break the populations by each type of Pop.~III explosions in Figure~\ref{fig:popiiitype}. Beyond CCSNe (first panel), HNe (second panel), we divide PISNe into low-mass progenitors ($140 \leq m_{\star} \leq 160 \, \Msol$; third panel) and high-mass progenitors ($160 < m_{\star} \leq 300 \, \Msol$; fourth panel). The precise value for this split is motivated by a strong evolution in carbon and iron production below and above this value (\citealt{Nomoto2013}), although the precise value is somewhat arbitrary since we sample a continuum of progenitor masses and (linearly) interpolate between progenitor mass bins of the yield tables. 

First, almost no \textsc{megatron} dwarf galaxies undergo Pop.~III CCSNe and HNe (first and second panels). This is due to our choice of a top-heavy Pop.~III IMF, favouring the formation of more massive progenitors (see \citealt{Storck2025MegP1}, fig.\ 1). As a result, the vast majority of explosions experienced by our dwarf galaxies are PISNe.

Dwarf galaxies that sit on the plateau of the mass-metallicity relation (brown line) almost exclusively host high-mass PISN explosions (fourth panel). In contrast, dwarf galaxies below the plateau (gold curve) show a mix of explosions and have a significant contribution from low-mass PISNe. This split is naturally explained when considering the iron yield of each of these progenitors. According to \citet{Nomoto2013}, an explosion with progenitor mass $m_{\star} = 140 \, \Msol$ yields $\approx 0.5 \, \Msol$ of iron, while a progenitor with $m_{\star} = 180 \, \Msol$ yields $\approx 5 \, \Msol$ of iron, with little evolution as progenitor mass keeps increasing. Thanks to their much more extended progenitor mass range ($160 < m_{\star} \leq 300 \, \Msol$), high-mass PISNe dominate the number of explosions once integrated over the IMF. In all cases, the explosion yields enough carbon, oxygen and other elements for the total metallicity of the ISM to transition to Pop.~II star formation, but the difference in iron leads to the position on, or below, the plateau.

These results clearly show that the low-mass end of the mass-metallicity is highly sensitive to the exact assumptions and details of Pop.~III physics (see also \citealt{Jeon2021FaintPopIIISNe, Sanati2023}). In particular, we find that, everything else being fixed, the number of dwarf galaxies on the plateau is tied to high-mass Pop.~III PISNe, while the number below the plateau (or lack thereof) are tied to low-mass Pop.~III PISNe. This is highly impactful, as one could leverage this knowledge and the distribution of observed iron metallicities into an effective constraint on the Pop.~III IMF and PISN iron yields. We leave this to future work and show next how, beyond iron, this dichotomy of PISN progenitors is reflected on the chemical abundances of our simulated dwarfs.

\subsection{Chemical abundances beyond the plateau} \label{sec:popiii:chemistry}

\begin{figure*}
  \centering
    \includegraphics[width=\textwidth]{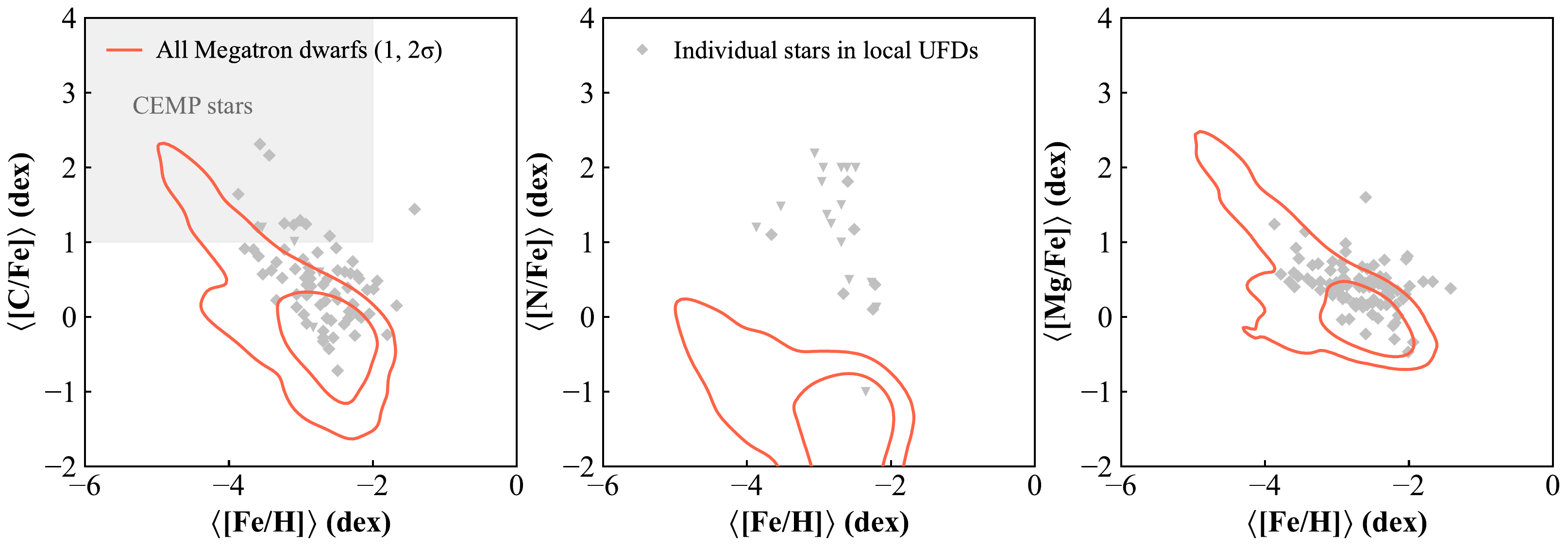}

    \caption{Individual abundances in simulated \textsc{megatron} dwarfs (red) compared to local data (gray), including carbon-enhanced metal-poor (CEMP) stars . There is reasonable agreement for carbon and magnesium, although simulated data favour lower $\cfe$ and $\mgfe$ abundances at $\feh \approx -2.5$. This reflects our choice of yield models for high-mass PISNe which dominate enrichment close to plateau metallicities (see text for discussion). Nitrogen is vastly under-produced in our simulated dwarfs since PISNe are the sole enrichment channel in \textsc{megatron}. Additional channels (stellar winds, rotation) could help alleviate this tension (Section~\ref{sec:popiii:chemistry}). 
    }
    \label{fig:chemistry}
\end{figure*}

Figure~\ref{fig:chemistry} shows the average carbon, nitrogen and magnesium content of all \textsc{megatron} dwarf galaxies (left, middle, right panels, respectively). The contours show the 2D distribution of simulated galaxies in average abundance versus $\averagefeh$. This is contrasted with the abundances of individual stars in ultra-faint dwarf galaxies (see Appendix~\ref{app:data} for exact citations). 

Carbon and magnesium show broad agreement between the simulated and observed data, providing encouraging support for the PISN enrichment scenario we highlighted above. Nitrogen, however, is under-produced in our simulated dwarf galaxies compared to observed stars. Most stars in UFDs only have upper limits in nitrogen content (triangles), but the few with detections (diamonds) are several order of magnitude above the simulated data. PISNe are expected to have strong odd-even anti-correlation in chemical yields (e.g.\ \citealt{Heger2002}), with the lack of nitrogen yield being a defining characteristic (e.g.\ \citealt{Salvadori2019}). This is directly reflected in our simulated data for which PISN explosions dominate the enrichment budget. 

Our results suggesting PISN-dominated enrichment in UFDs might appear surprising given the lack of observed metal-poor stars with strong odd-even abundance patterns. Early claims of Milky Way halo stars showing such patterns (e.g. \citealt{Aoki2014, Salvadori2019, Xing2023}) have been ruled out by subsequent analyses (e.g. \citealt{Takahashi2018, Thibodeaux2024, Skuladottir2024}). However, the contribution from UFD stars to the final Milky Way halo is expected to be small ($\leq 1\%$; \citealt{Ji2019}), so their non-detection does not rule out this possibility. Within UFDs themselves at $z=0$, nitrogen itself is difficult to measure and a lack of it could easily be masked by other channels not considered in our simulations. For example, the average nitrogen content of our galaxies is $\rm [N/H] \approx -6$, so even a small amount of additional pollution from other channels could significantly sway the average upwards and `fill' the `odd' gaps in abundance patterns (see discussion in \citealt{Ji2015}). 

\textsc{megatron} does not explicitly model many of these additional channels of nitrogen production. For example pre-explosion Pop.~III winds could significantly boost nitrogen production in such massive low-metallicity or metal-poor stars (e.g.\ \citealt{Hirschi2007, Liu2021a, Vink2023Nitrogen, Charbonnel2023}). We also assume nucleosynthetic yields from non-rotating stellar models, while rotation can increase the nitrogen yield of PISNe explosions by four orders of magnitude (see \citealt{Takahashi2018}, fig.\ 10 and 11). Lastly, we stress that high-redshift galaxies with surprisingly high nitrogen contents are currently being revealed by JWST (\citealt{Cameron2023GNz11, Topping2024NEmitterSample,Schaerer2002}), highlighting gaps in our understanding of nitrogen enrichment in metal-poor environments. Overall, the tension in nitrogen abundances between our simulated dwarfs and observed UFD stars is a strong motivation to implement additional channels of nitrogen production in future \textsc{megatron} simulations, as well as exploring other key  elements that telltale signs of PISN enrichment (e.g.\ Na, Zn; \citealt{Salvadori2019}).

Turning now to carbon, the shape of the distribution of carbon abundances matches well that of the observed data. However, the simulated data seems systematically shifted down, by $\approx 0.5$ dex. Such shifts are greater than typical uncertainties in observational data (e.g.\ non-local thermodynamic equilibrium corrections can shift observed abundances by up to $\approx 0.3$ dex;  discussion in \citealt{Koutsouridou2025}) but well within the uncertainties in nucleosynthetic yields from PISNe. For example, differences in the treatment of convection and mixing in 1D stellar evolution codes can change the carbon yields of PISNe by a factor of a few (e.g.\ \citealt{Heger2002, Nomoto2013, Takahashi2018}), and uncertainties in nuclear reaction rates directly impact predictions of the iron yields, the explosion energy and the window of explodability of PISNe (e.g.\ \citealt{Farmer2019, Kawashimo2024}; see also Appendix~\ref{app:popiii}).

Overall, we conclude that, given the chemical information available in \textsc{megatron}, the abundances produced in our simulations are not in strong tension with observed data and neither rule in or out a PISN enrichment scenario. \textsc{megatron} tracks chemical elements key for gas cooling which are primarily even. Given the sensitivity of our only odd element, nitrogen, to other enrichment channels, the tension we observe is a strong motivation for future \textsc{megatron} simulations to (i) track additional elements relevant for stellar spectroscopy and abundance constraints (e.g.\ Na, Al, Mn, Cu, Zn), and (ii) directly investigate the impact of winds and rotation by varying yield tables.

So far, we have established that Pop.~III PISNe are the key driver of the iron plateau and chemical abundances of UFDs in \textsc{megatron}. However, given the large explosion energies, metals from PISN explosions are easily expelled from minihaloes (e.g.\ \citealt{Bromm2003, Greif2007}) and can be ineffective in polluting metal-poor stars (e.g.\ \citealt{Cooke2014}). In the next section, we explore what enables \textsc{megatron} Pop.~III star formation sites to retain the signatures of PISN enrichment.

\subsection{The coupling between the plateau and Pop.~III star formation sites} \label{sec:popiii:yieldminihalo}

\begin{figure}
  \centering
    \includegraphics[width=\columnwidth]{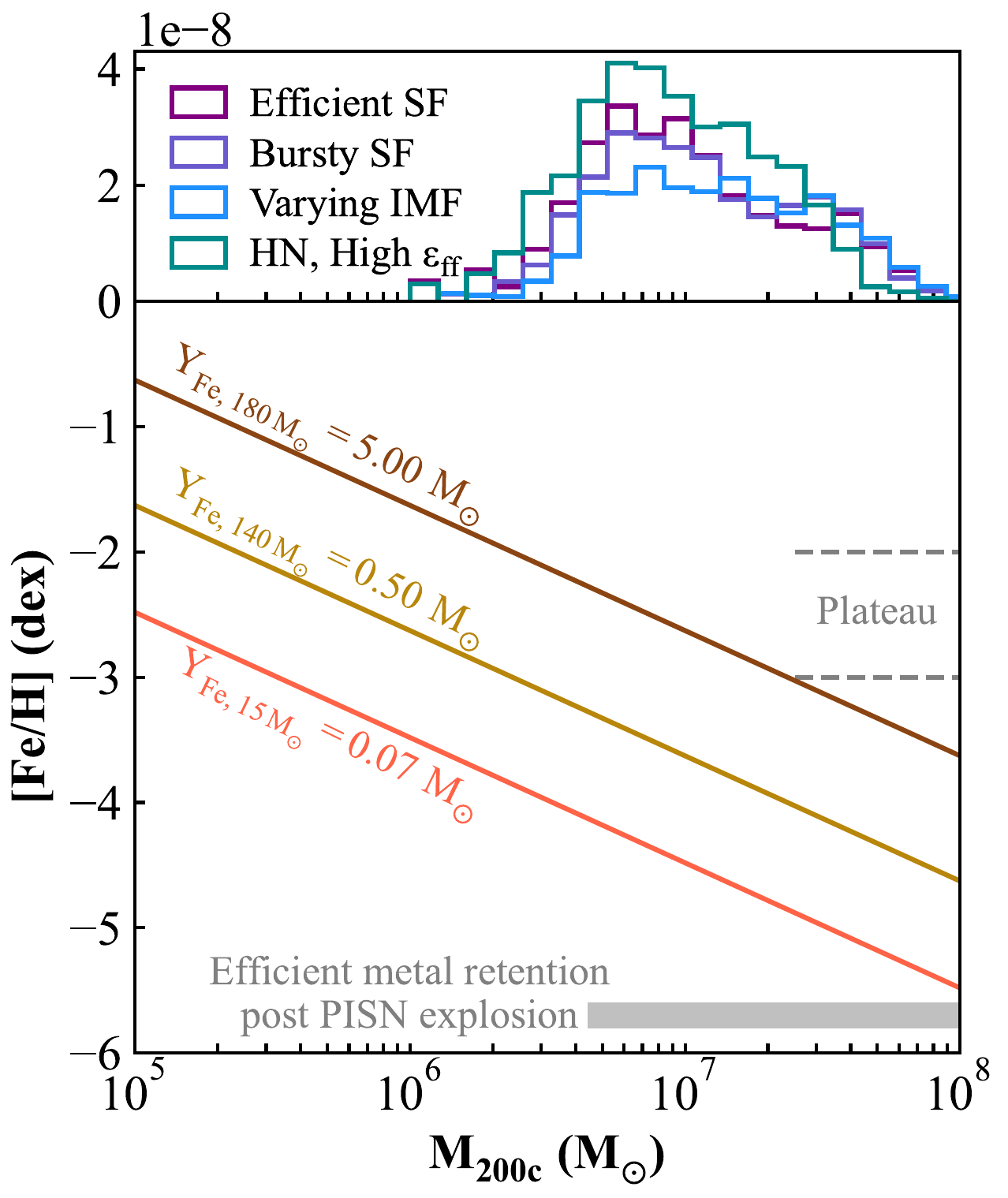}

    \caption{Top: Distribution of dark matter halo masses hosting a Pop.~III star formation event. Bottom: iron metallicity resulting from the dilution of an iron yield in the gas reservoir of a halo as a function of mass. In \textsc{megatron}, Pop.~III stars form in haloes with $\Mvir \geq \xMsol{5}{6}$. This mass allows the retention of metals from high-energy PISN explosions (bottom shading, see \citet{Cooke2014}, fig.\ 3 and text for discussion). Diluting the iron yield of a high-mass PISN (brown) in a $10^7 \Msol$ halo naturally leads to the emergence of the iron plateau at $\averagefeh \approx -2.5$ (horizontal dashed line). Low-mass PISNe (gold) and CCSNe (red) lead to lower iron metallicities, populating the tail of iron-poor galaxies below the plateau.
    }
    \label{fig:yieldminihalo}
\end{figure}

In this section, we connect our findings to the properties of Pop.~III star formation sites in \textsc{megatron}. Figure~\ref{fig:yieldminihalo} shows the distribution of dark matter halo masses hosting newborn Pop.~III stars\footnote{Our output cadence ($\approx$ 5 Myr) ensures that halo masses evolve minimally even if the Pop.~III star is born in-between outputs} (top panel) and the average iron metallicity when diluting typical Pop.~III iron yields in the gas reservoir of a halo of this mass (bottom panel). 

Focussing first on the top panel, we find that Pop.~III stars in \textsc{megatron} predominantly form in haloes with $\Mvir \geq \xMsol{5}{6}$, with a median of $\Mvir \approx 1-3 \times 10^7 \, \Msol$ (vertical coloured lines). In \citet{Storck2025MegP1}, we explore the physical mechanisms setting this mass-scale in detail. Briefly, we find that a strong LW radiation field sourced by the central galaxy ($J_{21} \gtrsim 10^{-1} \, \text{erg} \, \text{s}^{-1} \text{cm}^{-2} \, \text{Hz}^{-1} \, \text{sr}^{-1}$; their fig.\ 4; see also Figure~\ref{fig:hero}) suppresses $\hmol$ formation and cooling in small minihaloes and elevates the mass-scale of Pop.~III star formation (see also \citealt{Ahn2009, Xu2013, Wise2014, Xu2016, Zier2025a} for similar results).

Such $10^7 \, \Msol$ haloes can retain metals from much larger explosion energies ($\geq 10^{53}$ erg) due to the deeper gravitational potential wells and increased cooling losses (e.g.\citealt{Kitayama2005, Whalen2008, Cooke2014}). A single PISN is insufficient to evacuate metals, which are then readily re-incorporated in the next generation of (Pop.~II) stars. This aligns with our findings in Section~\ref{sec:popiii:types} and~\ref{sec:popiii:chemistry} that most \textsc{megatron} dwarf galaxies undergo exactly one Pop.~III explosion, retain the metals from this enrichment event, and form Pop.~II stars with the chemical signature of PISNe. 

We further connect this point to the mass-metallicity plateau in the bottom panel of Figure~\ref{fig:yieldminihalo} in which we show the iron metallicity obtained when diluting three typical iron yields into the primordial gas of a dark matter halo. Here, we convert a virial mass $\Mvir$ to a total hydrogen mass assuming a 100\% baryon fraction with $\Omega_{\text{b}} = 0.04916$ and a primordial mass fraction of hydrogen $Y_{\text{H}} = 0.76$ (e.g.\ \citealt{PlanckCollaboration2020}). We then uniformly dilute the iron yield across this gas reservoir assuming that the explosion shock is contained within the virial radius (e.g.\ \citealt{Whalen2008,Cooke2014}). We use $Y_{\text{Fe}} = 5 \, \Msol$, $0.5 \, \Msol$, $0.07 \, \Msol$ to represent a high-mass Pop.~III PISN progenitor, a low-mass Pop.~III PISN progenitor, and a Pop.~III or Pop.~II CCSN respectively. 

Diluting a fixed yield in a halo of increasing mass naturally leads to lower iron metallicities, as expected. But more importantly, diluting the yield of a high-mass PISN (brown line) in a $\Mvir \approx 10^7 \Msol$ minihalo naturally leads to $\averagefeh \approx -2.5$, the average position of the plateau in the observed stellar mass-metallicity relation. In contrast, diluting the yield of a low-mass PISN (gold line) in the same halo leads to $\averagefeh \approx -3.5$, while a Pop.~II CCSN (red line) leads to $\averagefeh \approx -4.5$. Since the bulk of Pop.~III star formation sites in \textsc{megatron} have $\Mvir \approx 10^7 \, \Msol$ haloes (top panel) and undergo PISN explosions (Section~\ref{sec:popiii:types}), this naturally explains the emergence of the plateau and the tail of iron-poor galaxies in \textsc{megatron}.

We stress that this is an approximate and heavily simplified calculation bypassing many of the complexities of metal mixing and retention in realistic haloes. In particular, in this analytical calculation (but not in the hydrodynamical simulation), we have assumed that all metals are retained within the virial radius and uniformly mixed within the gas reservoir. This is unlikely to be the case in reality (e.g.\ \citealt{Ritter2015}). Such additional effects are likely to explain the non-zero width of the distribution, but our first-order calculation readily explains how the plateau arises from the coupling between Pop.~III yields and their halo formation sites.

According to this same argument, PISNe explosions in higher-mass haloes are not the only solution to produce plateau-like metallicities. In particular, a single CCSN event (red) in a much smaller minihalo ($\Mvir \approx 10^5 \, \Msol$) could also yield $\averagefeh \approx -2.5$, especially since metal retention is also expected to be efficient given the lower explosion energy ($10^{51}$ erg; \citealt{Cooke2014}, fig.\ 3). This highlights the degeneracies between the Pop.~III IMF and yields (controlling iron production) and the halo mass-scale for Pop.~III star formation (controlling iron retention and dilution). Furthermore, these effects are coupled together -- as just one example, a more top-heavy Pop.~III IMF raises the effective iron yield, but also the amount of LW background and in turn the mass-scale for Pop.~III star formation in nearby haloes (see e.g.\ \citealt{Brauer2025PopIIIIMF}).

The emergence of the plateau in \textsc{megatron} is thus neither fine-tuned nor obvious, rather resulting from the choices of Pop.~III implementation we have made and the specific simulated environment of a Milky-Way-mass protogalaxy. Our results highlight the importance of accounting for these coupled effects self-consistently to predict the distribution of UFD metallicities and leverage local data to constrain Pop.~III microphysics. We continue to explore these prospects in the next section, quantifying how much of the features we have highlighted at $z\approx 8$ would remain observable at $z=0$ in local data.

\section{Should we expect agreement between local data and high-redshift simulations?} \label{sec:lowzconnection}

By comparing simulated dwarf galaxies at $z\approx8$ with observed dwarf galaxies at $z=0$, we have jumped $12 \, \Gyr$ of cosmic evolution. In this Section, we leverage a defining feature of \textsc{megatron} simulations -- that they follow the formation of a Milky-Way-like galaxy -- to show that bridging between these two epochs is quantitatively justified.   

\subsection{Linking high- and low-redshift star formation histories and stellar masses} \label{sec:lowzconnection:sfhs}

Stellar masses and star formation at the faint end of the stellar mass function of dwarf galaxies is expected to be primarily shaped by reionization quenching. In fact, all ultra-faint dwarfs observed at $z=0$ are quenched and have formed most of their stars $\geq 10 \, \Gyr$ ago (e.g.\ \citealt{Savino2025M31SFHs, Durbin2025}). By contrast, most dwarfs plotted in Fig~\ref{fig:mstar-feh-suite} are actively forming stars since none of the \textsc{megatron} simulations are fully reionized by $z\approx8$ (see Appendix~\ref{app:reionization} for more details). Catching these dwarfs before their `fossilization' by reionization quenching could mean that their properties significantly evolve after $z\approx8$ and they are not progenitors of local UFDs. 

However, reionization quenching is strongly modulated by the specific spatial location of a given dwarf. Dwarfs closer to photon-producing regions will be reached first by ionization fronts and see their halo gas reservoir climb in temperature (e.g.\ \citealt{Aubert2018, Katz2020,Ocvirk2020}). We thus expect a subpopulation of dwarfs in \textsc{megatron} to already be reionization-quenched by $z\approx8$, particularly those closer to the centre of the simulated volume where most star formation and photon production occurs (see temperature map in Figure~\ref{fig:hero}). 

\begin{figure}
  \centering
    \includegraphics[width=\columnwidth]{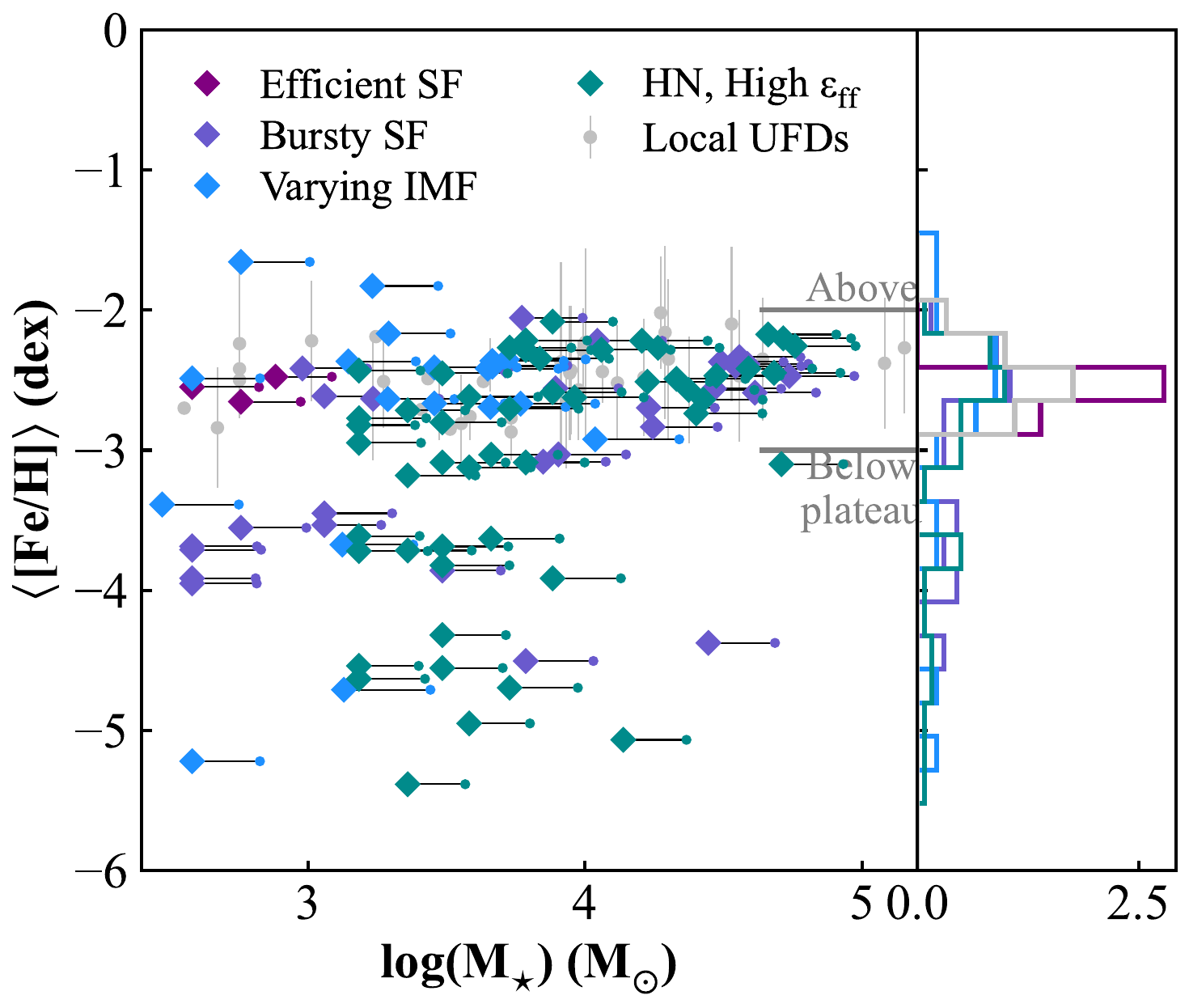}

    \caption{Same as Figure~\ref{fig:mstar-feh-suite} but for the subpopulation of dwarf galaxies already quenched by reionization at $z\approx8$. Dots show their stellar mass at $z\approx8$, while diamonds show the projected stellar mass at $z=0$ after accounting for a Hubble time of stellar mass loss.  The faint end of the mass-metallicity relation is unchanged by either effects. 
    }
    \label{fig:mstar-feh-z0}
\end{figure}

We extract this subpopulation by selecting central faint dwarf galaxies that have formed no new stars for the last 100 Myr (since $z\approx10$). Figure~\ref{fig:mstar-feh-z0} shows their mass-metallicity relation. (Restricting to dwarfs quenched for 200 Myr -- since $z\approx 11.5$ -- shows the same trend, but prevents statistical tests by decreasing the sample size to too few objects).  

As expected, the subpopulation of already-quenched dwarfs is much smaller than in Figure~\ref{fig:mstar-feh-suite}, but the over-abundance of dwarfs along the plateau remains (see histogram on the right-hand panel). The sample sizes decrease from $N \geq 250$ per simulation to 3, 28, 17, and 51 quenched dwarfs for the `Efficient SF', `Bursty SF', `Varying IMF` and `HN, $\epsilon_{\rm eff}$', respectively. The lowest number of quenched dwarfs in the `Efficient SF' simulation correlates with the fact that this simulation is the least reionized at $z\approx8$ (Figure~\ref{fig:reionization}). Combining all simulations together, we run a Kolmogorov-Smirnov test to compare the distribution of $\averagefeh$ in the quenched sample to that in the overall sample. We obtain $p \geq 0.1$, so we do not reject the null hypothesis that both samples are drawn from the same parent distribution. In other words, there is no evidence for a difference in the  distributions.

Importantly, Figure~\ref{fig:mstar-feh-z0} also shows dwarfs over a range of masses, up to $\Mstar \approx 10^5 \, \Msol$. \textsc{megatron} dwarfs have thus already undergone significant evolution and several star formation episodes. We confirm that the median distance of the quenched sample is at most 25\% away from the center of mass of the simulated volume. These dwarfs are also unlikely to restart star formation quickly, since their gas reservoir has a volume average temperature of $T = \xScientific{1.1}{4}\, \K$ (20\% increased compared to the median of the overall faint dwarf population) and a volume average $\hi$ fraction $\leq 0.5$ (30\% decreased). 

We conclude that reionization feedback is already acting on the subpopulation of dwarfs closer to the centre, where the protogalaxy is collapsing and most ionizing photons are produced. These reionization-quenched dwarfs show no difference in $\averagefeh$ compared to the rest of the population, as expected since the physics setting their iron content is primarily internal (Section~\ref{sec:popiii}). 

We further match these dwarfs to their counterpart haloes in the dark matter only simulation and find that their median peak halo mass over the whole Hubble time is $ \xScientific{8}{8} \, \Msol$. This is well below the threshold to re-accrete gas at later times once the UV background is homogeneous (see e.g.\ discussions in \citealt{Rey2020, Benitez-Llambay2020}). The dwarfs plotted in Figure~\ref{fig:mstar-feh-z0} are thus already `fossilized' at $z\approx8$, and would stay quenched at all times like observed UFDs at $z=0$. 

Another bias in the comparison between high and low redshift is that stellar evolution will induce mass-loss over the Hubble time. In Figure~\ref{fig:mstar-feh-z0}, dots show the stellar mass at $z\approx8$ while diamonds show the projected stellar mass at $z=0$ for quenched dwarf galaxies (calculated as explained in Section~\ref{sec:methods:tagging}). Accounting for this mass loss shifts stellar masses by $0.2-0.3$ dex, but makes little difference to the shape of the low-mass end of the mass-metallicity relation. 

\subsection{Future environmental post-processing by a Milky Way's tidal field} \label{sec:lowzconnection:mwtidal}

\begin{figure}
  \centering
    \includegraphics[width=\columnwidth]{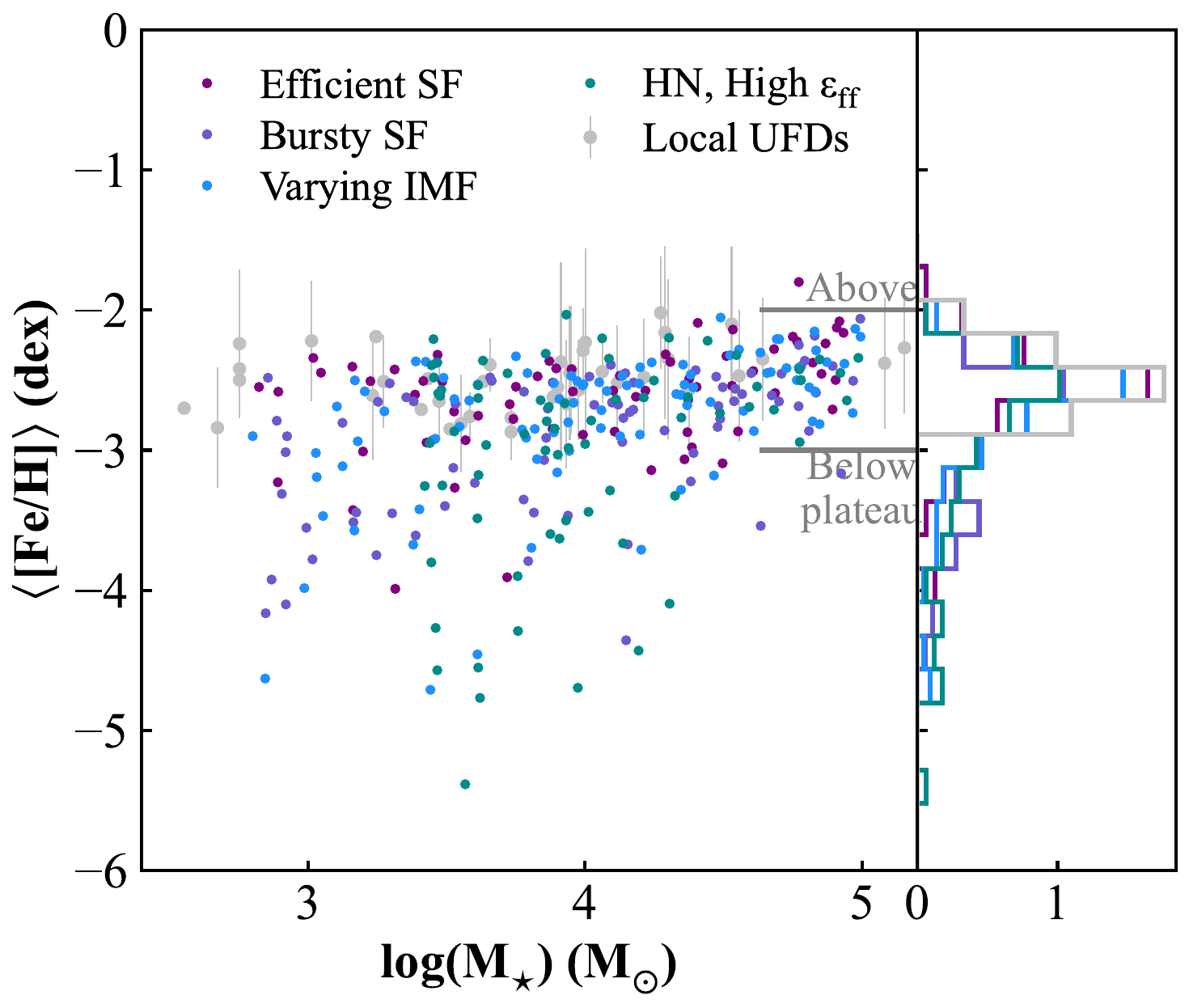}

    \caption{Same as Figure~\ref{fig:mstar-feh-suite} but for the subpopulation of dwarf galaxies that survive as $z=0$ bound satellites. The plateau at $\averagefeh \approx -2.5$ and the tail of iron-deficient dwarfs both survive in bound satellites around the host Milky Way at $z=0$.
    }
    \label{fig:massmetallicitytagging}
\end{figure}

The second major physical process at play between $z\approx8$ and $z=0$ is environmental processing of the dwarf galaxy population by the tidal field of the central galaxy. Figure~\ref{fig:massmetallicitytagging} shows the $z\approx8$ mass-metallicity relation of dwarf galaxies that can be identified at $z=0$ according to the procedure described in Section~\ref{sec:methods:tagging} (i.e.\ that have survived dynamical disruption). 

As when restricting to quenched dwarfs, the sample size significantly decreases. Overall, 25\% of the faint dwarf galaxies at $z\approx8$ can be identified as a bound remnant at $z=0$, with the rest being disrupted by the central object and instead populating the diffuse halo (Figure~\ref{fig:hero}). There is little variance in this reduction between individual \textsc{megatron} simulations, with numbers going from $N \geq 250$ to (66, 76, 92, 70) survivors for the `Efficient SF', `Bursty SF', `Varying IMF' and `HN, $\epsilon_{\rm eff}$', respectively. This is to be expected since the dark-matter-only history of the central object is the same in all cases. 

The shape of the distribution of $\averagefeh$ is unchanged by considering only identifiable dwarfs. As for the quenched sample, the Kolmogorov-Smirnov test does not indicate a significant difference between this subpopulation and the parent sample ($p=0.55$). Again, there is no evidence for a difference in the two distributions. Together with Section~\ref{sec:lowzconnection:sfhs}, this supports that the comparison between $z\approx8$ simulated dwarfs and $z=0$ observed dwarfs is quantitatively justified. Features in the mass-metallicity relation identified and explained at $z\approx8$ are not transient at high redshift, but rather robust predictions that would survive to $z=0$. 

Our analysis highlights the power of using highly-resolved and detailed high-redshift simulations to inform Galactic archaeology studies aiming to constrain how early chemical enrichment proceeded in the proto-Milky Way. This connection, however, remains incomplete. Particle-tagging techniques based on dark-matter-only simulations for dynamical forecasting are efficient, but do not account for a number of baryonic effects (e.g.\ tidal stripping by the galactic disk potential, cusp-core transitions and its connection to tidal stripping, see discussions in \citealt{Bailin2014, LeBret2017, Cooper2017}). In the future, we plan on strengthening our low-redshift forecasts by mapping onto another hydrodynamical simulation of the same initial conditions evolved to $z=0$ with the \textsc{vintergatan} model (\citealt{Agertz2020Vintergatan,Rey2022VintergatanGM}). This will allow us to account for the potential impact of disc formation on the survival of faint dwarfs (e.g.\ \citealt{DOnghia2010}).

Furthermore, there is no guarantee that the proto-Milky Way environment simulated in \textsc{megatron} precisely matches that of the Milky Way. As a result, the early star formation, chemical enrichment, and radiation background history may be different from that of our Galaxy. Future simulations varying the environment and assembly history in a controlled way will thus be essential to fill this gap, for example using the genetic modification or `splicing' techniques top introduce controlled changes (e.g. \citealt{Roth2016, Cadiou2021a}) or by performing similar simulations as \textsc{megatron} in a constrained environment resembling our Local Group (e.g. \citealt{Wempe2024}). Lastly, there remains a significant open question regarding the completeness of the observed sample of UFDs at $z=0$ and the accuracy of the inferred stellar masses and average metallicities, which we discuss in more detail in the next section.

\section{Discussion and conclusion} \label{sec:conclusion}

We have presented new results from the \textsc{megatron} simulation suite showing how Pop.~III stars drive the distribution of iron metallicities in ultra-faint dwarf galaxies. The \textsc{megatron} offers a new theoretical framework to connect high-redshift and low-redshift chemical enrichment data by (i) offering high resolution in the ISM of small dwarfs ($\approx 5 \pc$) to resolve Pop.~III and Pop.~II star formation sites; (ii) an extensive physical model (hydrodynamics, radiative transfer, non-equilibrium primordial and metal chemistry) to self-consistently capture the build-up of radiation backgrounds from the first stars to hydrogen reionization; and (iii) simulating the environment of a protogalaxy that will collapse into a Milky-Way-mass object to allow connections with local Galactic archaeology data. 

Out of four \textsc{megatron} simulations varying Pop.~II star formation and feedback models, dwarf galaxies in three simulations provide an excellent match to the observed mass-metallicity relation of UFDs at $z=0$, capturing the slope and normalisation of the relation over several decades in stellar mass (Figure~\ref{fig:mstar-feh-suite}). This agreement is non-trivial, as shown by the fourth simulation (`Efficient SF') which over-shoots in iron content for $\Mstar \geq 10^6 \, \Msol$ due to weaker galactic outflows. These results reaffirm that the mass-metallicity relation is a sensitive probe of the star formation and feedback assumptions in dwarf galaxy formation models (see also \citealt{Agertz2020EDGE,Prgomet2022, Sanati2023, Go2025, Rey2025}).

At the low-mass end of the relation ($\Mstar \leq 10^5 \, \Msol$), all \textsc{megatron} simulations showcase an over-abundance of galaxies with $\averagefeh \approx -2.5$, similar to the observed plateau at $z=0$ in the local data, and a tail of iron-deficient dwarf galaxies ($\averagefeh \leq -3.0$) that is yet to be observed in local data. Both features are robust to large variations in Pop.~II star formation and feedback variations and survive as observable features at $z=0$ (Figure~\ref{fig:mstar-feh-z0}, \ref{fig:massmetallicitytagging}).

We connect the iron plateau in \textsc{megatron} to internal enrichment from a single explosion from a high-mass Pop.~III PISNe ($\geq 180 \, \Msol$, Figure~\ref{fig:popiiinumbers}, \ref{fig:popiiitype}). The over-dense \textsc{megatron} environment drives a strong LW background, ensuring that such explosions occur in haloes ($\Mvir \approx 10^7 \, \Msol$) where metals are retained efficiently and incorporated into subsequent generations of stars (Figure~\ref{fig:yieldminihalo}). This scenario explains the normalisation of the plateau and provides qualitatively consistent carbon and magnesium abundances, although offset by $\sim 0.5$ dex relative to observations (Figure~\ref{fig:chemistry}). The tail of iron-poor galaxies is instead created by low-mass Pop.~III PISNe explosions ($\approx 170 \, \Msol$) and to a lesser extent by external pollution.

Our results support the interpretation that local UFDs are internally enriched at high redshift, with their chemical properties closely related to the microphysics of early chemical enrichment by the first stars (see also \citealt{Jeon2017, Sanati2023, Go2025}). As an example, following the \textsc{megatron} results, a complete census of iron metallicities in UFDs could be leveraged to constrain the Pop.~III IMF. In particular, the relative frequency of high-mass to low-mass PISNe would directly set the ratio between dwarf galaxies on the plateau and in the tail. A lack of such iron-poor tail might instead push the IMF towards extremely top-heavy values (i.e.\ only high-mass PISNe), or instead point to another origin to the UFD iron plateau (e.g.\ CCSNe in a lower-mass minihalo, see discussion in Section~\ref{sec:popiii:yieldminihalo}).

Current local data is yet to observe a UFD with galaxy-averaged $\averagefeh \leq -3.0$. If the underlying distribution of $\averagefeh$ in UFDs is similar to that predicted by \textsc{megatron}, we expect $\approx 20\%$ of UFDs to have such low iron contents. The probability of not observing any dwarfs with $\averagefeh \geq -3.0$ out of 38 independent samples is $(1-0.2)^{38} \approx 0.02\%$, i.e.\ disfavouring the existence of an iron-poor tail. However, the completeness of iron metallicity measurements in UFDs is sparse, both in surveying different objects and in the number of stars measured per galaxy. For example, \citet{Pace2025} recently reported the first characterization of Pictor II, finding $\averagefeh \approx -2.99$ pushing down towards the tail identified in \textsc{megatron}. Furthermore, in the same galaxy, \citet{Chiti2025} identified the first ultra iron-poor star ($\feh \leq -4.0$, not included in the average above), proving that (i) such low iron contents can be expected in UFDs; and (ii) that galaxy averages derived from low numbers of iron measurements could be biased high. To further illustrate this point, Appendix~\ref{app:mdfs} shows the composite metallicity distribution functions of \textsc{megatron} dwarfs. Even for galaxies with an average iron content close to the plateau, there is a significant spread in the metallicity distribution function ($\sigma_{\text{Fe}} \approx 0.4$ dex), with some stars reaching $\feh \leq -3.0$. In other words, even if the average iron content of UFDs is close to the plateau, individual stars can populate a long tail with $\feh \leq -3.0$ due to the combination of low-mass Pop.~III PISN and Pop.~II CCSN enrichment. The \textsc{megatron} tail of low $\averagefeh$ is just an extension in galactic average of a tail of iron-poor stars in all dwarfs that might currently be hidden by the limited number of stars observed per galaxy.

Our results thus provide strong theoretical incentives to complete the census of UFD iron metallicities. Major efforts are already underway in this direction and will likely yield fruits in the coming years. The now-online Vera C. Rubin Telescope is expected to propose many new photometric candidates of UFDs (e.g.\ \citealt{Mutlu-Pakdil2021}), while the next generation of extremely large telescopes will allow for spectroscopic follow-up of these faint objects and the measurement of their chemical abundances. Simulation efforts like \textsc{megatron} will be key to interpret these data and leverage them to constrain the physics of early chemical enrichment.

We stress that the ability of \textsc{megatron} to self-consistently model both Pop.~III and Pop.~II star formation and feedback in a cosmological-radiation hydrodynamics framework is key to our results. The emergence of the iron plateau is not a fine-tuned result, but rather stems from the combination of 
\begin{enumerate}
  \item a strong LW background in the proto- Milky Way forcing Pop.~III stars to form in higher-mass haloes that can retain metals,
  \item a top-heavy IMF favouring PISNe explosions,
  \item a high iron yield per PISNe explosion. 
\end{enumerate}
Each of these aspects carries significant uncertainties, and variations in any of them could lead to a different distribution of iron metallicities in UFDs.

Capturing the first aspect is only possible with cosmological-radiation hydrodynamics. Such setups have become increasingly common in high-redshift simulations (e.g.\ \citealt{Xu2013, Wise2014, Kimm2017, Brauer2025, Kang2025, Zier2025a}), although the requirement to resolve both Pop.~III and Pop.~II star formation sites over a large volume still poses computational challenges. Such simulations have repeatedly highlighted the inhomogeneity and environmental dependence of the LW background (e.g.\ \citealt{Xu2016, Zier2025a}). In the case of \textsc{megatron}, the LW background is always elevated due to the specific cosmic environment we have picked -- an overdense region collapsing early to form a disc galaxy at $z=0$ and which, as a result, contains more stars for its halo mass than the cosmic average at $z=8$ (\citealt{Katz2025MegP1}). As a result, irrespective of the four Pop.~II feedback models we have explored, Pop.~III stars always form in higher-mass haloes that can more efficiently retain PISN metals. This likely explains differences with simulations of isolated dwarfs, in which PISN metals are rapidly lost to the IGM (\citealt{Sanati2023}). Our results make it clear that accounting, at least partially, for the special environment of the proto Milky Way will be key for accurate predictions of early chemical enrichment observables in Galactic data.

Another important choice driving our results lies in a top-heavy Pop.~III IMF that favours PISN explosions (see Appendix~\ref{app:popiii}). This assumption is motivated by early simulations pointing out that the lack of metal cooling in primordial gas leads to gas clumps with characteristic masses $\geq 100 \, \Msol$ close to PISN progenitors (e.g.\ \citealt{Abel1998, Bromm2001, Nakamura2001, Abel2002, Bromm2002}). More recent models, however, show that hydrodynamical instabilities and turbulence during minihalo collapse likely drives additional gas fragmentation, potentially lowering the characteristic mass of primordial star forming clouds to $\leq 100 \, \Msol$ (e.g.\ \citealt{Turk2009, Greif2012, Sugimura2020, Wollenberg2020}). This in turn would lead to much less top-heavy IMF predictions than we have assumed here (e.g.\ \citealt{Susa2014, Stacy2016, Prole2022}). However, turbulence-amplified magnetic fields (e.g.\ \citealt{Sharda2020}) and the radiative input of newborn Pop.~III stars (e.g.\ \citealt{Hirano2014, Hirano2015}, though see also \citealt{Jaura2022}) can prevent fragmentation in the primordial minihalo and raise the characteristic mass again. 

Furthermore, in addition to uncertainties in the physics of primordial star formation and the IMF, there also remains significant uncertainties in the Pop.~III yields (Appendix~\ref{app:popiii}). In particular, the mass range for PISN progenitors is still debated, with some models predicting that stars as low as $\approx 140 \, \Msol$ can undergo PISN explosions (e.g.\ \citealt{Nomoto2013}), while others predict a higher threshold of $\approx 180 \, \Msol$ (e.g.\ \citealt{Heger2002}). The iron yield per PISN explosion is also uncertain, with different models predicting different yields for the same progenitor mass (see comparison in Appendix~\ref{app:popiii}), as is the explosion energy associated to each progenitor mass. These uncertainties then couple with the choice of IMF to regulate the effective iron in a minihalo which effectively sets the plateau (Section~\ref{sec:popiii:yieldminihalo}). Overall, the uncertainties surrounding predictions of the Pop.~III properties are significant (see \citealt{Klessen2023} for a more in-depth review), making our choice justified, but one among many.

Despite these uncertainties, our results show that some choices for the properties of Pop.~III stars can leave a clear imprint on the iron metallicity distribution of UFDs. Strengthening the connection between high-redshift simulations and local archaeological data as we are doing here is thus full of promises but will require a more systematic exploration of parameter space. We plan on addressing this in future work re-simulating our volume with a different Pop.~III IMF and studying the abundances and spatial distribution of all metal-poor stars, rather than just those inside dwarf galaxies. This will be highly complementary to galactic chemical enrichment models that aim to constrain the Pop.~III IMF from stellar abundances (e.g.\ \citealt{deBennassuti2017, Ishigaki2018, Rossi2021}), for example allowing to calibrate some of their free parameters (e.g.\ metal retention factors) from first principles. The end goal of this exercise will be to jointly leverage high-redshift emission line data and local chemical enrichment data to constrain the physics of early chemical enrichment. This is a timely and exciting endeavour since both JWST and Galactic spectroscopic surveys are maturing at the same time. 

\section*{Acknowledgements}
The authors thank the anonymous referees for their constructive report that improved the manuscript.  MR thanks Eric P. Andersson for useful comments on a preliminary version of this manuscript.
AJC acknowledges funding from the `FirstGalaxies' Advanced Grant from the European Research Council (ERC) under the European Union’s Horizon 2020 research and innovation programme (Grant agreement No.789056). AFS acknowledges support from the Science and Technology Facilities Council (STFC) for a PhD studentship. FRM is supported by the Kavli Institute for Cosmological physics at the University of Chicago through an endowment from the Kavli Foundation and its founder Fred Kavli. GCJ acknowledges support by the Science and Technology Facilities Council (STFC), by the ERC through Advanced Grant 695671 `QUENCH', and by the UKRI Frontier Research grant RISEandFALL. HK acknowledges support from FACCTS. The material in this manuscript is based upon work supported by NASA under award No. 80NSSC25K7009. KM acknowledges the Flemish Fund for Scientific Research (FWO-Vlaanderen), Grant number 1169822N. M.S. acknowledges the support from the Swiss National Science Foundation under Grant No. P500PT\_214488. NC acknowledges support from the Science and Technology Facilities Council (STFC) for a PhD studentship. OA acknowledges support from the Knut and Alice Wallenberg Foundation, the Swedish Research Council (grant 2019-04659), the Swedish National Space Agency (SNSA Dnr 2023-00164), and the LMK foundation. TK is supported by the National Research Foundation of Korea (RS-2022-NR070872 and RS-2025-00516961) and by the Yonsei Fellowship, funded by Lee Youn Jae.

This work made extensive use of the dp265, dp016, dp373, and dp379 projects on the DiRAC ecosystem. This work was performed using the DiRAC Data Intensive service at Leicester, operated by the University of Leicester IT Services, which forms part of the STFC DiRAC HPC Facility (www.dirac.ac.uk). The equipment was funded by BEIS capital funding via STFC capital grants ST/K000373/1 and ST/R002363/1 and STFC DiRAC Operations grant ST/R001014/1. 
This work used the DiRAC@Durham facility managed by the Institute for Computational Cosmology on behalf of the STFC DiRAC HPC Facility (www.dirac.ac.uk). The equipment was funded by BEIS capital funding via STFC capital grants ST/P002293/1, ST/R002371/1 and ST/S002502/1, Durham University and STFC operations grant ST/R000832/1. 
This work was performed using resources provided by the Cambridge Service for Data Driven Discovery (CSD3) operated by the University of Cambridge Research Computing Service (www.csd3.cam.ac.uk), provided by Dell EMC and Intel using Tier-2 funding from the Engineering and Physical Sciences Research Council (capital grant EP/T022159/1), and DiRAC funding from the Science and Technology Facilities Council (www.dirac.ac.uk). DiRAC is part of the National e-Infrastructure.
This work has made use of the Infinity Cluster hosted by Institut d'Astrophysique de Paris. We thank Stephane Rouberol for running smoothly this cluster for us.
The authors thank Jonathan Patterson for smoothly running the Glamdring Cluster hosted by the University of Oxford, where part of the data processing was performed.
The authors also acknowledge financial support from Oriel College’s Research Fund.

We thank the developers and maintainers of \textsc{pynbody} (\citealt{Pontzen2013}), \textsc{tangos} (\citealt{Pontzen2018}), \textsc{yt} (\citealt{Turk2011}), \textsc{NumPy} (\citealt{vanderWalt2011, Harris2020}), \textsc{SciPy} (\citealt{Virtanen2020}), \textsc{jupyter} (\citealt{Ragan-Kelley2014}), \textsc{matplotlib} (\citealt{Hunter2007}), the Astrophysics Data Service and the arXiv preprint repository for providing open-source software that was used extensively in this work. Data from the \textsc{megatron} simulation suite will be shared upon reasonable request to the corresponding author, along with the accompanying documentation and code to read and analyse the data.




\bibliographystyle{mn2e}
\bibliography{Meg-Mstar-Fe}

\begin{thebibliography}{287}
\expandafter\ifx\csname natexlab\endcsname\relax\def\natexlab#1{#1}\fi

\bibitem[{Abel {et~al}\mbox{.}(1998)Abel, Anninos, Norman, \& Zhang}]{Abel1998}
Abel T., Anninos P., Norman M.~L., Zhang Y., 1998, ApJ, 508, 518

\bibitem[{Abel, Bryan \& Norman(2002)Abel, Bryan, \& Norman}]{Abel2002}
Abel T., Bryan G.~L., Norman M.~L., 2002, Science, 295, 93

\bibitem[{Agertz {et~al}\mbox{.}(2020)Agertz, Pontzen, Read, Rey, Orkney, Rosdahl, Teyssier, Verbeke, Kretschmer, \& Nickerson}]{Agertz2020EDGE}
Agertz O. {et~al.}, 2020, MNRAS, 491, 1656

\bibitem[{Agertz {et~al}\mbox{.}(2021)Agertz, Renaud, Feltzing, Read, Ryde, Andersson, Rey, Bensby, \& Feuillet}]{Agertz2020Vintergatan}
Agertz O. {et~al.}, 2021, MNRAS, 503, 5826

\bibitem[{Ahn {et~al}\mbox{.}(2009)Ahn, Shapiro, Iliev, Mellema, \& Pen}]{Ahn2009}
Ahn K., Shapiro P.~R., Iliev I.~T., Mellema G., Pen U.-L., 2009, ApJ, 695, 1430

\bibitem[{Ahvazi {et~al}\mbox{.}(2024)Ahvazi, Benson, Sales, Nadler, Weerasooriya, Du, \& Bovill}]{Ahvazi2024}
Ahvazi N., Benson A., Sales L.~V., Nadler E.~O., Weerasooriya S., Du X., Bovill M.~S., 2024, MNRAS, 529, 3387

\bibitem[{Andersson {et~al}\mbox{.}(2025)Andersson, Rey, Pontzen, Cadiou, Agertz, Read, \& Martin}]{Andersson2025}
Andersson E.~P., Rey M.~P., Pontzen A., Cadiou C., Agertz O., Read J.~I., Martin N.~F., 2025, ApJ, 978, 129

\bibitem[{Aoki {et~al}\mbox{.}(2014)Aoki, Tominaga, Beers, Honda, \& Lee}]{Aoki2014}
Aoki W., Tominaga N., Beers T.~C., Honda S., Lee Y.~S., 2014, Science, 345, 912

\bibitem[{Applebaum {et~al}\mbox{.}(2021)Applebaum, Brooks, Christensen, Munshi, Quinn, Shen, \& Tremmel}]{Applebaum2021}
Applebaum E., Brooks A.~M., Christensen C.~R., Munshi F., Quinn T.~R., Shen S., Tremmel M., 2021, ApJ, 906, 96

\bibitem[{Asplund {et~al}\mbox{.}(2009)Asplund, Grevesse, Sauval, \& Scott}]{Asplund2009}
Asplund M., Grevesse N., Sauval A.~J., Scott P., 2009, ARA\&A, 47, 481

\bibitem[{Attard {et~al}\mbox{.}(2026)Attard, Conaboy, Libeskind, Pillipenko, Dixon, \& Iliev}]{Attard2026}
Attard D., Conaboy L., Libeskind N., Pillipenko S., Dixon K., Iliev I.~T., 2026, MNRAS, 547, stag467

\bibitem[{Aubert {et~al}\mbox{.}(2018)Aubert, Deparis, Ocvirk, Shapiro, Iliev, Yepes, Gottl{\"o}ber, Hoffman, \& Teyssier}]{Aubert2018}
Aubert D. {et~al.}, 2018, ApJL, 856, L22

\bibitem[{Bailin {et~al}\mbox{.}(2014)Bailin, Bell, Valluri, Stinson, Debattista, Couchman, \& Wadsley}]{Bailin2014}
Bailin J., Bell E.~F., Valluri M., Stinson G.~S., Debattista V.~P., Couchman H. M.~P., Wadsley J., 2014, ApJ, 783, 95

\bibitem[{Behroozi, Wechsler \& Wu(2013)Behroozi, Wechsler, \& Wu}]{Behroozi2013haloes}
Behroozi P.~S., Wechsler R.~H., Wu H.-Y., 2013, ApJ, 762, 109

\bibitem[{Bellazzini, Gennari \& Ferraro(2005)Bellazzini, Gennari, \& Ferraro}]{Bellazzini2005}
Bellazzini M., Gennari N., Ferraro F.~R., 2005, MNRAS, 360, 185

\bibitem[{Belokurov {et~al}\mbox{.}(2007)Belokurov, Zucker, Evans, Kleyna, Koposov, Hodgkin, Irwin, Gilmore, Wilkinson, Fellhauer, Bramich, Hewett, Vidrih, De~Jong, Smith, Rix, Bell, Wyse, Newberg, Mayeur, Yanny, Rockosi, Gnedin, Schneider, Beers, Barentine, Brewington, Brinkmann, Harvanek, Kleinman, Krzesinski, Long, Nitta, \& Snedden}]{Belokurov2007}
Belokurov V. {et~al.}, 2007, ApJ, 654, 897

\bibitem[{{Benitez-Llambay} \& Frenk(2020)}]{Benitez-Llambay2020}
{Benitez-Llambay} A., Frenk C., 2020, MNRAS, 498, 4887

\bibitem[{Benson {et~al}\mbox{.}(2002)Benson, Frenk, Lacey, Baugh, \& Cole}]{Benson2002}
Benson A.~J., Frenk C.~S., Lacey C.~G., Baugh C.~M., Cole S., 2002, MNRAS, 333, 177

\bibitem[{Bhardwaj {et~al}\mbox{.}(2024)Bhardwaj, Rejkuba, Ngeow, Marconi, Ripepi, Samantaray, \& Singh}]{Bhardwaj2024}
Bhardwaj A., Rejkuba M., Ngeow C.-C., Marconi M., Ripepi V., Samantaray A.~S., Singh H.~P., 2024, AJ, 167, 247

\bibitem[{{Bland-Hawthorn}, Sutherland \& Webster(2015){Bland-Hawthorn}, Sutherland, \& Webster}]{Bland-Hawthorn2015}
{Bland-Hawthorn} J., Sutherland R., Webster D., 2015, ApJ, 807, 154

\bibitem[{Boettcher {et~al}\mbox{.}(2013)Boettcher, Willman, Fadely, Strader, Baker, Hopkins, Tasnim~Ananna, Cunningham, Douglas, Gilbert, Preston, \& Sturner}]{Boettcher2013}
Boettcher E. {et~al.}, 2013, AJ, 146, 94

\bibitem[{Bonifacio {et~al}\mbox{.}(2025)Bonifacio, Caffau, Fran{\c c}ois, \& Spite}]{Bonifacio2025}
Bonifacio P., Caffau E., Fran{\c c}ois P., Spite M., 2025, ARA\&A, 33, 2

\bibitem[{Bovill \& Ricotti(2009)}]{Bovill2009}
Bovill M.~S., Ricotti M., 2009, ApJ, 693, 1859

\bibitem[{Brauer {et~al}\mbox{.}(2025{\natexlab{a}})Brauer, Emerick, Mead, Ji, Wise, Bryan, Mac~Low, Cote, Andersson, \& Frebel}]{Brauer2025}
Brauer K. {et~al.}, 2025{\natexlab{a}}, ApJ, 980, 41

\bibitem[{Brauer {et~al}\mbox{.}(2025{\natexlab{b}})Brauer, Mead, Wise, Bryan, Low, Ji, Emerick, Andersson, Frebel, \& C{\^o}t{\'e}}]{Brauer2025PopIIIIMF}
Brauer K. {et~al.}, 2025{\natexlab{b}}, ApJ, 993, 2

\bibitem[{Bromm, Coppi \& Larson(2002)Bromm, Coppi, \& Larson}]{Bromm2002}
Bromm V., Coppi P.~S., Larson R.~B., 2002, ApJ, 564, 23

\bibitem[{Bromm {et~al}\mbox{.}(2001)Bromm, Ferrara, Coppi, \& Larson}]{Bromm2001}
Bromm V., Ferrara A., Coppi P.~S., Larson R.~B., 2001, MNRAS, 328, 969

\bibitem[{Bromm, Yoshida \& Hernquist(2003)Bromm, Yoshida, \& Hernquist}]{Bromm2003}
Bromm V., Yoshida N., Hernquist L., 2003, ApJ, 596, L135

\bibitem[{Brown {et~al}\mbox{.}(2014)Brown, Tumlinson, Geha, Simon, Vargas, VandenBerg, Kirby, Kalirai, Avila, Gennaro, Ferguson, Mu{\~n}oz, Guhathakurta, \& Renzini}]{Brown2014}
Brown T.~M. {et~al.}, 2014, ApJ, 796, 91

\bibitem[{Bruce {et~al}\mbox{.}(2023)Bruce, Li, Pace, Heiger, Song, \& Simon}]{Bruce2023}
Bruce J., Li T.~S., Pace A.~B., Heiger M., Song Y.-Y., Simon J.~D., 2023, ApJ, 950, 167

\bibitem[{Bullock \& Johnston(2005)}]{Bullock2005}
Bullock J.~S., Johnston K.~V., 2005, ApJ, 635, 931

\bibitem[{Bullock, Kravtsov \& Weinberg(2000)Bullock, Kravtsov, \& Weinberg}]{Bullock2000}
Bullock J.~S., Kravtsov A.~V., Weinberg D.~H., 2000, ApJ, 539, 517

\bibitem[{Cadiou, Dubois \& Pichon(2019)Cadiou, Dubois, \& Pichon}]{Cadiou2019}
Cadiou C., Dubois Y., Pichon C., 2019, A\&A, 621, A96

\bibitem[{Cadiou {et~al}\mbox{.}(2026)Cadiou, Katz, Rey, Agertz, Blaizot, Cameron, Choustikov, Devriendt, Hauk, Jones, Kimm, Laseter, {\'A}lvarez, Matsumoto, Nyhagen, Pearce, Montero, Rosdahl, Pastor, Sanati, Saxena, Slyz, Stiskalek, Storck, \& Yee}]{Cadiou2025MegP1}
Cadiou C. {et~al.}, 2026, OpJA, Accepted

\bibitem[{Cadiou {et~al}\mbox{.}(2021)Cadiou, Pontzen, Peiris, \& {Lucie-Smith}}]{Cadiou2021a}
Cadiou C., Pontzen A., Peiris H.~V., {Lucie-Smith} L., 2021, MNRAS, 508, 1189

\bibitem[{Cameron {et~al}\mbox{.}(2023)Cameron, Katz, Rey, \& Saxena}]{Cameron2023GNz11}
Cameron A.~J., Katz H., Rey M.~P., Saxena A., 2023, MNRAS, 523, 3516

\bibitem[{Cantu {et~al}\mbox{.}(2021)Cantu, Pace, Marshall, Strigari, Crnojevic, Simon, {Drlica-Wagner}, Bechtol, {Mart{\'i}nez-V{\'a}zquez}, Santiago, Amara, Stringer, Diehl, Aguena, Allam, Avila, Brooks, Carnero~Rosell, Carrasco~Kind, Carretero, Costanzi, Da~Costa, De~Vicente, Desai, Doel, Eifler, Everett, Frieman, {Garc{\'i}a-Bellido}, Gaztanaga, Gruen, Gruendl, Gschwend, Gutierrez, Hinton, Hollowood, Honscheid, James, Kuehn, Maia, Menanteau, Miquel, Palmese, {Paz-Chinch{\'o}n}, Plazas, Sanchez, Scarpine, Schubnell, Serrano, {Sevilla-Noarbe}, Smith, {Soares-Santos}, Suchyta, Swanson, Tarle, Walker, Wilkinson, \& Collaboration}]{Cantu2021}
Cantu S.~A. {et~al.}, 2021, ApJ, 916, 81

\bibitem[{Carlin {et~al}\mbox{.}(2009)Carlin, Grillmair, Mu{\~n}oz, Nidever, \& Majewski}]{Carlin2009}
Carlin J.~L., Grillmair C.~J., Mu{\~n}oz R.~R., Nidever D.~L., Majewski S.~R., 2009, ApJL, 702, L9

\bibitem[{Carlin \& Sand(2018)}]{Carlin2018}
Carlin J.~L., Sand D.~J., 2018, ApJ, 865, 7

\bibitem[{Carlin {et~al}\mbox{.}(2017)Carlin, Sand, Mu{\~n}oz, Spekkens, Willman, Crnojevi{\'c}, Forbes, Hargis, Kirby, Peter, Romanowsky, \& Strader}]{Carlin2017}
Carlin J.~L. {et~al.}, 2017, AJ, 154, 267

\bibitem[{Casey {et~al}\mbox{.}(2025)Casey, {Mutlu-Pakdil}, Sand, Pace, Crnojevic, {Doliva-Dolinsky}, Cerny, Heiger, Riley, Ji, Limberg, Marin, {Mart{\'i}nez-V{\'a}zquez}, Medina, Li, Campana, Chaturvedi, Sakowska, Zenteno, {Carballo-Bello}, Navabi, \& Bom}]{Casey2025}
Casey Q.~O. {et~al.}, 2025, ApJ, 984, 148

\bibitem[{Cerny {et~al}\mbox{.}(2025)Cerny, Chiti, Geha, {Mutlu-Pakdil}, {Drlica-Wagner}, Tan, Adam{\'o}w, Pace, Simon, Sand, Ji, Li, Vivas, Bell, Carlin, {Carballo-Bello}, Chaturvedi, Choi, {Doliva-Dolinsky}, Gnedin, Limberg, {Mart{\'i}nez-V{\'a}zquez}, Mau, Medina, Navabi, No{\"e}l, Placco, Riley, Roederer, Stringfellow, Bom, Ferguson, James, {Mart{\'i}nez-Delgado}, Massana, Nidever, Sakowska, {Santana-Silva}, Sherman, \& Tollerud}]{Cerny2025}
Cerny W. {et~al.}, 2025, ApJ, 979, 164

\bibitem[{Cerny {et~al}\mbox{.}(2023{\natexlab{a}})Cerny, {Drlica-Wagner}, Li, Pace, Olsen, No{\"e}l, {van der Marel}, Carlin, Choi, Erkal, Geha, James, {Mart{\'i}nez-V{\'a}zquez}, Massana, Medina, Miller, {Mutlu-Pakdil}, Nidever, Sakowska, Stringfellow, {Carballo-Bello}, Ferguson, Kuropatkin, Mau, Tollerud, \& Vivas}]{Cerny2023DELVE6}
Cerny W. {et~al.}, 2023{\natexlab{a}}, ApJL, 953, L21

\bibitem[{Cerny {et~al}\mbox{.}(2023{\natexlab{b}})Cerny, {Mart{\'i}nez-V{\'a}zquez}, {Drlica-Wagner}, Pace, {Mutlu-Pakdil}, Li, Riley, Crnojevi{\'c}, Bom, {Carballo-Bello}, Carlin, Chiti, Choi, Collins, {Darragh-Ford}, Ferguson, Geha, {Mart{\'i}nez-Delgado}, Massana, Mau, Medina, Mu{\~n}oz, Nadler, Olsen, Pieres, Sakowska, Simon, Stringfellow, Vivas, Walker, \& Wechsler}]{Cerny2023DELVE}
Cerny W. {et~al.}, 2023{\natexlab{b}}, ApJ, 953, 1

\bibitem[{Cerny {et~al}\mbox{.}(2021{\natexlab{a}})Cerny, Pace, {Drlica-Wagner}, Ferguson, Mau, Adam{\'o}w, Carlin, Choi, Erkal, Johnson, Li, {Mart{\'i}nez-V{\'a}zquez}, {Mutlu-Pakdil}, Nidever, Olsen, Pieres, Simon, Tollerud, Vivas, James, Kuropatkin, Majewski, {Mart{\'i}nez-Delgado}, Massana, Miller, No{\"e}l, Riley, Sand, {Santana-Silva}, Stringfellow, Neilsen, \& Tucker}]{Cerny2021a}
Cerny W. {et~al.}, 2021{\natexlab{a}}, ApJ, 910, 18

\bibitem[{Cerny {et~al}\mbox{.}(2021{\natexlab{b}})Cerny, Pace, {Drlica-Wagner}, Koposov, Vivas, Mau, Riley, Bom, Carlin, Choi, Erkal, Ferguson, James, Li, {Mart{\'i}nez-Delgado}, {Mart{\'i}nez-V{\'a}zquez}, Munoz, {Mutlu-Pakdil}, Olsen, Pieres, Sakowska, Sand, Simon, Smercina, Stringfellow, Tollerud, Adam{\'o}w, {Hernandez-Lang}, Kuropatkin, {Santana-Silva}, Tucker, Zenteno, \& Collaboration}]{Cerny2021}
Cerny W. {et~al.}, 2021{\natexlab{b}}, ApJL, 920, L44

\bibitem[{Cerny {et~al}\mbox{.}(2023{\natexlab{c}})Cerny, Simon, Li, {Drlica-Wagner}, Pace, {Mart{\'i}nez-V{\'a}zquez}, Riley, {Mutlu-Pakdil}, Mau, Ferguson, Erkal, Munoz, Bom, Carlin, Carollo, Choi, Ji, Manwadkar, {Mart{\'i}nez-Delgado}, Miller, No{\"e}l, Sakowska, Sand, Stringfellow, Tollerud, Vivas, {Carballo-Bello}, {Hernandez-Lang}, James, Nidever, Castellon, Olsen, Zenteno, \& {Delve Collaboration}}]{Cerny2023PegIV}
Cerny W. {et~al.}, 2023{\natexlab{c}}, ApJ, 942, 111

\bibitem[{Charbonnel {et~al}\mbox{.}(2023)Charbonnel, Schaerer, Prantzos, {Ram{\'i}rez-Galeano}, Fragos, Kuruvandothi, {Marques-Chaves}, \& Gieles}]{Charbonnel2023}
Charbonnel C., Schaerer D., Prantzos N., {Ram{\'i}rez-Galeano} L., Fragos T., Kuruvandothi A., {Marques-Chaves} R., Gieles M., 2023, A\&A, 673, L7

\bibitem[{Chiti {et~al}\mbox{.}(2023)Chiti, Frebel, Ji, Mardini, Ou, Simon, Rasmussen, Jerjen, Kim, \& Norris}]{Chiti2023}
Chiti A. {et~al.}, 2023, AJ, 165, 55

\bibitem[{Chiti {et~al}\mbox{.}(2021)Chiti, Frebel, Simon, Erkal, Chang, Necib, Ji, Jerjen, Kim, \& Norris}]{Chiti2021}
Chiti A. {et~al.}, 2021, Nat Ast, 5, 392

\bibitem[{Chiti {et~al}\mbox{.}(2026)Chiti, Placco, Pace, Ji, Prabhu, Cerny, Limberg, Stringfellow, {Drlica-Wagner}, Atzberger, Choi, Crnojevi{\'c}, Ferguson, Kallivayalil, No{\"e}l, Riley, Sand, Simon, Walker, Bom, {Carballo-Bello}, James, {Mart{\'i}nez-V{\'a}zquez}, Medina, \& Vivas}]{Chiti2025}
Chiti A. {et~al.}, 2026, Nat Ast

\bibitem[{Chiti {et~al}\mbox{.}(2022)Chiti, Simon, Frebel, Pace, Ji, \& Li}]{Chiti2022GrusI}
Chiti A., Simon J.~D., Frebel A., Pace A.~B., Ji A.~P., Li T.~S., 2022, ApJ, 939, 41

\bibitem[{Choustikov {et~al}\mbox{.}(2026)Choustikov, Katz, Cameron, Saxena, Devriendt, Slyz, Rey, Cadiou, Blaizot, Kimm, Laseter, Matsumoto, \& Rosdahl}]{Choustikov2025MegP1}
Choustikov N. {et~al.}, 2026, OpJA, 9, 58199

\bibitem[{Cicu{\'e}ndez {et~al}\mbox{.}(2018)Cicu{\'e}ndez, Battaglia, Irwin, {Bermejo-Climent}, McMonigal, Bate, Lewis, Conn, {de Boer}, Gallart, Guglielmo, Ibata, McConnachie, Tolstoy, \& Fernando}]{Cicuendez2018}
Cicu{\'e}ndez L. {et~al.}, 2018, A\&A, 609, A53

\bibitem[{Cooke \& Madau(2014)}]{Cooke2014}
Cooke R.~J., Madau P., 2014, ApJ, 791, 116

\bibitem[{Cooper {et~al}\mbox{.}(2017)Cooper, Cole, Frenk, Le~Bret, \& Pontzen}]{Cooper2017}
Cooper A.~P., Cole S., Frenk C.~S., Le~Bret T., Pontzen A., 2017, MNRAS, 469, 1691

\bibitem[{Cooper {et~al}\mbox{.}(2010)Cooper, Cole, Frenk, White, Helly, Benson, De~Lucia, Helmi, Jenkins, Navarro, Springel, \& Wang}]{Cooper2010}
Cooper A.~P. {et~al.}, 2010, MNRAS, 406, 744

\bibitem[{Correnti, Bellazzini \& Ferraro(2009)Correnti, Bellazzini, \& Ferraro}]{Correnti2009}
Correnti M., Bellazzini M., Ferraro F.~R., 2009, MNRAS, 397, L26

\bibitem[{Crnojevi{\'c} {et~al}\mbox{.}(2016)Crnojevi{\'c}, Sand, Zaritsky, Spekkens, Willman, \& Hargis}]{Crnojevic2016}
Crnojevi{\'c} D., Sand D.~J., Zaritsky D., Spekkens K., Willman B., Hargis J.~R., 2016, ApJ, 824, L14

\bibitem[{Cullen {et~al}\mbox{.}(2021)Cullen, Shapley, McLure, Dunlop, Sanders, Topping, Reddy, Amor{\'i}n, Begley, Bolzonella, Calabr{\`o}, Carnall, Castellano, Cimatti, Cirasuolo, Cresci, Fontana, Fontanot, Garilli, Guaita, Hamadouche, Hathi, Mannucci, McLeod, Pentericci, Saxena, Talia, \& Zamorani}]{Cullen2021}
Cullen F. {et~al.}, 2021, MNRAS, 505, 903

\bibitem[{Curti {et~al}\mbox{.}(2023)Curti, D'Eugenio, Carniani, Maiolino, Sandles, Witstok, Baker, Bennett, Piotrowska, Tacchella, Charlot, Nakajima, Maheson, Mannucci, Arribas, Belfiore, Bonaventura, Bunker, Chevallard, Cresci, {Curtis-Lake}, {Hayden-Pawson}, Kumari, Laseter, Looser, Marconi, Maseda, Jones, Scholtz, Smit, Ubler, \& Wallace}]{Curti2023Te}
Curti M. {et~al.}, 2023, MNRAS, 518, 425

\bibitem[{Curti {et~al}\mbox{.}(2024)Curti, Maiolino, Carniani, D'Eugenio, Chevallard, {Curtis-Lake}, Looser, Scholtz, {\"U}bler, Witstok, Cameron, Charlot, Laseter, Sandles, Arribas, Bunker, Giardino, Maseda, Rawle, Del~Pino, Smit, Willott, Eisenstein, Hausen, Johnson, Rieke, Robertson, Tacchella, Williams, Willmer, Baker, Bhatawdekar, Boyett, Egami, Helton, Ji, Kumari, Shivaei, \& Sun}]{Curti2024LowMassEnd}
Curti M. {et~al.}, 2024, A\&A, 684, A75

\bibitem[{Dall'Ora {et~al}\mbox{.}(2006)Dall'Ora, Clementini, Kinemuchi, Ripepi, Marconi, Di~Fabrizio, Greco, Rodgers, Kuehn, \& Smith}]{DallOra2006}
Dall'Ora M. {et~al.}, 2006, ApJ, 653, L109

\bibitem[{Dall'Ora {et~al}\mbox{.}(2012)Dall'Ora, Kinemuchi, Ripepi, Rodgers, Clementini, Di~Fabrizio, Smith, Marconi, Musella, Greco, Kuehn, Catelan, Pritzl, \& Beers}]{DallOra2012}
Dall'Ora M. {et~al.}, 2012, ApJ, 752, 42

\bibitem[{{de Bennassuti} {et~al}\mbox{.}(2017){de Bennassuti}, Salvadori, Schneider, Valiante, \& Omukai}]{deBennassuti2017}
{de Bennassuti} M., Salvadori S., Schneider R., Valiante R., Omukai K., 2017, MNRAS, 465, 926

\bibitem[{De~Lucia \& Helmi(2008)}]{DeLucia2008}
De~Lucia G., Helmi A., 2008, MNRAS, 391, 14

\bibitem[{Dixon {et~al}\mbox{.}(2018)Dixon, Iliev, Gottl{\"o}ber, Yepes, Knebe, Libeskind, \& Hoffman}]{Dixon2018}
Dixon K.~L., Iliev I.~T., Gottl{\"o}ber S., Yepes G., Knebe A., Libeskind N., Hoffman Y., 2018, MNRAS, 477, 867

\bibitem[{D'Onghia {et~al}\mbox{.}(2010)D'Onghia, Springel, Hernquist, \& Keres}]{DOnghia2010}
D'Onghia E., Springel V., Hernquist L., Keres D., 2010, ApJ, 709, 1138

\bibitem[{{Drlica-Wagner} {et~al}\mbox{.}(2015){Drlica-Wagner}, Bechtol, Rykoff, Luque, Queiroz, Mao, Wechsler, Simon, Santiago, Yanny, Balbinot, Dodelson, Fausti~Neto, James, Li, Maia, Marshall, Pieres, Stringer, Walker, Abbott, Abdalla, Allam, {Benoit-L{\'e}vy}, Bernstein, Bertin, Brooks, {Buckley-Geer}, Burke, Carnero~Rosell, Carrasco~Kind, Carretero, Crocce, {da Costa}, Desai, Diehl, Dietrich, Doel, Eifler, Evrard, Finley, Flaugher, Fosalba, Frieman, Gaztanaga, Gerdes, Gruen, Gruendl, Gutierrez, Honscheid, Kuehn, Kuropatkin, Lahav, Martini, Miquel, Nord, Ogando, Plazas, Reil, Roodman, Sako, Sanchez, Scarpine, Schubnell, {Sevilla-Noarbe}, Smith, {Soares-Santos}, Sobreira, Suchyta, Swanson, Tarle, Tucker, Vikram, Wester, Zhang, Zuntz, \& {DES Collaboration}}]{Drlica-Wagner2015}
{Drlica-Wagner} A. {et~al.}, 2015, ApJ, 813, 109

\bibitem[{Durbin {et~al}\mbox{.}(2025)Durbin, Choi, Savino, Weisz, Dolphin, Dalcanton, Jeon, Kallivayalil, Li, Pace, Patel, Sacchi, Skillman, Sohn, van~der Marel, Wetzel, \& Williams}]{Durbin2025}
Durbin M.~J. {et~al.}, 2025, MNRAS, 992, 106

\bibitem[{Efstathiou(1992)}]{Efstathiou1992}
Efstathiou G., 1992, MNRAS, 256, 43P

\bibitem[{Escala {et~al}\mbox{.}(2018)Escala, Wetzel, Kirby, Hopkins, Ma, Wheeler, Kere{\v s}, {Faucher-Gigu{\`e}re}, \& Quataert}]{Escala2018}
Escala I. {et~al.}, 2018, MNRAS, 474, 2194

\bibitem[{Farmer {et~al}\mbox{.}(2019)Farmer, Renzo, {de Mink}, Marchant, \& Justham}]{Farmer2019}
Farmer R., Renzo M., {de Mink} S.~E., Marchant P., Justham S., 2019, ApJ, 887, 53

\bibitem[{Frebel \& Norris(2015)}]{Frebel2015}
Frebel A., Norris J.~E., 2015, ARA\&A, 53, 631

\bibitem[{Fritz {et~al}\mbox{.}(2019)Fritz, Carrera, Battaglia, \& Taibi}]{Fritz2019}
Fritz T.~K., Carrera R., Battaglia G., Taibi S., 2019, A\&A, 623, A129

\bibitem[{Fu {et~al}\mbox{.}(2023)Fu, Weisz, Starkenburg, Martin, Savino, {Boylan-Kolchin}, Cote, Dolphin, Ji, Longeard, Mateo, Patel, \& Sandford}]{Fu2023}
Fu S.~W. {et~al.}, 2023, ApJ, 958, 167

\bibitem[{Garofalo {et~al}\mbox{.}(2025)Garofalo, Clementini, Cusano, Muraveva, \& Monti}]{Garofalo2025}
Garofalo A., Clementini G., Cusano F., Muraveva T., Monti L., 2025, A\&A, 695, A88

\bibitem[{Garofalo {et~al}\mbox{.}(2013)Garofalo, Cusano, Clementini, Ripepi, Dall'Ora, Moretti, Coppola, Musella, \& Marconi}]{Garofalo2013}
Garofalo A. {et~al.}, 2013, ApJ, 767, 62

\bibitem[{Gnedin(2000)}]{Gnedin2000}
Gnedin N.~Y., 2000, ApJ, 542, 535

\bibitem[{Go {et~al}\mbox{.}(2025)Go, Jeon, Choi, Kallivayalil, Sohn, Besla, Richstein, Fu, Jeong, \& Shin}]{Go2025}
Go M. {et~al.}, 2025, ApJ, 986, 214

\bibitem[{Greco {et~al}\mbox{.}(2008)Greco, Dall'Ora, Clementini, Ripepi, Di~Fabrizio, Kinemuchi, Marconi, Musella, Smith, Rodgers, Kuehn, Beers, Catelan, \& Pritzl}]{Greco2008}
Greco C. {et~al.}, 2008, ApJ, 675, L73

\bibitem[{Greif {et~al}\mbox{.}(2011)Greif, White, Klessen, \& Springel}]{Greif2011}
Greif T., White S., Klessen R., Springel V., 2011, ApJ, 736, 147

\bibitem[{Greif {et~al}\mbox{.}(2012)Greif, Bromm, Clark, Glover, Smith, Klessen, Yoshida, \& Springel}]{Greif2012}
Greif T.~H., Bromm V., Clark P.~C., Glover S. C.~O., Smith R.~J., Klessen R.~S., Yoshida N., Springel V., 2012, MNRAS, 424, 399

\bibitem[{Greif {et~al}\mbox{.}(2007)Greif, Johnson, Bromm, \& Klessen}]{Greif2007}
Greif T.~H., Johnson J.~L., Bromm V., Klessen R.~S., 2007, ApJ, 670, 1

\bibitem[{Griffen {et~al}\mbox{.}(2018)Griffen, Dooley, Ji, O'Shea, G{\'o}mez, \& Frebel}]{Griffen2018}
Griffen B.~F., Dooley G.~A., Ji A.~P., O'Shea B.~W., G{\'o}mez F.~A., Frebel A., 2018, MNRAS, 474, 443

\bibitem[{Hansen {et~al}\mbox{.}(2024)Hansen, Simon, Li, Sharkey, Ji, Thompson, Reggiani, \& Galarza}]{Hansen2024}
Hansen T.~T., Simon J.~D., Li T.~S., Sharkey D., Ji A.~P., Thompson I.~B., Reggiani H.~M., Galarza J.~Y., 2024, ApJ, 968, 21

\bibitem[{Harris {et~al}\mbox{.}(2020)Harris, Millman, {van der Walt}, Gommers, Virtanen, Cournapeau, Wieser, Taylor, Berg, Smith, Kern, Picus, Hoyer, {van Kerkwijk}, Brett, Haldane, {del R{\'i}o}, Wiebe, Peterson, {G{\'e}rard-Marchant}, Sheppard, Reddy, Weckesser, Abbasi, Gohlke, \& Oliphant}]{Harris2020}
Harris C.~R. {et~al.}, 2020, Nature, 585, 357

\bibitem[{Hartwig {et~al}\mbox{.}(2015)Hartwig, Bromm, Klessen, \& Glover}]{Hartwig2015}
Hartwig T., Bromm V., Klessen R.~S., Glover S. C.~O., 2015, MNRAS, 447, 3892

\bibitem[{Heger \& Woosley(2002)}]{Heger2002}
Heger A., Woosley S.~E., 2002, ApJ, 567, 532

\bibitem[{Heiger {et~al}\mbox{.}(2024)Heiger, Li, Pace, Simon, Ji, Chiti, Bom, {Carballo-Bello}, Carlin, Cerny, Choi, {Drlica-Wagner}, James, {Mart{\'i}nez-V{\'a}zquez}, Medina, {Mutlu-Pakdil}, Navabi, No{\"e}l, Sakowska, \& Stringfellow}]{Heiger2024}
Heiger M.~E. {et~al.}, 2024, ApJ, 961, 234

\bibitem[{Heintz {et~al}\mbox{.}(2023)Heintz, Brammer, {Gim{\'e}nez-Arteaga}, Lagos, Vijayan, Matthee, Watson, Mason, Hutter, Toft, Fynbo, Oesch, \& Strait}]{Heintz2023a}
Heintz K.~E. {et~al.}, 2023, Nat Ast, 7, 1517

\bibitem[{Hirano {et~al}\mbox{.}(2015)Hirano, Hosokawa, Yoshida, Omukai, \& Yorke}]{Hirano2015}
Hirano S., Hosokawa T., Yoshida N., Omukai K., Yorke H.~W., 2015, MNRAS, 448, 568

\bibitem[{Hirano {et~al}\mbox{.}(2014)Hirano, Hosokawa, Yoshida, Umeda, Omukai, Chiaki, \& Yorke}]{Hirano2014}
Hirano S., Hosokawa T., Yoshida N., Umeda H., Omukai K., Chiaki G., Yorke H.~W., 2014, ApJ, 781, 60

\bibitem[{Hirschi(2007)}]{Hirschi2007}
Hirschi R., 2007, A\&A, 461, 571

\bibitem[{Homma {et~al}\mbox{.}(2019)Homma, Chiba, Komiyama, Tanaka, Okamoto, Tanaka, Ishigaki, Hayashi, Arimoto, Carlsten, Lupton, Strauss, Miyazaki, Torrealba, Wang, \& Murayama}]{Homma2019}
Homma D. {et~al.}, 2019, PASJ, 71, 94

\bibitem[{Homma {et~al}\mbox{.}(2024)Homma, Chiba, Komiyama, Tanaka, Okamoto, Tanaka, Ishigaki, Hayashi, Arimoto, Lupton, Strauss, Miyazaki, Wang, \& Murayama}]{Homma2024}
Homma D. {et~al.}, 2024, PASJ, 76, 733

\bibitem[{Homma {et~al}\mbox{.}(2018)Homma, Chiba, Okamoto, Komiyama, Tanaka, Tanaka, Ishigaki, Hayashi, Arimoto, Garmilla, Lupton, Strauss, Miyazaki, Wang, \& Murayama}]{Homma2018}
Homma D. {et~al.}, 2018, PASJ, 70, S18

\bibitem[{Hsiao {et~al}\mbox{.}(2025)Hsiao, Topping, Coe, Chisholm, Berg, Abdurro'uf, {\'A}lvarez-M{\'a}rquez, Maiolino, Dayal, \& Furtak}]{Hsiao2024Carbon}
Hsiao T. Y.-Y. {et~al.}, 2025, ApJ, 993, 70

\bibitem[{Hunter(2007)}]{Hunter2007}
Hunter J.~D., 2007, CiSE, 9, 90

\bibitem[{Ishigaki {et~al}\mbox{.}(2018)Ishigaki, Tominaga, Kobayashi, \& Nomoto}]{Ishigaki2018}
Ishigaki M.~N., Tominaga N., Kobayashi C., Nomoto K., 2018, ApJ, 857, 46

\bibitem[{Isobe {et~al}\mbox{.}(2023)Isobe, Ouchi, Tominaga, Watanabe, Nakajima, Umeda, Yajima, Harikane, Fukushima, Xu, Ono, \& Zhang}]{Isobe2023LowC/N}
Isobe Y. {et~al.}, 2023, ApJ, 959, 100

\bibitem[{Jaura {et~al}\mbox{.}(2022)Jaura, Glover, Wollenberg, Klessen, Geen, \& Haemmerl{\'e}}]{Jaura2022}
Jaura O., Glover S. C.~O., Wollenberg K. M.~J., Klessen R.~S., Geen S., Haemmerl{\'e} L., 2022, MNRAS, 512, 116

\bibitem[{Jenkins {et~al}\mbox{.}(2021)Jenkins, Li, Pace, Ji, Koposov, \& {Mutlu-Pakdil}}]{Jenkins2021}
Jenkins S.~A., Li T.~S., Pace A.~B., Ji A.~P., Koposov S.~E., {Mutlu-Pakdil} B., 2021, ApJ, 920, 92

\bibitem[{Jeon, Besla \& Bromm(2017)Jeon, Besla, \& Bromm}]{Jeon2017}
Jeon M., Besla G., Bromm V., 2017, ApJ, 848, 85

\bibitem[{Jeon {et~al}\mbox{.}(2021)Jeon, Bromm, Besla, Yoon, \& Choi}]{Jeon2021FaintPopIIISNe}
Jeon M., Bromm V., Besla G., Yoon J., Choi Y., 2021, MNRAS, 502, 1

\bibitem[{Jessop {et~al}\mbox{.}(2026)Jessop, Jenkins, Pontzen, Schaye, Schaller, \& Helly}]{Jessop2026}
Jessop O., Jenkins A., Pontzen A., Schaye J., Schaller M., Helly J.~C., 2026, MNRAS, 547, 519

\bibitem[{Ji, Frebel \& Bromm(2015)Ji, Frebel, \& Bromm}]{Ji2015}
Ji A.~P., Frebel A., Bromm V., 2015, MNRAS, 454, 659

\bibitem[{Ji {et~al}\mbox{.}(2021)Ji, Koposov, Li, Erkal, Pace, Simon, Belokurov, Cullinane, Da~Costa, Kuehn, Lewis, Mackey, Shipp, Simpson, Zucker, Hansen, {Bland-Hawthorn}, \& Collaboration}]{Ji2021}
Ji A.~P. {et~al.}, 2021, ApJ, 921, 32

\bibitem[{Ji {et~al}\mbox{.}(2019)Ji, Simon, Frebel, Venn, \& Hansen}]{Ji2019}
Ji A.~P., Simon J.~D., Frebel A., Venn K.~A., Hansen T.~T., 2019, ApJ, 870, 83

\bibitem[{Jiang {et~al}\mbox{.}(2024)Jiang, Zhao, Li, \& Xing}]{Jiang2024b}
Jiang R., Zhao G., Li H., Xing Q., 2024, ApJ, 976, 68

\bibitem[{Joshi {et~al}\mbox{.}(2025)Joshi, Pontzen, Agertz, Read, \& Rey}]{Joshi2025WholeSuite}
Joshi G.~D., Pontzen A., Agertz O., Read J., Rey M.~P., 2025, MNRAS, 544, 2811

\bibitem[{Kang {et~al}\mbox{.}(2025)Kang, Kimm, Han, Katz, Devriendt, Slyz, \& Teyssier}]{Kang2025}
Kang C., Kimm T., Han D., Katz H., Devriendt J., Slyz A., Teyssier R., 2025, A\&A, 693, A149

\bibitem[{Karczmarek {et~al}\mbox{.}(2015)Karczmarek, Pietrzy{\'n}ski, Gieren, Suchomska, Konorski, G{\'o}rski, Pilecki, Graczyk, \& Wielg{\'o}rski}]{Karczmarek2015}
Karczmarek P. {et~al.}, 2015, AJ, 150, 90

\bibitem[{Katz(2022)}]{Katz2022RTZ}
Katz H., 2022, MNRAS, 512, 348

\bibitem[{Katz {et~al}\mbox{.}(2023)Katz, Kimm, Ellis, Devriendt, \& Slyz}]{Katz2023PopIII}
Katz H., Kimm T., Ellis R.~S., Devriendt J., Slyz A., 2023, MNRAS, 524, 351

\bibitem[{Katz {et~al}\mbox{.}(2022)Katz, Liu, Kimm, Rey, Andersson, Cameron, {Rodriguez-Montero}, Agertz, Devriendt, \& Slyz}]{Katz2022PRISM}
Katz H. {et~al.}, 2022, arXiv e-prints, 2211.04626

\bibitem[{Katz {et~al}\mbox{.}(2020)Katz, Ramsoy, Rosdahl, Kimm, Blaizot, Haehnelt, {Michel-Dansac}, Garel, Laigle, Devriendt, \& Slyz}]{Katz2020}
Katz H. {et~al.}, 2020, MNRAS, 494, 2200

\bibitem[{Katz {et~al}\mbox{.}(2026{\natexlab{a}})Katz, Rey, Cadiou, Agertz, Blaizot, Cameron, Choustikov, Devriendt, Hauk, Jones, Kimm, Laseter, {Martin-Alvarez}, Matsumoto, Pearce, Montero, Rosdahl, Sanati, Saxena, Slyz, Stiskalek, Storck, Veenema, \& Yee}]{Katz2025MegP1}
Katz H. {et~al.}, 2026{\natexlab{a}}, OpJA, Accepted

\bibitem[{Katz {et~al}\mbox{.}(2026{\natexlab{b}})Katz, Rey, Cadiou, Kimm, \& Agertz}]{Katz2024MegPilot}
Katz H., Rey M.~P., Cadiou C., Kimm T., Agertz O., 2026{\natexlab{b}}, OpJA, 9, 56097

\bibitem[{Kawashimo {et~al}\mbox{.}(2024)Kawashimo, Sawada, Suwa, Moriya, Tanikawa, \& Tominaga}]{Kawashimo2024}
Kawashimo H., Sawada R., Suwa Y., Moriya T.~J., Tanikawa A., Tominaga N., 2024, MNRAS, 531, 2786

\bibitem[{Kim \& Jerjen(2015)}]{Kim2015Horologium}
Kim D., Jerjen H., 2015, ApJ, 808, L39

\bibitem[{Kim {et~al}\mbox{.}(2016)Kim, Jerjen, Geha, Chiti, Milone, Da~Costa, Mackey, Frebel, \& Conn}]{Kim2016PegIII}
Kim D. {et~al.}, 2016, ApJ, 833, 16

\bibitem[{Kimm {et~al}\mbox{.}(2017)Kimm, Katz, Haehnelt, Rosdahl, Devriendt, \& Slyz}]{Kimm2017}
Kimm T., Katz H., Haehnelt M., Rosdahl J., Devriendt J., Slyz A., 2017, MNRAS, 466, 4826

\bibitem[{Kirby {et~al}\mbox{.}(2013)Kirby, Cohen, Guhathakurta, Cheng, Bullock, \& Gallazzi}]{Kirby2013}
Kirby E.~N., Cohen J.~G., Guhathakurta P., Cheng L., Bullock J.~S., Gallazzi A., 2013, ApJ, 779, 102

\bibitem[{Kirby {et~al}\mbox{.}(2015)Kirby, Cohen, Simon, \& Guhathakurta}]{Kirby2015}
Kirby E.~N., Cohen J.~G., Simon J.~D., Guhathakurta P., 2015, ApJ, 814, L7

\bibitem[{Kirby {et~al}\mbox{.}(2017)Kirby, Rizzi, Held, Cohen, Cole, Manning, Skillman, \& Weisz}]{Kirby2017}
Kirby E.~N., Rizzi L., Held E.~V., Cohen J.~G., Cole A.~A., Manning E.~M., Skillman E.~D., Weisz D.~R., 2017, ApJ, 834, 9

\bibitem[{Kitayama \& Yoshida(2005)}]{Kitayama2005}
Kitayama T., Yoshida N., 2005, ApJ, 630, 675

\bibitem[{Klessen \& Glover(2023)}]{Klessen2023}
Klessen R.~S., Glover S. C.~O., 2023, ARA\&A, 61, 65

\bibitem[{Kobayashi {et~al}\mbox{.}(2014)Kobayashi, Ishigaki, Tominaga, \& Nomoto}]{Kobayashi2014}
Kobayashi C., Ishigaki M.~N., Tominaga N., Nomoto K., 2014, ApJL, 785, L5

\bibitem[{Kobayashi \& Nakasato(2011)}]{Kobayashi2011}
Kobayashi C., Nakasato N., 2011, ApJ, 729

\bibitem[{Koposov {et~al}\mbox{.}(2015{\natexlab{a}})Koposov, Belokurov, Torrealba, \& Evans}]{Koposov2015}
Koposov S.~E., Belokurov V., Torrealba G., Evans N.~W., 2015{\natexlab{a}}, ApJ, 805, 130

\bibitem[{Koposov {et~al}\mbox{.}(2015{\natexlab{b}})Koposov, Casey, Belokurov, Lewis, Gilmore, Worley, Hourihane, Randich, Bensby, Bragaglia, Bergemann, Carraro, Costado, Flaccomio, Francois, Heiter, Hill, Jofre, Lando, Lanzafame, {de Laverny}, Monaco, Morbidelli, Sbordone, Mikolaitis, \& Ryde}]{Koposov2015RetII}
Koposov S.~E. {et~al.}, 2015{\natexlab{b}}, ApJ, 811, 62

\bibitem[{Koposov {et~al}\mbox{.}(2011)Koposov, Gilmore, Walker, Belokurov, Evans, Fellhauer, Gieren, Geisler, Monaco, Norris, Okamoto, Pe{\~n}arrubia, Wilkinson, Wyse, \& Zucker}]{Koposov2011}
Koposov S.~E. {et~al.}, 2011, ApJ, 736, 146

\bibitem[{Koposov {et~al}\mbox{.}(2018)Koposov, Walker, Belokurov, Casey, {Geringer-Sameth}, Mackey, Da~Costa, Erkal, Jethwa, Mateo, Olszewski, \& Bailey}]{Koposov2018}
Koposov S.~E. {et~al.}, 2018, MNRAS, 479, 5343

\bibitem[{Koutsouridou {et~al}\mbox{.}(2023)Koutsouridou, Salvadori, Sk{\'u}lad{\'o}ttir, Rossi, Vanni, \& Pagnini}]{Koutsouridou2023a}
Koutsouridou I., Salvadori S., Sk{\'u}lad{\'o}ttir {\'A}., Rossi M., Vanni I., Pagnini G., 2023, MNRAS, 525, 190

\bibitem[{Koutsouridou, Sk{\'u}lad{\'o}ttir \& Salvadori(2025)Koutsouridou, Sk{\'u}lad{\'o}ttir, \& Salvadori}]{Koutsouridou2025}
Koutsouridou I., Sk{\'u}lad{\'o}ttir {\'A}., Salvadori S., 2025, A\&A, 699, 32

\bibitem[{Kroupa(2001)}]{Kroupa2001}
Kroupa P., 2001, MNRAS, 322, 231

\bibitem[{Kuehn {et~al}\mbox{.}(2008)Kuehn, Kinemuchi, Ripepi, Clementini, Dall'Ora, Di~Fabrizio, Rodgers, Greco, Marconi, Musella, Smith, Catelan, Beers, \& Pritzl}]{Kuehn2008}
Kuehn C. {et~al.}, 2008, ApJL, 674, L81

\bibitem[{Kulkarni {et~al}\mbox{.}(2019)Kulkarni, Keating, Haehnelt, Bosman, Puchwein, Chardin, \& Aubert}]{Kulkarni2019ReionizationHistory}
Kulkarni G., Keating L.~C., Haehnelt M.~G., Bosman S. E.~I., Puchwein E., Chardin J., Aubert D., 2019, MNRAS, 485, L24

\bibitem[{Laseter {et~al}\mbox{.}(2024)Laseter, Maseda, Curti, Maiolino, D'Eugenio, Cameron, Looser, Arribas, Baker, Bhatawdekar, Boyett, Bunker, Carniani, Charlot, Chevallard, {Curtis-lake}, Egami, Eisenstein, Hainline, Hausen, Ji, Kumari, Perna, Rawle, Rix, Robertson, Del~Pino, Sandles, Scholtz, Smit, Tacchella, {\"U}bler, Williams, Willott, \& Witstok}]{Laseter2024Calibration}
Laseter I.~H. {et~al.}, 2024, A\&A, 681, A70

\bibitem[{Le~Bret {et~al}\mbox{.}(2017)Le~Bret, Pontzen, Cooper, Frenk, Zolotov, Brooks, Governato, \& Parry}]{LeBret2017}
Le~Bret T., Pontzen A., Cooper A.~P., Frenk C., Zolotov A., Brooks A.~M., Governato F., Parry O.~H., 2017, MNRAS, 468, 3212

\bibitem[{Lee {et~al}\mbox{.}(2009)Lee, Yuk, Park, Harris, \& Zaritsky}]{Lee2009}
Lee M.~G., Yuk I.-S., Park H.~S., Harris J., Zaritsky D., 2009, ApJ, 703, 692

\bibitem[{Li {et~al}\mbox{.}(2017)Li, Simon, {Drlica-Wagner}, Bechtol, Wang, {Garc{\'i}a-Bellido}, Frieman, Marshall, James, Strigari, Pace, Balbinot, Zhang, Abbott, Allam, {Benoit-L{\'e}vy}, Bernstein, Bertin, Brooks, Burke, Carnero~Rosell, Carrasco~Kind, Carretero, Cunha, D'Andrea, {da Costa}, DePoy, Desai, Diehl, Eifler, Flaugher, Goldstein, Gruen, Gruendl, Gschwend, Gutierrez, Krause, Kuehn, Lin, Maia, March, Menanteau, Miquel, Plazas, Romer, Sanchez, Santiago, Schubnell, {Sevilla-Noarbe}, Smith, Sobreira, Suchyta, Tarle, Thomas, Tucker, Walker, Wechsler, Wester, Yanny, \& {DES Collaboration}}]{Li2017}
Li T.~S. {et~al.}, 2017, ApJ, 838, 8

\bibitem[{Li {et~al}\mbox{.}(2018)Li, Simon, Pace, Torrealba, Kuehn, {Drlica-Wagner}, Bechtol, Vivas, {van der Marel}, Wood, Yanny, Belokurov, Jethwa, Zucker, Lewis, Kron, Nidever, {S{\'a}nchez-Conde}, Ji, Conn, James, Martin, {Martinez-Delgado}, No{\"e}l, \& {MagLiteS Collaboration}}]{Li2018}
Li T.~S. {et~al.}, 2018, ApJ, 857, 145

\bibitem[{Li {et~al}\mbox{.}(2025)Li, Kakiichi, Christensen, Cai, Dekel, Fan, Farina, Jun, Li, Li, Pudoka, Sun, Trebitsch, Walter, Wang, Yang, Zhang, \& Zou}]{Li2025MZR}
Li Z. {et~al.}, 2025, A\&A, 703, A106

\bibitem[{Limongi \& Chieffi(2018)}]{Limongi2018}
Limongi M., Chieffi A., 2018, ApJS, 237, 13

\bibitem[{Liu {et~al}\mbox{.}(2021)Liu, Sibony, Meynet, \& Bromm}]{Liu2021a}
Liu B., Sibony Y., Meynet G., Bromm V., 2021, MNRAS, 506, 5247

\bibitem[{Longeard {et~al}\mbox{.}(2018)Longeard, Martin, Starkenburg, Ibata, Collins, Geha, Laevens, Rich, Aguado, Arentsen, Carlberg, C{\^o}t{\'e}, Hill, Jablonka, Gonz{\'a}lez~Hern{\'a}ndez, Navarro, {S{\'a}nchez-Janssen}, Tolstoy, Venn, \& Youakim}]{Longeard2018}
Longeard N. {et~al.}, 2018, MNRAS, 480, 2609

\bibitem[{Madau {et~al}\mbox{.}(2024)Madau, Giallongo, Grazian, \& Haardt}]{Madau2024}
Madau P., Giallongo E., Grazian A., Haardt F., 2024, ApJ, 971, 75

\bibitem[{Magg {et~al}\mbox{.}(2018)Magg, Hartwig, Agarwal, Frebel, Glover, Griffen, \& Klessen}]{Magg2018}
Magg M., Hartwig T., Agarwal B., Frebel A., Glover S. C.~O., Griffen B.~F., Klessen R.~S., 2018, MNRAS, 473, 5308

\bibitem[{Maoz \& Graur(2017)}]{Maoz2017}
Maoz D., Graur O., 2017, ApJ, 848, 25

\bibitem[{Marks {et~al}\mbox{.}(2012)Marks, Kroupa, Dabringhausen, \& Pawlowski}]{Marks2012}
Marks M., Kroupa P., Dabringhausen J., Pawlowski M.~S., 2012, MNRAS, 422, 2246

\bibitem[{{Mart{\'i}nez-V{\'a}zquez} {et~al}\mbox{.}(2021){Mart{\'i}nez-V{\'a}zquez}, Cerny, Vivas, {Drlica-Wagner}, Pace, Simon, Munoz, Walker, Allam, Tucker, Adam{\'o}w, Carlin, Choi, Ferguson, Ji, Kuropatkin, Li, {Mart{\'i}nez-Delgado}, Mau, {Mutlu-Pakdil}, Nidever, Riley, Sakowska, Sand, \& Stringfellow}]{Martinez-Vazquez2021}
{Mart{\'i}nez-V{\'a}zquez} C.~E. {et~al.}, 2021, AJ, 162, 253

\bibitem[{{Mart{\'i}nez-V{\'a}zquez} {et~al}\mbox{.}(2015){Mart{\'i}nez-V{\'a}zquez}, Monelli, Bono, Stetson, Ferraro, Bernard, Gallart, Fiorentino, Iannicola, \& Udalski}]{Martinez-Vazquez2015}
{Mart{\'i}nez-V{\'a}zquez} C.~E. {et~al.}, 2015, MNRAS, 454, 1509

\bibitem[{{Mart{\'i}nez-V{\'a}zquez} {et~al}\mbox{.}(2019){Mart{\'i}nez-V{\'a}zquez}, Vivas, Gurevich, Walker, McCarthy, Pace, Stringer, Santiago, Hounsell, Macri, Li, Bechtol, Riley, Kim, Simon, {Drlica-Wagner}, Nadler, Marshall, Annis, Avila, Bertin, Brooks, {Buckley-Geer}, Burke, Carnero~Rosell, Carrasco~Kind, {da Costa}, De~Vicente, Desai, Diehl, Doel, Everett, Frieman, {Garc{\'i}a-Bellido}, Gaztanaga, Gruen, Gruendl, Gschwend, Gutierrez, Hollowood, Honscheid, James, Kuehn, Kuropatkin, Maia, Menanteau, Miller, Miquel, {Paz-Chinch{\'o}n}, Plazas, Sanchez, Scarpine, Serrano, {Sevilla-Noarbe}, Smith, {Soares-Santos}, Sobreira, Swanson, Tarle, Vikram, \& Collaboration}]{Martinez-Vazquez2019}
{Mart{\'i}nez-V{\'a}zquez} C.~E. {et~al.}, 2019, MNRAS, 490, 2183

\bibitem[{Mateo, Olszewski \& Walker(2008)Mateo, Olszewski, \& Walker}]{Mateo2008}
Mateo M., Olszewski E.~W., Walker M.~G., 2008, ApJ, 675, 201

\bibitem[{McConnachie(2012)}]{McConnachie2012}
McConnachie A.~W., 2012, AJ, 144, 4

\bibitem[{Medina {et~al}\mbox{.}(2018)Medina, Mu{\~n}oz, Vivas, Carlin, F{\"o}rster, Mart{\'i}nez, Galbany, {Gonz{\'a}lez-Gait{\'a}n}, Hamuy, {de Jaeger}, Maureira, \& San~Mart{\'i}n}]{Medina2018}
Medina G.~E. {et~al.}, 2018, ApJ, 855, 43

\bibitem[{Moskowitz \& Walker(2020)}]{Moskowitz2020}
Moskowitz A.~G., Walker M.~G., 2020, ApJ, 892, 27

\bibitem[{Mu{\~n}oz {et~al}\mbox{.}(2018)Mu{\~n}oz, C{\^o}t{\'e}, Santana, Geha, Simon, Oyarz{\'u}n, Stetson, \& Djorgovski}]{Munoz2018}
Mu{\~n}oz R.~R., C{\^o}t{\'e} P., Santana F.~A., Geha M., Simon J.~D., Oyarz{\'u}n G.~A., Stetson P.~B., Djorgovski S.~G., 2018, ApJ, 860, 66

\bibitem[{Munshi {et~al}\mbox{.}(2019)Munshi, Brooks, Christensen, Applebaum, {Holley-Bockelmann}, Quinn, \& Wadsley}]{Munshi2019}
Munshi F., Brooks A.~M., Christensen C., Applebaum E., {Holley-Bockelmann} K., Quinn T.~R., Wadsley J., 2019, ApJ, 874, 40

\bibitem[{Musella {et~al}\mbox{.}(2009)Musella, Ripepi, Clementini, Dall'Ora, Kinemuchi, {di Fabrizio}, Greco, Marconi, Smith, Radovich, \& Beers}]{Musella2009}
Musella I. {et~al.}, 2009, ApJ, 695, L83

\bibitem[{{Mutlu-Pakdil} {et~al}\mbox{.}(2018){Mutlu-Pakdil}, Sand, Carlin, Spekkens, Caldwell, Crnojevi{\'c}, Hughes, Willman, \& Zaritsky}]{Mutlu-Pakdil2018}
{Mutlu-Pakdil} B. {et~al.}, 2018, ApJ, 863, 25

\bibitem[{{Mutlu-Pakdil} {et~al}\mbox{.}(2021){Mutlu-Pakdil}, Sand, Crnojevi{\'c}, {Drlica-Wagner}, Caldwell, Guhathakurta, Seth, Simon, Strader, \& Toloba}]{Mutlu-Pakdil2021}
{Mutlu-Pakdil} B. {et~al.}, 2021, ApJ, 918, 88

\bibitem[{{Mutlu-Pakdil} {et~al}\mbox{.}(2020){Mutlu-Pakdil}, Sand, Crnojevi{\'c}, Olszewski, Zaritsky, Strader, Collins, Seth, \& Willman}]{Mutlu-Pakdil2020}
{Mutlu-Pakdil} B. {et~al.}, 2020, ApJ, 902, 106

\bibitem[{Nakamura \& Umemura(2001)}]{Nakamura2001}
Nakamura F., Umemura M., 2001, ApJ, 548, 19

\bibitem[{Nakane {et~al}\mbox{.}(2025)Nakane, Ouchi, Nakajima, Ono, Harikane, Isobe, Nomoto, Ishigaki, Yanagisawa, Kashino, Tominaga, Takahashi, Nishigaki, Takeda, \& Watanabe}]{Nakane2025}
Nakane M. {et~al.}, 2025, ApJ, 994, 65

\bibitem[{Naoz, Yoshida \& Gnedin(2012)Naoz, Yoshida, \& Gnedin}]{Naoz2012}
Naoz S., Yoshida N., Gnedin N.~Y., 2012, ApJ, 747, 128

\bibitem[{Naoz, Yoshida \& Gnedin(2013)Naoz, Yoshida, \& Gnedin}]{Naoz2013}
Naoz S., Yoshida N., Gnedin N.~Y., 2013, ApJ, 763, 27

\bibitem[{Noh \& McQuinn(2014)}]{Noh2014}
Noh Y., McQuinn M., 2014, MNRAS, 444, 503

\bibitem[{Nomoto, Kobayashi \& Tominaga(2013)Nomoto, Kobayashi, \& Tominaga}]{Nomoto2013}
Nomoto K., Kobayashi C., Tominaga N., 2013, ARA\&A, 51, 457

\bibitem[{Nomoto {et~al}\mbox{.}(2006)Nomoto, Tominaga, Umeda, Kobayashi, \& Maeda}]{Nomoto2006}
Nomoto K., Tominaga N., Umeda H., Kobayashi C., Maeda K., 2006, NuPhA, 777, 424

\bibitem[{Oakes {et~al}\mbox{.}(2022)Oakes, Hoyt, Freedman, Madore, Tran, Cerny, Beaton, \& Seibert}]{Oakes2022}
Oakes E.~K., Hoyt T.~J., Freedman W.~L., Madore B.~F., Tran Q.~H., Cerny W., Beaton R.~L., Seibert M., 2022, ApJ, 929, 116

\bibitem[{Ocvirk {et~al}\mbox{.}(2020)Ocvirk, Aubert, Sorce, Shapiro, Deparis, Dawoodbhoy, Lewis, Teyssier, Yepes, Gottl{\"o}ber, Ahn, Iliev, \& Hoffman}]{Ocvirk2020}
Ocvirk P. {et~al.}, 2020, MNRAS, 496, 4087

\bibitem[{Ocvirk {et~al}\mbox{.}(2014)Ocvirk, Gillet, Aubert, Knebe, Libeskind, Chardin, Gottl{\"o}ber, Yepes, \& Hoffman}]{Ocvirk2014}
Ocvirk P. {et~al.}, 2014, ApJ, 794, 20

\bibitem[{Okamoto {et~al}\mbox{.}(2012)Okamoto, Arimoto, Yamada, \& Onodera}]{Okamoto2012}
Okamoto S., Arimoto N., Yamada Y., Onodera M., 2012, ApJ, 744, 96

\bibitem[{O'Leary \& McQuinn(2012)}]{OLeary2012}
O'Leary R.~M., McQuinn M., 2012, ApJ, 760, 4

\bibitem[{Omukai {et~al}\mbox{.}(2005)Omukai, Tsuribe, Schneider, \& Ferrara}]{Omukai2005}
Omukai K., Tsuribe T., Schneider R., Ferrara A., 2005, ApJ, 626, 627

\bibitem[{O{\~n}orbe {et~al}\mbox{.}(2015)O{\~n}orbe, {Boylan-Kolchin}, Bullock, Hopkins, Kere{\v s}, {Faucher-Gigu{\`e}re}, Quataert, \& Murray}]{Onorbe2015}
O{\~n}orbe J., {Boylan-Kolchin} M., Bullock J.~S., Hopkins P.~F., Kere{\v s} D., {Faucher-Gigu{\`e}re} C.-A., Quataert E., Murray N., 2015, MNRAS, 454, 2092

\bibitem[{Paardekooper, Khochfar \& Dalla~Vecchia(2015)Paardekooper, Khochfar, \& Dalla~Vecchia}]{Paardekooper2015}
Paardekooper J.-P., Khochfar S., Dalla~Vecchia C., 2015, MNRAS, 451, 2544

\bibitem[{Paardekooper {et~al}\mbox{.}(2011)Paardekooper, Pelupessy, Altay, \& Kruip}]{Paardekooper2011}
Paardekooper J.-P., Pelupessy F.~I., Altay G., Kruip C. J.~H., 2011, A\&A, 530, A87

\bibitem[{Pace(2025)}]{Pace2025LVDatabase}
Pace A.~B., 2025, OpJA, 8

\bibitem[{Pace {et~al}\mbox{.}(2020)Pace, Kaplinghat, Kirby, Simon, Tollerud, Mu{\~n}oz, C{\^o}t{\'e}, Djorgovski, \& Geha}]{Pace2020}
Pace A.~B. {et~al.}, 2020, MNRAS, 495, 3022

\bibitem[{Pace {et~al}\mbox{.}(2025)Pace, Li, Ji, Simon, Cerny, Senkevich, {Drlica-Wagner}, Bechtol, Tan, Chiti, Erkal, {Mart{\'i}nez-V{\'a}zquez}, Ferguson, Kron, Atzberger, Chaturvedi, Frieman, Kallivayalil, Limberg, Medina, Placco, Riley, Sand, Stringfellow, {van der Marel}, {Carballo-Bello}, Choi, Crnojevi{\'c}, Massana, {Mutlu-Pakdil}, Navabi, No{\"e}l, \& Sakowska}]{Pace2025}
Pace A.~B. {et~al.}, 2025, OpJA, 8, 112

\bibitem[{Pillepich {et~al}\mbox{.}(2018)Pillepich, Springel, Nelson, Genel, Naiman, Pakmor, Hernquist, Torrey, Vogelsberger, Weinberger, \& Marinacci}]{Pillepich2018TNGModel}
Pillepich A. {et~al.}, 2018, MNRAS, 473, 4077

\bibitem[{{Planck Collaboration} {et~al}\mbox{.}(2020){Planck Collaboration}, Aghanim, Akrami, Ashdown, Aumont, Baccigalupi, Ballardini, Banday, Barreiro, Bartolo, Basak, Battye, Benabed, Bernard, Bersanelli, Bielewicz, Bock, Bond, Borrill, Bouchet, Boulanger, Bucher, Burigana, Butler, Calabrese, Cardoso, Carron, Challinor, Chiang, Chluba, Colombo, Combet, Contreras, Crill, Cuttaia, {de Bernardis}, {de Zotti}, Delabrouille, Delouis, Di~Valentino, Diego, Dor{\'e}, Douspis, Ducout, Dupac, Dusini, Efstathiou, Elsner, En{\ss}lin, Eriksen, Fantaye, Farhang, Fergusson, {Fernandez-Cobos}, Finelli, Forastieri, Frailis, Fraisse, Franceschi, Frolov, Galeotta, Galli, Ganga, {G{\'e}nova-Santos}, Gerbino, Ghosh, {Gonz{\'a}lez-Nuevo}, G{\'o}rski, Gratton, Gruppuso, Gudmundsson, Hamann, Handley, Hansen, Herranz, Hildebrandt, Hivon, Huang, Jaffe, Jones, Karakci, Keih{\"a}nen, Keskitalo, Kiiveri, Kim, Kisner, Knox, Krachmalnicoff, Kunz, {Kurki-Suonio}, Lagache, Lamarre, Lasenby, Lattanzi, Lawrence, Le~Jeune, Lemos, Lesgourgues, Levrier, Lewis, Liguori, Lilje, Lilley, Lindholm, {L{\'o}pez-Caniego}, Lubin, Ma, {Mac{\'i}as-P{\'e}rez}, Maggio, Maino, Mandolesi, Mangilli, {Marcos-Caballero}, Maris, Martin, Martinelli, {Mart{\'i}nez-Gonz{\'a}lez}, Matarrese, Mauri, McEwen, Meinhold, Melchiorri, Mennella, Migliaccio, Millea, Mitra, {Miville-Desch{\^e}nes}, Molinari, Montier, Morgante, Moss, Natoli, {N{\o}rgaard-Nielsen}, Pagano, Paoletti, Partridge, Patanchon, Peiris, Perrotta, Pettorino, Piacentini, Polastri, Polenta, Puget, Rachen, Reinecke, Remazeilles, Renzi, Rocha, Rosset, Roudier, {Rubi{\~n}o-Mart{\'i}n}, {Ruiz-Granados}, Salvati, Sandri, Savelainen, Scott, Shellard, Sirignano, Sirri, Spencer, Sunyaev, {Suur-Uski}, Tauber, Tavagnacco, Tenti, Toffolatti, Tomasi, Trombetti, Valenziano, Valiviita, Van~Tent, Vibert, Vielva, Villa, Vittorio, Wandelt, Wehus, White, White, Zacchei, \& Zonca}]{PlanckCollaboration2020}
{Planck Collaboration} {et~al.}, 2020, A\&A, 641, A6

\bibitem[{Pontzen {et~al}\mbox{.}(2021)Pontzen, Rey, Cadiou, Agertz, Teyssier, Read, \& Orkney}]{Pontzen2021}
Pontzen A., Rey M.~P., Cadiou C., Agertz O., Teyssier R., Read J., Orkney M. D.~A., 2021, MNRAS, 501, 1755

\bibitem[{Pontzen {et~al}\mbox{.}(2013)Pontzen, Ro{\v s}kar, Stinson, \& Woods}]{Pontzen2013}
Pontzen A., Ro{\v s}kar R., Stinson G., Woods R., 2013, Astrophysics Source Code Library, ascl:1305.002

\bibitem[{Pontzen \& Tremmel(2018)}]{Pontzen2018}
Pontzen A., Tremmel M., 2018, ApJS, 237, 23

\bibitem[{Prgomet {et~al}\mbox{.}(2022)Prgomet, Rey, Andersson, Segovia~Otero, Agertz, Renaud, Pontzen, \& Read}]{Prgomet2022}
Prgomet M., Rey M.~P., Andersson E.~P., Segovia~Otero A., Agertz O., Renaud F., Pontzen A., Read J.~I., 2022, MNRAS, 513, 2326

\bibitem[{Prole {et~al}\mbox{.}(2022)Prole, Clark, Klessen, \& Glover}]{Prole2022}
Prole L.~R., Clark P.~C., Klessen R.~S., Glover S. C.~O., 2022, MNRAS, 510, 4019

\bibitem[{{Ragan-Kelley} {et~al}\mbox{.}(2014){Ragan-Kelley}, Perez, Granger, Kluyver, Ivanov, Frederic, \& Bussonnier}]{Ragan-Kelley2014}
{Ragan-Kelley} M., Perez F., Granger B., Kluyver T., Ivanov P., Frederic J., Bussonnier M., 2014, Am. Geophys. Un., 2014, H44D

\bibitem[{Rasera \& Teyssier(2006)}]{Rasera2006}
Rasera Y., Teyssier R., 2006, A\&A, 445, 1

\bibitem[{Rey {et~al}\mbox{.}(2023)Rey, Agertz, Starkenburg, Renaud, Joshi, Pontzen, Martin, Feuillet, \& Read}]{Rey2022VintergatanGM}
Rey M.~P. {et~al.}, 2023, MNRAS, 521, 995

\bibitem[{Rey \& Pontzen(2018)}]{Rey2018}
Rey M.~P., Pontzen A., 2018, MNRAS, 474, 45

\bibitem[{Rey {et~al}\mbox{.}(2020)Rey, Pontzen, Agertz, Orkney, Read, \& Rosdahl}]{Rey2020}
Rey M.~P., Pontzen A., Agertz O., Orkney M. D.~A., Read J.~I., Rosdahl J., 2020, MNRAS, 497, 1508

\bibitem[{Rey \& Starkenburg(2022)}]{Rey2022}
Rey M.~P., Starkenburg T.~K., 2022, MNRAS, 510, 4208

\bibitem[{Rey {et~al}\mbox{.}(2025)Rey, Taylor, Gray, Kim, Andersson, Pontzen, Agertz, Read, Cadiou, Yates, Orkney, Scholte, Saintonge, Breneman, McQuinn, Muni, \& Das}]{Rey2025}
Rey M.~P. {et~al.}, 2025, MNRAS, 541, 1195

\bibitem[{Richstein {et~al}\mbox{.}(2024)Richstein, Kallivayalil, Simon, Garling, Wetzel, Warfield, {van der Marel}, Jeon, Rose, Torrey, Engelhardt, Besla, Choi, Geha, Guhathakurta, Kirby, Patel, Sacchi, \& Sohn}]{Richstein2024}
Richstein H. {et~al.}, 2024, ApJ, 967, 72

\bibitem[{Richstein {et~al}\mbox{.}(2022)Richstein, Patel, Kallivayalil, Simon, Zivick, Tollerud, Fritz, Warfield, Besla, {van der Marel}, Wetzel, Choi, Deason, Geha, Guhathakurta, Jeon, Kirby, Libralato, Sacchi, \& Sohn}]{Richstein2022}
Richstein H. {et~al.}, 2022, ApJ, 933, 217

\bibitem[{Ricotti \& Gnedin(2005)}]{Ricotti2005}
Ricotti M., Gnedin N.~Y., 2005, ApJ, 629, 259

\bibitem[{Ritter {et~al}\mbox{.}(2015)Ritter, Sluder, {Safranek-Shrader}, Milosavljevi{\'c}, \& Bromm}]{Ritter2015}
Ritter J.~S., Sluder A., {Safranek-Shrader} C., Milosavljevi{\'c} M., Bromm V., 2015, MNRAS, 451, 1190

\bibitem[{Rosdahl {et~al}\mbox{.}(2013)Rosdahl, Blaizot, Aubert, Stranex, \& Teyssier}]{Rosdahl2013}
Rosdahl J., Blaizot J., Aubert D., Stranex T., Teyssier R., 2013, MNRAS, 436, 2188

\bibitem[{Rosdahl {et~al}\mbox{.}(2018)Rosdahl, Katz, Blaizot, Kimm, {Michel-Dansac}, Garel, Haehnelt, Ocvirk, \& Teyssier}]{Rosdahl2018}
Rosdahl J. {et~al.}, 2018, MNRAS, 479, 994

\bibitem[{Rossi, Salvadori \& Sk{\'u}lad{\'o}ttir(2021)Rossi, Salvadori, \& Sk{\'u}lad{\'o}ttir}]{Rossi2021}
Rossi M., Salvadori S., Sk{\'u}lad{\'o}ttir {\'A}., 2021, MNRAS, 503, 6026

\bibitem[{Rossi {et~al}\mbox{.}(2023)Rossi, Salvadori, Sk{\'u}lad{\'o}ttir, \& Vanni}]{Rossi2023}
Rossi M., Salvadori S., Sk{\'u}lad{\'o}ttir {\'A}., Vanni I., 2023, MNRAS, 522, L1

\bibitem[{Rossi {et~al}\mbox{.}(2025)Rossi, Salvadori, Sk{\'u}lad{\'o}ttir, Vanni, \& Koutsouridou}]{Rossi2025}
Rossi M., Salvadori S., Sk{\'u}lad{\'o}ttir {\'A}., Vanni I., Koutsouridou I., 2025, ApJ, 987, 121

\bibitem[{Roth, Pontzen \& Peiris(2016)Roth, Pontzen, \& Peiris}]{Roth2016}
Roth N., Pontzen A., Peiris H.~V., 2016, MNRAS, 455, 974

\bibitem[{Salvadori {et~al}\mbox{.}(2019)Salvadori, Bonifacio, Caffau, Korotin, Andreevsky, Spite, \& Skuladottir}]{Salvadori2019}
Salvadori S., Bonifacio P., Caffau E., Korotin S., Andreevsky S., Spite M., Skuladottir A., 2019, MNRAS, 487, 4261

\bibitem[{Salvadori \& Ferrara(2009)}]{Salvadori2009}
Salvadori S., Ferrara A., 2009, MNRAS, 395, L6

\bibitem[{Sanati {et~al}\mbox{.}(2023)Sanati, Jeanquartier, Revaz, \& Jablonka}]{Sanati2023}
Sanati M., Jeanquartier F., Revaz Y., Jablonka P., 2023, A\&A, 669, A94

\bibitem[{Sand {et~al}\mbox{.}(2012)Sand, Strader, Willman, Zaritsky, McLeod, Caldwell, Seth, \& Olszewski}]{Sand2012}
Sand D.~J., Strader J., Willman B., Zaritsky D., McLeod B., Caldwell N., Seth A., Olszewski E., 2012, ApJ, 756, 79

\bibitem[{Sanders {et~al}\mbox{.}(2024)Sanders, Shapley, Topping, Reddy, \& Brammer}]{Sanders2023DirectTe}
Sanders R.~L., Shapley A.~E., Topping M.~W., Reddy N.~A., Brammer G.~B., 2024, ApJ, 962, 24

\bibitem[{Sandford {et~al}\mbox{.}(2026)Sandford, Li, Koposov, Hayashi, Pace, Erkal, Bovy, Costa, Cullinane, Ji, Kuehn, Zucker, Limberg, Medina, Simon, Yang, \& Collaboration}]{Sandford2025}
Sandford N.~R. {et~al.}, 2026, ApJ, 998, 47

\bibitem[{Savino {et~al}\mbox{.}(2025)Savino, Weisz, Skillman, Dolphin, Cole, Kallivayalil, Wetzel, Anderson, Besla, {Boylan-Kolchin}, Brown, Bullock, Collins, Cooper, Deason, Dotter, Fardal, Ferguson, Fritz, Geha, Gilbert, Guhathakurta, Ibata, Irwin, Jeon, Kirby, Lewis, Mackey, Majewski, Martin, McConnachie, Patel, Rich, Simon, Sohn, Tollerud, \& {van der Marel}}]{Savino2025M31SFHs}
Savino A. {et~al.}, 2025, ApJ, 979, 205

\bibitem[{Schaerer(2002)}]{Schaerer2002}
Schaerer D., 2002, A\&A, 382, 28

\bibitem[{Schaerer {et~al}\mbox{.}(2024)Schaerer, {Marques-Chaves}, Xiao, \& Korber}]{Schaerer2024}
Schaerer D., {Marques-Chaves} R., Xiao M., Korber D., 2024, A\&A, 687, L11

\bibitem[{Schauer {et~al}\mbox{.}(2023)Schauer, {Boylan-Kolchin}, Colston, Sameie, Bromm, Bullock, \& Wetzel}]{Schauer2022}
Schauer A. T.~P., {Boylan-Kolchin} M., Colston K., Sameie O., Bromm V., Bullock J.~S., Wetzel A., 2023, ApJ, 950, 20

\bibitem[{Schauer {et~al}\mbox{.}(2019)Schauer, Glover, Klessen, \& Ceverino}]{Schauer2019}
Schauer A. T.~P., Glover S. C.~O., Klessen R.~S., Ceverino D., 2019, MNRAS, 484, 3510

\bibitem[{Scholte {et~al}\mbox{.}(2025)Scholte, Cullen, Carnall, {Arellano-C{\'o}rdova}, Stanton, Barrufet, Donnan, Dunlop, Leung, McLeod, McLure, Moustakas, Pollock, Shapley, Stevenson, \& Zou}]{Scholte2025MZR}
Scholte D. {et~al.}, 2025, MNRAS, 540, 1800

\bibitem[{Seitenzahl {et~al}\mbox{.}(2013)Seitenzahl, {Ciaraldi-Schoolmann}, R{\"o}pke, Fink, Hillebrandt, Kromer, Pakmor, Ruiter, Sim, \& Taubenberger}]{Seitenzahl2013}
Seitenzahl I.~R. {et~al.}, 2013, MNRAS, 429, 1156

\bibitem[{Shapiro, Giroux \& Babul(1994)Shapiro, Giroux, \& Babul}]{Shapiro1994}
Shapiro P.~R., Giroux M.~L., Babul A., 1994, ApJ, 427, 25

\bibitem[{Sharda, Federrath \& Krumholz(2020)Sharda, Federrath, \& Krumholz}]{Sharda2020}
Sharda P., Federrath C., Krumholz M.~R., 2020, MNRAS, 497, 336

\bibitem[{Simon(2019)}]{Simon2019}
Simon J.~D., 2019, ARA\&A, 57, 375

\bibitem[{Simon {et~al}\mbox{.}(2015)Simon, {Drlica-Wagner}, Li, Nord, Geha, Bechtol, Balbinot, {Buckley-Geer}, Lin, Marshall, Santiago, Strigari, Wang, Wechsler, Yanny, Abbott, Bauer, Bernstein, Bertin, Brooks, Burke, Capozzi, Carnero~Rosell, Carrasco~Kind, D'Andrea, {da Costa}, DePoy, Desai, Diehl, Dodelson, Cunha, Estrada, Evrard, Fausti~Neto, Fernandez, Finley, Flaugher, Frieman, Gaztanaga, Gerdes, Gruen, Gruendl, Honscheid, James, Kent, Kuehn, Kuropatkin, Lahav, Maia, March, Martini, Miller, Miquel, Ogando, Romer, Roodman, Rykoff, Sako, Sanchez, Schubnell, Sevilla, Smith, {Soares-Santos}, Sobreira, Suchyta, Swanson, Tarle, Thaler, Tucker, Vikram, Walker, Wester, \& Collaboration}]{Simon2015}
Simon J.~D. {et~al.}, 2015, ApJ, 808, 95

\bibitem[{Simon \& Geha(2007)}]{Simon2007}
Simon J.~D., Geha M., 2007, ApJ, 670, 313

\bibitem[{Simon {et~al}\mbox{.}(2011)Simon, Geha, Minor, Martinez, Kirby, Bullock, Kaplinghat, Strigari, Willman, Choi, Tollerud, \& Wolf}]{Simon2011}
Simon J.~D. {et~al.}, 2011, ApJ, 733, 46

\bibitem[{Simon {et~al}\mbox{.}(2017)Simon, Li, {Drlica-Wagner}, Bechtol, Marshall, James, Wang, Strigari, Balbinot, Kuehn, Walker, Abbott, Allam, Annis, {Benoit-L{\'e}vy}, Brooks, {Buckley-Geer}, Burke, Carnero~Rosell, Carrasco~Kind, Carretero, Cunha, D'Andrea, {da Costa}, DePoy, Desai, Doel, Fernandez, Flaugher, Frieman, {Garc{\'i}a-Bellido}, Gaztanaga, Goldstein, Gruen, Gutierrez, Kuropatkin, Maia, Martini, Menanteau, Miller, Miquel, Neilsen, Nord, Ogando, Plazas, Romer, Rykoff, Sanchez, Santiago, Scarpine, Schubnell, {Sevilla-Noarbe}, Smith, Sobreira, Suchyta, Swanson, Tarle, Whiteway, Yanny, \& Collaboration}]{Simon2017}
Simon J.~D. {et~al.}, 2017, ApJ, 838, 11

\bibitem[{Simon {et~al}\mbox{.}(2020)Simon, Li, Erkal, Pace, {Drlica-Wagner}, James, Marshall, Bechtol, Hansen, Kuehn, Lidman, Allam, Annis, Avila, Bertin, Brooks, Burke, Rosell, Carrasco~Kind, Carretero, {da Costa}, De~Vicente, Desai, Doel, Eifler, Everett, Fosalba, Frieman, {Garc{\'i}a-Bellido}, Gaztanaga, Gerdes, Gruen, Gruendl, Gschwend, Gutierrez, Hollowood, Honscheid, Krause, Kuropatkin, MacCrann, Maia, March, Miquel, Palmese, {Paz-Chinch{\'o}n}, Plazas, Reil, Roodman, Sanchez, Santiago, Scarpine, Schubnell, Serrano, Smith, Suchyta, Tarle, Walker, \& Collaboration}]{Simon2020}
Simon J.~D. {et~al.}, 2020, ApJ, 892, 137

\bibitem[{Sk{\'u}lad{\'o}ttir {et~al}\mbox{.}(2024)Sk{\'u}lad{\'o}ttir, Koutsouridou, Vanni, Amarsi, Lucchesi, Salvadori, \& Aguado}]{Skuladottir2024}
Sk{\'u}lad{\'o}ttir {\'A}., Koutsouridou I., Vanni I., Amarsi A.~M., Lucchesi R., Salvadori S., Aguado D.~S., 2024, ApJL, 968, L23

\bibitem[{Smith {et~al}\mbox{.}(2023)Smith, Jensen, Roediger, Sestito, Hayes, McConnachie, Cuillandre, Gwyn, Magnier, Chambers, Hammer, Hudson, Martin, Navarro, \& Scott}]{Smith2023BootesV}
Smith S. E.~T. {et~al.}, 2023, AJ, 166, 76

\bibitem[{Somerville(2002)}]{Somerville2002}
Somerville R.~S., 2002, ApJ, 572, L23

\bibitem[{Spencer {et~al}\mbox{.}(2018)Spencer, Mateo, Olszewski, Walker, McConnachie, \& Kirby}]{Spencer2018}
Spencer M.~E., Mateo M., Olszewski E.~W., Walker M.~G., McConnachie A.~W., Kirby E.~N., 2018, AJ, 156, 257

\bibitem[{Spencer {et~al}\mbox{.}(2017)Spencer, Mateo, Walker, \& Olszewski}]{Spencer2017}
Spencer M.~E., Mateo M., Walker M.~G., Olszewski E.~W., 2017, ApJ, 836, 202

\bibitem[{Stacy, Bromm \& Lee(2016)Stacy, Bromm, \& Lee}]{Stacy2016}
Stacy A., Bromm V., Lee A.~T., 2016, MNRAS, 462, 1307

\bibitem[{Stacy, Bromm \& Loeb(2011)Stacy, Bromm, \& Loeb}]{Stacy2011}
Stacy A., Bromm V., Loeb A., 2011, ApJ, 730, L1

\bibitem[{Stanton {et~al}\mbox{.}(2025)Stanton, Cullen, Carnall, Scholte, {Arellano-C{\'o}rdova}, McLeod, Begley, Donnan, Dunlop, Hamadouche, McLure, Shapley, Bondestam, \& Stevenson}]{Stanton2025}
Stanton T.~M. {et~al.}, 2025, MNRAS, 537, 1735

\bibitem[{Stanton {et~al}\mbox{.}(2024)Stanton, Cullen, McLure, Shapley, {Arellano-C{\'o}rdova}, Begley, Amor{\'i}n, Barrufet, Calabr{\`o}, Carnall, Cirasuolo, Dunlop, Donnan, Hamadouche, Liu, McLeod, Pentericci, Pozzetti, Sanders, Scholte, \& Topping}]{Stanton2024b}
Stanton T.~M. {et~al.}, 2024, MNRAS, 532, 3102

\bibitem[{Stanway, Eldridge \& Becker(2016)Stanway, Eldridge, \& Becker}]{Stanway2016}
Stanway E.~R., Eldridge J.~J., Becker G.~D., 2016, MNRAS, 456, 485

\bibitem[{Stetson {et~al}\mbox{.}(2014)Stetson, Fiorentino, Bono, Bernard, Monelli, Iannicola, Gallart, \& Ferraro}]{Stetson2014}
Stetson P.~B., Fiorentino G., Bono G., Bernard E.~J., Monelli M., Iannicola G., Gallart C., Ferraro I., 2014, PASP, 126, 616

\bibitem[{Stiavelli {et~al}\mbox{.}(2023)Stiavelli, Morishita, Chiaberge, Grillo, Rosati, Schuldt, Trenti, \& Treu}]{Stiavelli2023}
Stiavelli M., Morishita T., Chiaberge M., Grillo C., Rosati P., Schuldt S., Trenti M., Treu T., 2023, ApJL, 957, L18

\bibitem[{Stopyra {et~al}\mbox{.}(2021)Stopyra, Pontzen, Peiris, Roth, \& Rey}]{Stopyra2021}
Stopyra S., Pontzen A., Peiris H., Roth N., Rey M.~P., 2021, ApJS, 252, 28

\bibitem[{Storck {et~al}\mbox{.}(2026)Storck, Katz, Devriendt, Slyz, Cadiou, Choustikov, Rey, Saxena, Agertz, \& Kimm}]{Storck2025MegP1}
Storck A. {et~al.}, 2026, MNRAS, 548, stag529

\bibitem[{Sugimura {et~al}\mbox{.}(2020)Sugimura, Matsumoto, Hosokawa, Hirano, \& Omukai}]{Sugimura2020}
Sugimura K., Matsumoto T., Hosokawa T., Hirano S., Omukai K., 2020, ApJL, 892, L14

\bibitem[{Susa, Hasegawa \& Tominaga(2014)Susa, Hasegawa, \& Tominaga}]{Susa2014}
Susa H., Hasegawa K., Tominaga N., 2014, ApJ, 792, 32

\bibitem[{Susa \& Umemura(2004)}]{Susa2004}
Susa H., Umemura M., 2004, ApJ, 600, 1

\bibitem[{Takahashi, Yoshida \& Umeda(2018)Takahashi, Yoshida, \& Umeda}]{Takahashi2018}
Takahashi K., Yoshida T., Umeda H., 2018, ApJ, 857, 111

\bibitem[{Tan {et~al}\mbox{.}(2025)Tan, Cerny, {Drlica-Wagner}, Pace, Geha, Ji, Li, Adam{\'o}w, Anbajagane, Bom, {Carballo-Bello}, Carlin, Chang, Choi, Collins, {Doliva-Dolinsky}, Ferguson, Gruendl, James, Limberg, Navabi, {Mart{\'i}nez-Delgado}, {Mart{\'i}nez-V{\'a}zquez}, Medina, {Mutlu-Pakdil}, Nidever, No{\"e}l, Riley, Sakowska, Sand, Sharp, Stringfellow, Tolley, \& Vivas}]{Tan2025}
Tan C.~Y. {et~al.}, 2025, ApJ, 979, 176

\bibitem[{Tarumi, Hartwig \& Magg(2020)Tarumi, Hartwig, \& Magg}]{Tarumi2020}
Tarumi Y., Hartwig T., Magg M., 2020, ApJ, 897

\bibitem[{Teyssier(2002)}]{Teyssier2002}
Teyssier R., 2002, A\&A, 385, 337

\bibitem[{Thibodeaux {et~al}\mbox{.}(2024)Thibodeaux, Ji, Cerny, Kirby, \& Simon}]{Thibodeaux2024}
Thibodeaux P., Ji A.~P., Cerny W., Kirby E.~N., Simon J.~D., 2024, OpJA, 7, 66

\bibitem[{Tolstoy, Hill \& Tosi(2009)Tolstoy, Hill, \& Tosi}]{Tolstoy2009}
Tolstoy E., Hill V., Tosi M., 2009, ARA\&A, 47, 371

\bibitem[{Topping {et~al}\mbox{.}(2025)Topping, Stark, Senchyna, Chen, Zitrin, Endsley, Charlot, Furtak, Maseda, Plat, Smit, Mainali, Chevallard, Molyneux, \& Rigby}]{Topping2024NEmitterSample2}
Topping M.~W. {et~al.}, 2025, ApJ, 980, 225

\bibitem[{Topping {et~al}\mbox{.}(2024)Topping, Stark, Senchyna, Plat, Zitrin, Endsley, Charlot, Furtak, Maseda, Smit, Mainali, Chevallard, Molyneux, \& Rigby}]{Topping2024NEmitterSample}
Topping M.~W. {et~al.}, 2024, MNRAS, 529, 3301

\bibitem[{Torrealba {et~al}\mbox{.}(2018)Torrealba, Belokurov, Koposov, Bechtol, {Drlica-Wagner}, Olsen, Vivas, Yanny, Jethwa, Walker, Li, Allam, Conn, Gallart, Gruendl, James, Johnson, Kuehn, Kuropatkin, Martin, {Martinez-Delgado}, Nidever, No{\"e}l, Simon, Stringfellow, \& Tucker}]{Torrealba2018}
Torrealba G. {et~al.}, 2018, MNRAS, 475, 5085

\bibitem[{Torrealba {et~al}\mbox{.}(2016{\natexlab{a}})Torrealba, Koposov, Belokurov, \& Irwin}]{Torrealba2016}
Torrealba G., Koposov S.~E., Belokurov V., Irwin M., 2016{\natexlab{a}}, MNRAS, 459, 2370

\bibitem[{Torrealba {et~al}\mbox{.}(2016{\natexlab{b}})Torrealba, Koposov, Belokurov, Irwin, Collins, Spencer, Ibata, Mateo, Bonaca, \& Jethwa}]{Torrealba2016Aqua2}
Torrealba G. {et~al.}, 2016{\natexlab{b}}, MNRAS, 463, 712

\bibitem[{Trebitsch {et~al}\mbox{.}(2017)Trebitsch, Blaizot, Rosdahl, Devriendt, \& Slyz}]{Trebitsch2017}
Trebitsch M., Blaizot J., Rosdahl J., Devriendt J., Slyz A., 2017, MNRAS, 470, 224

\bibitem[{Tseliakhovich \& Hirata(2010)}]{Tseliakhovich2010}
Tseliakhovich D., Hirata C., 2010, Phys. Rev. D, 82, 083520

\bibitem[{Turk, Abel \& O'Shea(2009)Turk, Abel, \& O'Shea}]{Turk2009}
Turk M.~J., Abel T., O'Shea B., 2009, Science, 325, 601

\bibitem[{Turk {et~al}\mbox{.}(2011)Turk, Smith, Oishi, Skory, Skillman, Abel, \& Norman}]{Turk2011}
Turk M.~J., Smith B.~D., Oishi J.~S., Skory S., Skillman S.~W., Abel T., Norman M.~L., 2011, ApJS, 192, 9

\bibitem[{{van der Walt}, Colbert \& Varoquaux(2011){van der Walt}, Colbert, \& Varoquaux}]{vanderWalt2011}
{van der Walt} S., Colbert S.~C., Varoquaux G., 2011, Comput. Sci. Eng., 13, 22

\bibitem[{Vink(2023)}]{Vink2023Nitrogen}
Vink J.~S., 2023, A\&A, 679, L9

\bibitem[{Virtanen {et~al}\mbox{.}(2020)Virtanen, Gommers, Oliphant, Haberland, Reddy, Cournapeau, Burovski, Peterson, Weckesser, Bright, {van der Walt}, Brett, Wilson, Millman, Mayorov, Nelson, Jones, Kern, Larson, Carey, Polat, Feng, Moore, VanderPlas, Laxalde, Perktold, Cimrman, Henriksen, Quintero, Harris, Archibald, Ribeiro, Pedregosa, \& {van Mulbregt}}]{Virtanen2020}
Virtanen P. {et~al.}, 2020, Nat Methods, 17, 261

\bibitem[{Vivas, {Mart{\'i}nez-V{\'a}zquez} \& Walker(2020)Vivas, {Mart{\'i}nez-V{\'a}zquez}, \& Walker}]{Vivas2020}
Vivas A.~K., {Mart{\'i}nez-V{\'a}zquez} C., Walker A.~R., 2020, ApJS, 247, 35

\bibitem[{Vivas {et~al}\mbox{.}(2022)Vivas, {Mart{\'i}nez-V{\'a}zquez}, Walker, Belokurov, Li, \& Erkal}]{Vivas2022}
Vivas A.~K., {Mart{\'i}nez-V{\'a}zquez} C.~E., Walker A.~R., Belokurov V., Li T.~S., Erkal D., 2022, ApJ, 926, 78

\bibitem[{Vivas {et~al}\mbox{.}(2016)Vivas, Olsen, Blum, Nidever, Walker, Martin, Besla, Gallart, {van der Marel}, Majewski, Kaleida, Mu{\~n}oz, Saha, Conn, \& Jin}]{Vivas2016}
Vivas A.~K. {et~al.}, 2016, AJ, 151, 118

\bibitem[{Walker {et~al}\mbox{.}(2015)Walker, Mateo, Olszewski, Bailey, Koposov, Belokurov, \& Evans}]{Walker2015a}
Walker M.~G., Mateo M., Olszewski E.~W., Bailey J.~I., Koposov S.~E., Belokurov V., Evans N.~W., 2015, ApJ, 808, 108

\bibitem[{Walker {et~al}\mbox{.}(2009)Walker, Mateo, Olszewski, Sen, \& Woodroofe}]{Walker2009}
Walker M.~G., Mateo M., Olszewski E.~W., Sen B., Woodroofe M., 2009, AJ, 137, 3109

\bibitem[{Walker, Olszewski \& Mateo(2015)Walker, Olszewski, \& Mateo}]{Walker2015}
Walker M.~G., Olszewski E.~W., Mateo M., 2015, MNRAS, 448, 2717

\bibitem[{Walsh {et~al}\mbox{.}(2008)Walsh, Willman, Sand, Harris, Seth, Zaritsky, \& Jerjen}]{Walsh2008}
Walsh S.~M., Willman B., Sand D., Harris J., Seth A., Zaritsky D., Jerjen H., 2008, ApJ, 688, 245

\bibitem[{Wang {et~al}\mbox{.}(2019)Wang, {de Boer}, Pieres, Li, {Drlica-Wagner}, Koposov, Vivas, Pace, Santiago, Walker, Tucker, Strigari, Marshall, Yanny, DePoy, Bechtol, Roodman, Abbott, Abdalla, Allam, Annis, Avila, Bertin, Brooks, Burke, Carnero~Rosell, Carrasco~Kind, Cunha, D'Andrea, {da Costa}, De~Vicente, Desai, Eifler, Estrada, Flaugher, Frieman, {Garc{\'i}a-Bellido}, Gerdes, Gruen, Gruendl, Gutierrez, Hollowood, Honscheid, James, Kuehn, Kuropatkin, Lahav, Maia, Miquel, Sanchez, Scarpine, {Sevilla-Noarbe}, Smith, Smith, Sobreira, Suchyta, Swanson, Tarle, \& Collaboration}]{Wang2019}
Wang M.~Y. {et~al.}, 2019, ApJ, 881, 118

\bibitem[{Weisz {et~al}\mbox{.}(2014)Weisz, Dolphin, Skillman, Holtzman, Gilbert, Dalcanton, \& Williams}]{Weisz2014Reionization}
Weisz D.~R., Dolphin A.~E., Skillman E.~D., Holtzman J., Gilbert K.~M., Dalcanton J.~J., Williams B.~F., 2014, ApJ, 789, 148

\bibitem[{Wempe {et~al}\mbox{.}(2024)Wempe, Lavaux, White, Helmi, Jasche, \& Stopyra}]{Wempe2024}
Wempe E., Lavaux G., White S. D.~M., Helmi A., Jasche J., Stopyra S., 2024, A\&A, 691

\bibitem[{Whalen {et~al}\mbox{.}(2008)Whalen, {van Veelen}, O'Shea, \& Norman}]{Whalen2008}
Whalen D., {van Veelen} B., O'Shea B.~W., Norman M.~L., 2008, ApJ, 682, 49

\bibitem[{Wheeler {et~al}\mbox{.}(2019)Wheeler, Hopkins, Pace, {Garrison-Kimmel}, {Boylan-Kolchin}, Wetzel, Bullock, Kere{\v s}, {Faucher-Gigu{\`e}re}, \& Quataert}]{Wheeler2019}
Wheeler C. {et~al.}, 2019, MNRAS, 490, 4447

\bibitem[{Wheeler {et~al}\mbox{.}(2025)Wheeler, Kravtsov, Chiti, Katz, \& Semenov}]{Wheeler2025IGMPlateau}
Wheeler V., Kravtsov A., Chiti A., Katz H., Semenov V.~A., 2025, OpJA, 8, 151

\bibitem[{Willman {et~al}\mbox{.}(2011)Willman, Geha, Strader, Strigari, Simon, Kirby, Ho, \& Warres}]{Willman2011}
Willman B., Geha M., Strader J., Strigari L.~E., Simon J.~D., Kirby E., Ho N., Warres A., 2011, AJ, 142, 128

\bibitem[{Willman {et~al}\mbox{.}(2006)Willman, Masjedi, Hogg, Dalcanton, {Martinez-Delgado}, Blanton, West, Dotter, \& Chaboyer}]{Willman2006}
Willman B. {et~al.}, 2006, arXiv e-prints, astro

\bibitem[{Wise \& Cen(2009)}]{Wise2009}
Wise J.~H., Cen R., 2009, ApJ, 693, 984

\bibitem[{Wise {et~al}\mbox{.}(2014)Wise, Demchenko, Halicek, Norman, Turk, Abel, \& Smith}]{Wise2014}
Wise J.~H., Demchenko V.~G., Halicek M.~T., Norman M.~L., Turk M.~J., Abel T., Smith B.~D., 2014, MNRAS, 442, 2560

\bibitem[{Wise {et~al}\mbox{.}(2012)Wise, Turk, Norman, \& Abel}]{Wise2012}
Wise J.~H., Turk M.~J., Norman M.~L., Abel T., 2012, ApJ, 745, 50

\bibitem[{Wollenberg {et~al}\mbox{.}(2020)Wollenberg, Glover, Clark, \& Klessen}]{Wollenberg2020}
Wollenberg K. M.~J., Glover S. C.~O., Clark P.~C., Klessen R.~S., 2020, MNRAS, 494, 1871

\bibitem[{Xing {et~al}\mbox{.}(2023)Xing, Zhao, Liu, Heger, Han, Aoki, Chen, Ishigaki, Li, \& Zhao}]{Xing2023}
Xing Q.-F. {et~al.}, 2023, Nature, 618, 712

\bibitem[{Xu {et~al}\mbox{.}(2016)Xu, Norman, O'Shea, \& Wise}]{Xu2016}
Xu H., Norman M.~L., O'Shea B.~W., Wise J.~H., 2016, ApJ, 823, 140

\bibitem[{Xu, Wise \& Norman(2013)Xu, Wise, \& Norman}]{Xu2013}
Xu H., Wise J.~H., Norman M.~L., 2013, ApJ, 773, 83

\bibitem[{Zier {et~al}\mbox{.}(2025)Zier, Kannan, Smith, Puchwein, Vogelsberger, Borrow, Garaldi, Keating, McClymont, Shen, \& Hernquist}]{Zier2025a}
Zier O. {et~al.}, 2025, MNRAS, 544, 410

\end{thebibliography}



\appendix

\section{A. Pop.~III IMF variations} \label{app:popiii}

\textsc{megatron} simulations assume a Pop.~III IMF according to the functional form from \citet{Wise2012} given by
\begin{equation}
\frac{\mathrm{d}N}{\mathrm{d}\log M} \propto M^{-1.3} \exp \left[ -\left( \frac{M_{\rm char}}{M} \right)^{1.6} \right].
\end{equation}
As in \citet{Wise2012}, in this work, we use a characteristic mass of $M_{\rm char} = 100 \, \Msol$. When a star formation site is flagged suitable for star formation, we randomly draw from the IMF following a Poisson process first described in \citet{Rasera2006} (see also \citealt{Kang2025} for a detailed discussion). 

\begin{figure}
  \centering
    \includegraphics[width=0.49\columnwidth]{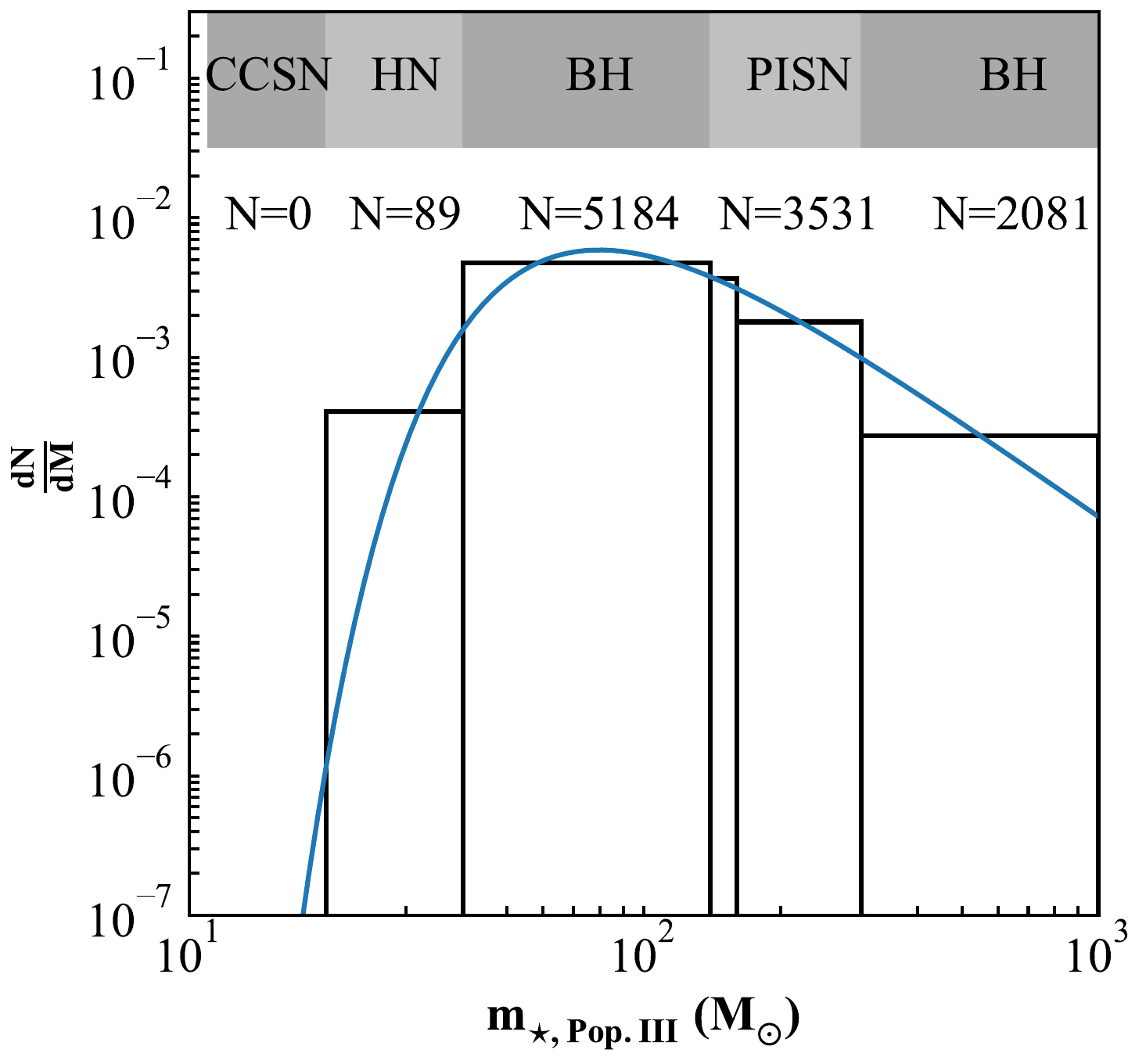}
    \includegraphics[width=0.45\columnwidth]{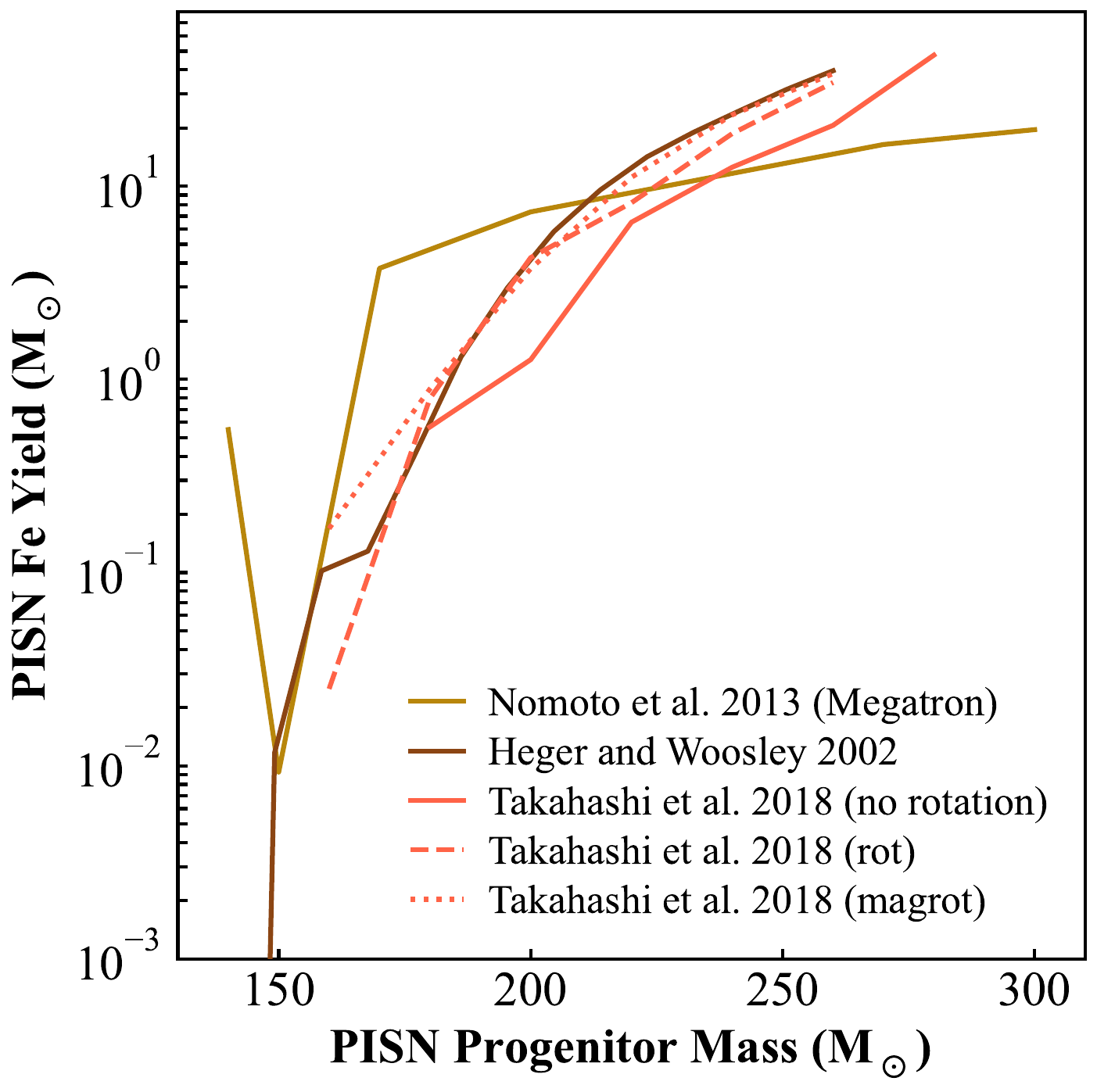}

    \caption{(Left): Pop.~III IMF assumed in \textsc{megatron} simulations, following the functional form from \citet{Wise2012} with a characteristic mass of $M_{\rm char} = 100 \, \Msol$. The line shows the analytical form, while the bins shows the realised IMF across all \textsc{megatron} simulations and galaxies. The IMF peaks at $\approx 100 \, \Msol$, with a long tail towards higher masses. With this choice of IMF and stellar evolution model, most Pop.~III stars end up collapsing directly into black holes or exploding as PISNe. (Right): Comparison of Fe yields from PISNe in different nucleosynthesis models (\citealt{Nomoto2013, Heger2002, Takahashi2018}). 
    }
    \label{fig:popiii}
\end{figure}

Figure~\ref{fig:popiii} shows the analytical form of this IMF, alongside the binned numbers of Pop.~III explosions across all \textsc{megatron} simulations and galaxies. As expected from the analytical form, the \textsc{megatron} Pop.~III IMF peaks at $\approx 100 \, \Msol$, with a long tail towards higher masses. Given the assumption of our stellar evolution model, most Pop.~III stars end up collapsing directly into black holes (central bins) without contributing to chemical enrichment or mechanical feedback. The second most common outcome are PISNe explosions, with Pop.~III CCSN and HNe being much rarer. These findings across all galaxies are in line with the results in Section~\ref{sec:popiii}. 

In Figure~\ref{fig:popiii}, we also show a comparison of the yields of Pop.~III PISNe in the model we have adopted in \textsc{megatron} (\citealt{Nomoto2013}) with other models in the literature (e.g.\ \citealt{Heger2002, Takahashi2018}, the latter showing their models with and without rotation, and with magnetic fields). Note that the models do not always cover the same range of progenitor masses, as different stellar evolution assumptions will lead to different explodability criteria. Furthermore, models do not necessarily lead to the same explosion energy for a given progenitor mass, thus affecting metal retention and the effective yield retained by a minihalo (Section~\ref{sec:popiii:yieldminihalo}).

Nonetheless, the comparison is useful to highlight the key uncertainties associated to our findings. For example, our chosen yields produce more iron per PISN explosion than other models for low-mass progenitors, although the trend reverses for high-mass progenitors. For our choice of IMF which preferentially produces low-mass PISNe, we thus hypothesize that other literature models (e.g.\ \citealt{Heger2002}) would likely lead to a more pronounced tail of iron-poor galaxies and a less pronounced plateau. 

The exact balance between the plateau and the tail is thus likely to be sensitive to this coupling between choices in star formation (e.g.\ the IMF) and stellar evolution (e.g.\ progenitor type, explodability and nucleosynthesis). Given the non-linearity of these processes, unravelling this requires a more systematic exploration of parameter space and detailed simulations that we plan to tackle in future work.

\section{B. Data compilation} \label{app:data}

Dwarf galaxy data presented here is drawn from the Local Volume Database \citep{Pace2025LVDatabase} which compiles UFD data (including the LMC system) from \citet{Bellazzini2005, Belokurov2007,  Bhardwaj2024, Boettcher2013, Bruce2023, Cantu2021, Carlin2009, Carlin2017, Carlin2018, Casey2025, Cerny2021a, Cerny2021, Cerny2023DELVE, Cerny2023DELVE6, Cerny2023PegIV, Cerny2025, Chiti2021, Chiti2023, Chiti2022GrusI, Cicuendez2018, Correnti2009, Crnojevic2016, DallOra2006, DallOra2012, Drlica-Wagner2015, Fritz2019, Garofalo2013, Garofalo2025, Greco2008, Hansen2024, Heiger2024, Homma2018, Homma2019, Homma2024, Jenkins2021, Ji2021, Karczmarek2015, Kim2015Horologium, Kim2016PegIII, Kirby2013, Kirby2015, Kirby2017, Koposov2011, Koposov2015, Koposov2015RetII, Koposov2018, Kuehn2008, Lee2009, Li2017, Li2018, Longeard2018, Martinez-Vazquez2015, Martinez-Vazquez2019, Martinez-Vazquez2021, Mateo2008, McConnachie2012, Medina2018, Moskowitz2020, Munoz2018, Musella2009, Mutlu-Pakdil2018, Mutlu-Pakdil2020, Oakes2022, Pace2020, Richstein2022, Richstein2024, Sand2012, Simon2007, Simon2011, Simon2015, Simon2017, Simon2019, Simon2020, Smith2023BootesV, Spencer2017, Spencer2018, Stetson2014, Tan2025, Torrealba2016Aqua2, Torrealba2016, Torrealba2018, Vivas2016, Vivas2020, Vivas2022, Walker2009, Walker2015, Walker2015a, Walsh2008, Wang2019, Willman2006, Willman2011}.

\section{C. Externally-enriched dwarfs} \label{app:externalpollution}

Figure~\ref{fig:popiiinumbers} shows that a small fraction of faint dwarfs have transitioned to Pop.~II star formation without ever hosting a Pop.~III star. 

\begin{figure}
  \centering
    \includegraphics[width=\columnwidth]{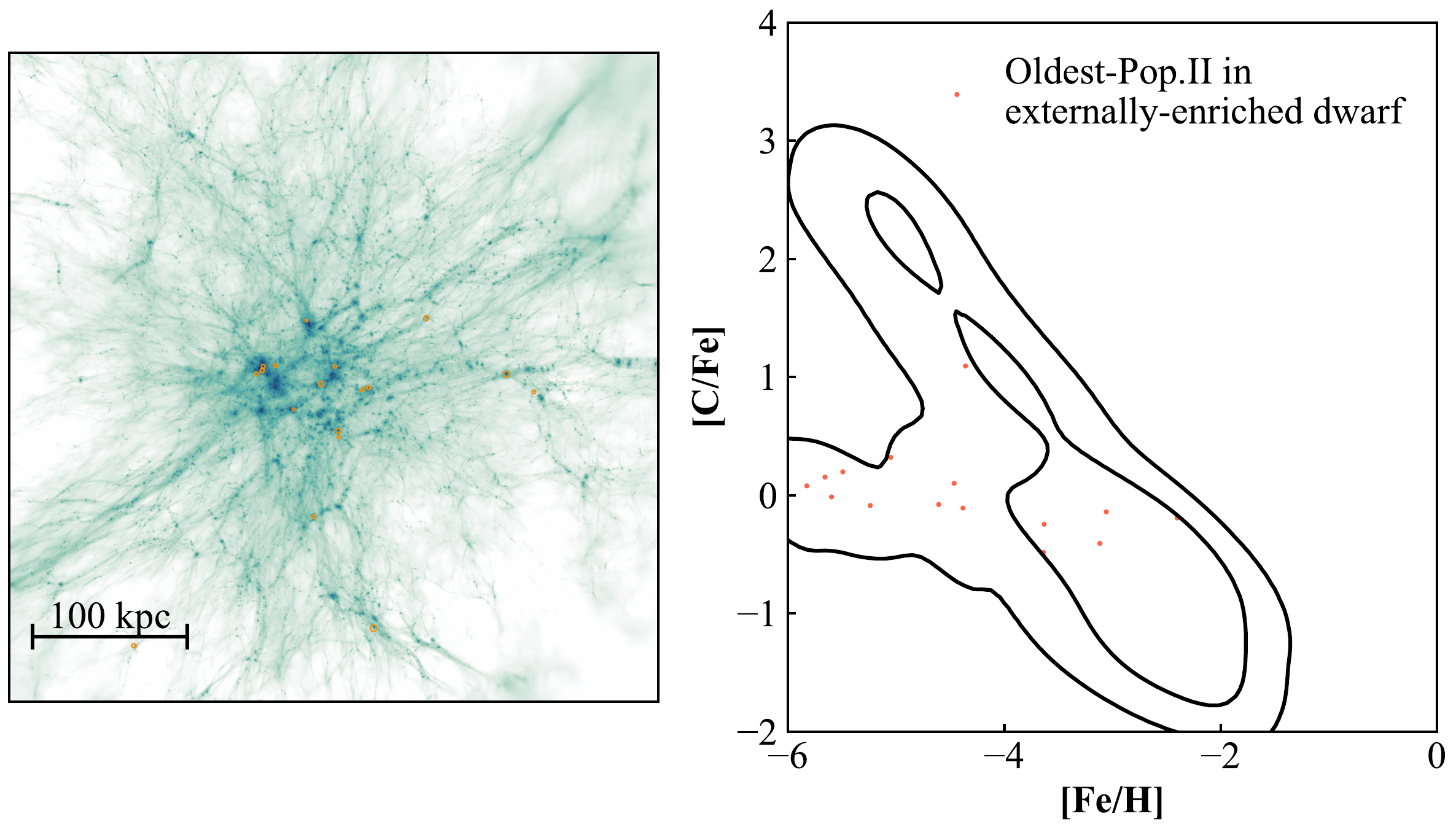}
    \caption{The spatial (left) and chemical (right) distribution of externally-enriched dwarfs that contain no Pop.~III stars within them. These dwarfs are compatible with the rest of the population in both spaces, reflecting the stochasticity of external enrichment by either Pop.~III- or Pop.~II-driven outflows.
    }
    \label{fig:externalpollution}
\end{figure}

Figure~\ref{fig:externalpollution} shows their spatial distribution across the simulated volume (left panel) and the carbon abundances of the oldest Pop.~II stellar population in these externally-enriched dwarfs (i.e.\ their first star formed since no Pop.~III are present; right panel). Black contours show the rest of the \textsc{megatron} population. Our externally-enriched dwarfs are not particularly clustered in space, tracking the broader large-scale structure. The carbon abundances of their oldest Pop.~II stars (red dots) are also representative of the rest of the dwarf population (black contour), with most stars having $ -0.5 \leq \cfe \approx 0.5$. With our assumptions, such ratios are typical for yields of low-mass Pop.~III PISNe ($\cfe \approx -0.5$) and low-metallicity Pop.~II CCSNe ($\cfe \approx 0.5$). We conclude that, occasionally, a star-forming galaxy can enrich a neighbouring minihalo to the critical metallicity for Pop.~II star formation before it forms its own Pop.~III star. Determining whether the ejecta is primarily from Pop.~II or Pop.~III stars will require more detailed tracking of the metal flows across our simulated volume, which we plan to tackle in future work using our Lagrangian particle tracers (\citealt{Cadiou2019}). 

Using the same methodology as in Section~\ref{sec:lowzconnection:mwtidal} and across all four simulations, we quantify that only 3 externally-enriched dwarfs out of 22 will survive at $z=0$ after processing by the tidal field of the Milky-Way-like host. This fraction ($13\%$) is small but not negligible, suggesting that several observed UFDs around the Milky Way could have been externally-enriched at high redshift. However, since their chemical abundances are indistinguishable from the rest of the population, the observational prospects to identify them would be challenging.

\section{D. Reionization history of \textsc{megatron} simulations} \label{app:reionization}

\begin{figure}
  \centering
    \includegraphics[width=0.5\textwidth]{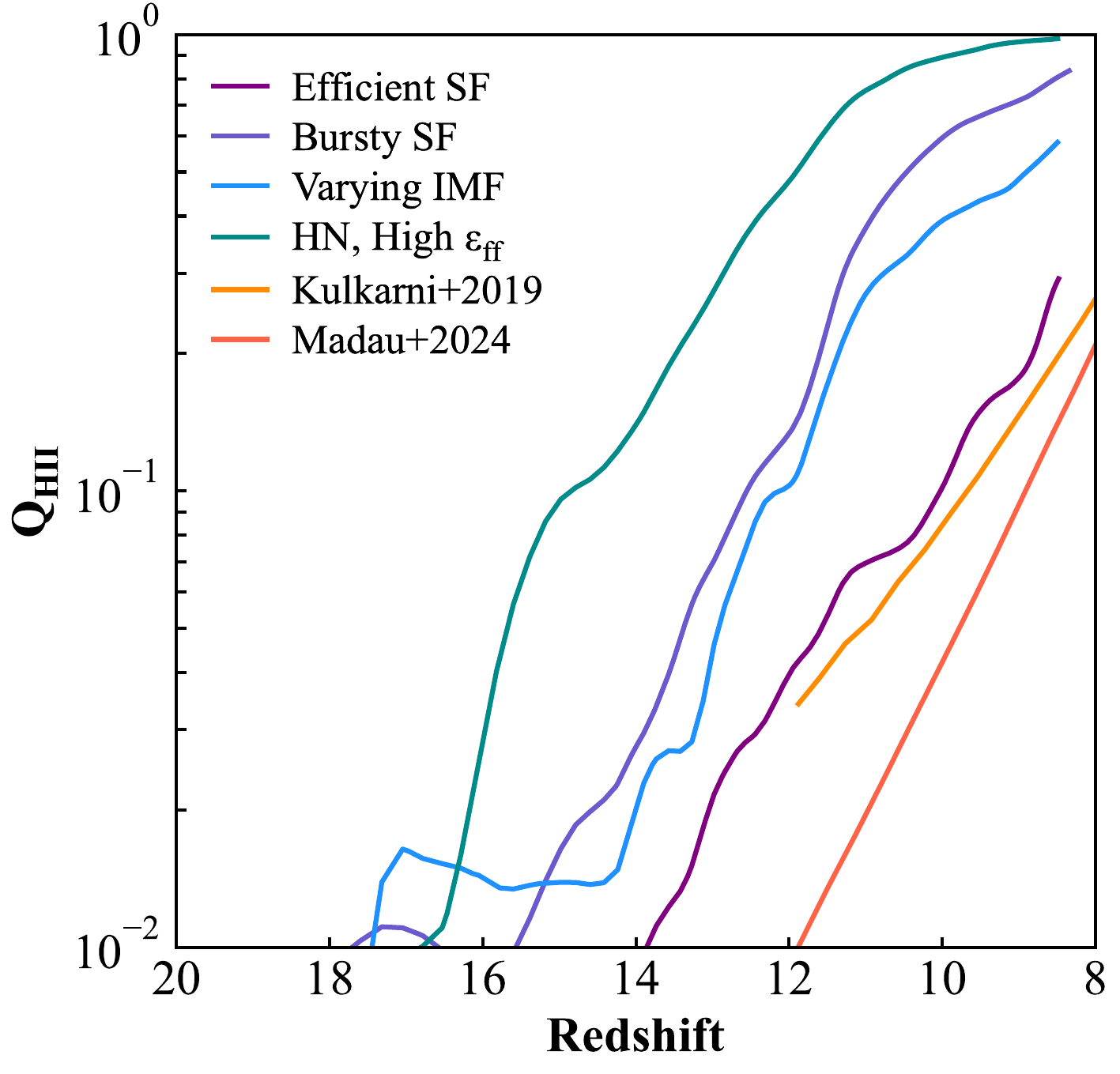}
    \caption{The volume-averaged ionized hydrogen fraction as a function of redshift in the four \textsc{megatron} simulations. All four simulations are more reionized than the cosmic average (e.g.\ \citealt{Kulkarni2019ReionizationHistory, Madau2024}), with the differences between models being driven by feedback-modulated escape fractions of ionizing photons.
    }
    \label{fig:reionization}
\end{figure}

We show the reionization history of the four \textsc{megatron} simulations in Figure~\ref{fig:reionization}. We compute the ionized fraction of the gas within the high-resolution region of the simulation and show the volume-weighted ionized fractions as a function of redshift. All simulations are more reionized than the cosmic average inferred from data (e.g.\ \citealt{Kulkarni2019ReionizationHistory, Madau2024}), as expected from sampling the overdense environment that will collapse into a Milky-Way-like galaxy (e.g. \citealt{Ocvirk2014, Dixon2018, Attard2026}).

We observe significant differences between models, with the `Efficient SF' simulation being the least reionized at a given redshift, and the `HN, $\epsilon_{\rm eff}$' simulation being completely reionized by $z\approx8$. These differences are not driven by the number of ionizing photons produced, but by feedback-modulated escape fractions. For example, the `Efficient SF' simulation produces significantly more ionizing photons due to its higher star formation rates (\citealt{Katz2025MegP1}, fig.\ 6). But the higher-density ISM drives a much lower escape fraction of these ionizing photons (see \citealt{Choustikov2025MegP1}, fig.\ 10), in turn reducing the ability to reionize the surrounding IGM. These findings align with other numerical simulations predicting that the escape of ionizing photons is feedback regulated and strongly dependent on the ISM structure, with mechanical feedback from SNe playing a larger role than radiative feedback (e.g.\ \citealt{Wise2009, Paardekooper2011, Paardekooper2015, Trebitsch2017, Kimm2017, Rosdahl2018}). Differences in reionization history explains why the fraction of dwarf galaxies that are reionization-quenched varies across simulations in Section~\ref{sec:lowzconnection:sfhs} and is highest in the `HN, $\epsilon_{\rm eff}$' simulation.

\section{E. Iron distribution functions} \label{app:mdfs}

Figure~\ref{fig:ufdplateau} and Figure~\ref{fig:chemistry} shows the galaxy-averaged abundances of faint dwarfs. However, each galaxy contains a metal distribution function (MDF) from which this average is derived. The shape of this distribution and the number of stars for which we have spectroscopic abundance measurements can introduce biases in the measurement of $\averagefeh$ (see e.g.\ \citealt{Andersson2025} and \citealt{Sandford2025} for theoretical and observational perspectives, respectively). 
 
\begin{figure}
  \centering
    \includegraphics[width=\textwidth]{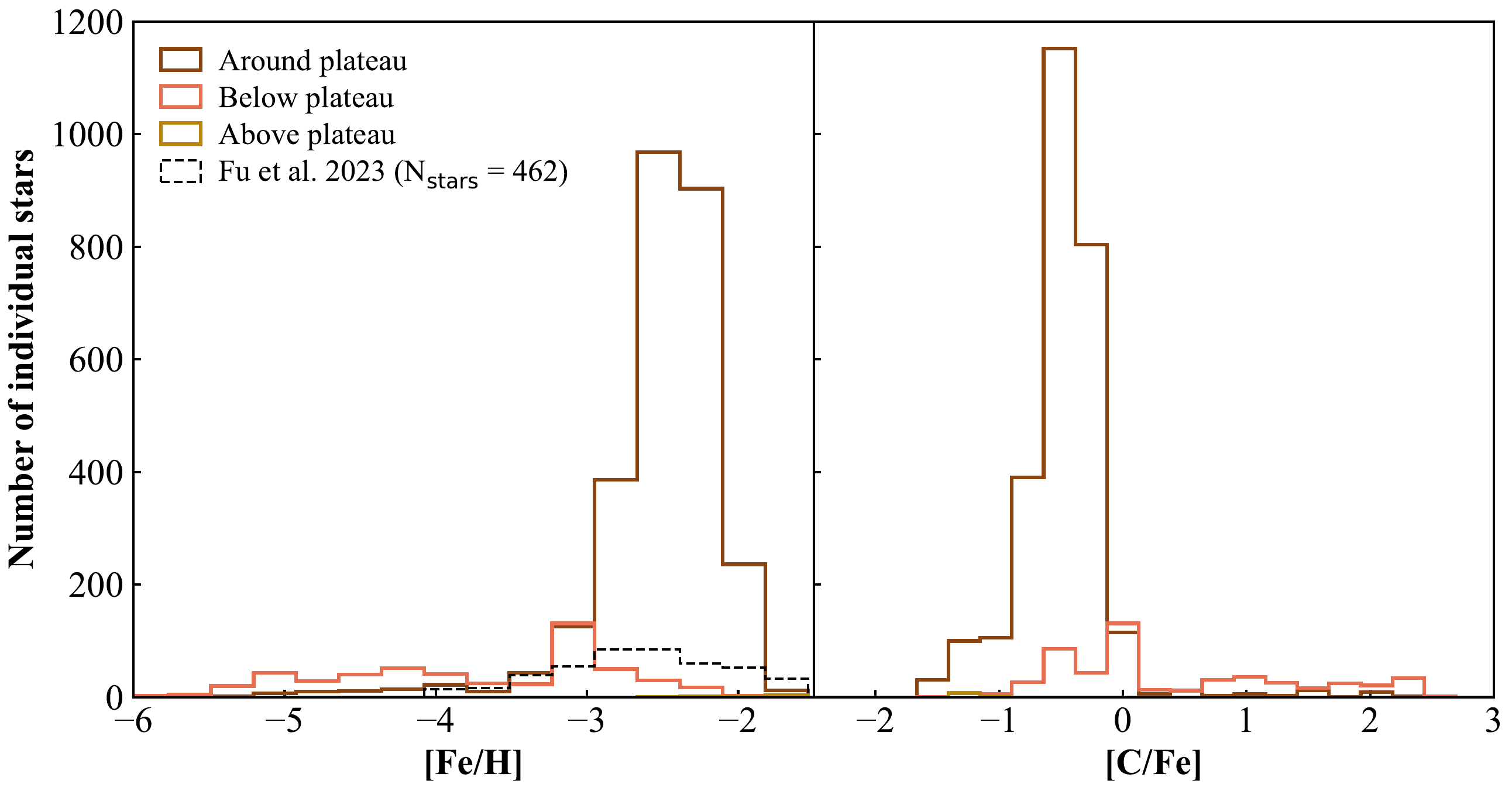}
    \caption{The iron distribution functions of individual stars within faint dwarf galaxies on, above and below the plateau. As expected, dwarfs on the plateau have a peak around $\feh \approx -2.5$, but also show a tail of iron-poor stars down to $\feh \leq -4.0$. Similarly, dwarfs with an average metallicity below the plateau can still have individual stars with higher iron contents.
    }

    \label{fig:mdfs}
\end{figure}

Figure~\ref{fig:mdfs} exemplifies this point, showing the composite iron (left) and carbon (right) distribution functions of individual stars within dwarfs with $\Mstar \leq 10^5 \, \Msol$ that are on, above, and below the iron plateau, alongside the data from \citet{Fu2023}. The shape of the distribution of individual stars across all galaxies clearly follows the shape of the distribution of galaxy-averaged $\averagefeh$ in Figure~\ref{fig:ufdplateau}, as expected given the large numbers of both stars and galaxies involved. Also expected, dwarfs on the plateau have a peak around $\feh \approx -2.5$. But the spread around this mean is also consistent with observed data in the Milky Way (dashed line). The standard deviation in the composite MDF of dwarfs on the plateau to have $\sigma_{\text{Fe}} = 0.43$ is the same as for all dwarfs combined and is in excellent qualitative agreement with observed standard deviations which vary between $\approx 0.3$ dex to $\approx 0.5$ dex from dwarf to dwarf (e.g.\ \citealt{Simon2019, Fu2023}).

Figure~\ref{fig:mdfs} also shows that plateau dwarfs have a tail of iron-poor stars down to $\feh \leq -4.0$ similar to the object recently discovered by \citet{Chiti2025}. Similarly, dwarfs with an average metallicity below the plateau can still have individual stars with plateau-like iron contents. The same picture holds for $\cfe$ (right panel), where significant overlap exists between the distributions of dwarfs on, above and below the plateau. 

From this, it is clear that a complete census of MDFs and a careful unpicking of their distribution over the full population of dwarfs will be key to accurately measure their galaxy-averaged $\averagefeh$ and capture the exact relationship between Pop.~III enrichment and UFD chemistry. Small-sample statistics and individual detections could otherwise bias the interpretation.

\label{lastpage}
\end{document}